\documentclass[3p]{elsarticle}
\usepackage{aasmacros}
\usepackage{amsmath, amssymb}
\usepackage{multicol}
\usepackage{xcolor}
\usepackage{xspace}
\usepackage{enumerate}
\usepackage{tablefootnote}
\usepackage{threeparttable}
\usepackage{booktabs}
\usepackage[english]{babel}
\usepackage[section]{placeins}
\usepackage[pdfpagelabels, plainpages=false, linktocpage=true]{hyperref}

\hypersetup{
   colorlinks   = true, 
   urlcolor     = blue, 
   linkcolor    = MidnightBlue, 
   citecolor    = red 
}

\journal{Astroparticle Physics}

\makeatletter
\let\oldtheequation\theequation
\renewcommand\tagform@[1]{\maketag@@@{\ignorespaces#1\unskip\@@italiccorr}}
\renewcommand\theequation{(\oldtheequation)}

\makeatother
\addto\extrasenglish{%
  \def\equationautorefname{Eq{.}}%
  \def\tableautorefname{Table}%
  \def\figureautorefname{Fig{.}}%
}

\newcommand{\Autoref}[1]{%
  \begingroup%
  \renewcommand\equationautorefname{Equation}%
  \renewcommand\tableautorefname{Table}%
  \renewcommand\figureautorefname{Figure}%
  \autoref{#1}%
  \endgroup%
}

\begin{document}

\title{Scrutinizing FR~0 Radio Galaxies as Ultra-High-Energy Cosmic Ray Source Candidates}

\author[1]{Lukas Merten\corref{cor1}}
\ead{lukas.merten@uibk.ac.at}
\author[1]{Margot Boughelilba}
\ead{margot.boughelilba@uibk.ac.at}
\author[1]{Anita Reimer}
\ead{anita.reimer@uibk.ac.at}
\author[1]{Paolo Da Vela}
\ead{paolo.da-vela@uibk.ac.at}
\author[2]{Serguei Vorobiov}
\ead{serguei.vorobiov@gmail.com}
\author[3]{Fabrizio Tavecchio}
\ead{fabrizio.tavecchio@inaf.it}
\author[4, 5]{Giacomo Bonnoli}
\ead{bonnoli@iaa.es}
\author[2]{Jon Paul Lundquist}
\ead{jplundquist@gmail.com}
\author[3]{Chiara Righi}
\ead{chiara.righi@inaf.it}

\cortext[cor1]{Corresponding author}

\address[1]{University of Innsbruck, Institute for Astro and Particle Physics \\
Technikerstraße 25, 6020 Innsbruck, Austria}
\address[2]{Center for Astrophysics and Cosmology (CAC), University of Nova Gorica\\
 Vipavska 13, SI-5000 Nova Gorica, Slovenia}
\address[3]{Astronomical Observatory of Brera\\
Via Brera 28, 20121 Milano, Italy}
\address[4]{Università degli Studi di Siena\\
Banchi di Sotto 55, 53100 Siena, Italy}
\address[5]{now at: Instituto de Astrofísica de Andalucía (CSIC)\\
Apartado 3004, E-18080 Granada, Spain }

\begin{abstract}
Fanaroff-Riley (FR) 0 radio galaxies compose a new class of radio galaxies, which are usually weaker but much more numerous than the well-established class of FR~1 and FR~2 galaxies. The latter classes have been proposed as sources of the ultra-high-energy cosmic rays (UHECRs) with energies reaching up to ${\sim}10^{20}$~eV. Based on this conjecture, the possibility of UHECR acceleration and survival in an FR 0 source environment is examined in this work.

In doing so, an average spectral energy distribution (SED) based on data from the FR~0 catalog (FR0\emph{CAT}) is compiled. The resulting photon fields are used as targets for UHECRs, which suffer from electromagnetic pair production, photo-disintegration, photo-meson production losses, and synchrotron radiation. Multiple mechanisms are discussed to assess the UHECR acceleration probability, including Fermi-I order and gradual shear accelerations, and particle escape from the source region.

This work shows that in a hybrid scenario, combining Fermi and shear accelerations, FR~0 galaxies can contribute to the observed UHECR flux, as long as~$\Gamma_\mathrm{j}\gtrsim 1.6$, where shear acceleration starts to dominate over escape. Even in less optimistic scenarios, FR 0s can be expected to contribute to the cosmic-ray flux between the knee and the ankle. Our results are relatively robust with respect to the realized magnetic turbulence model and the speed of the accelerating shocks.
\end{abstract}


\maketitle

\section{Introduction} 
\label{sec:intro}

The idea of radio galaxies~(RGs) contributing to the overall flux of ultra-high-energy cosmic rays~(UHECRs), as measured on Earth, is several decades old and has been revisited many times in the past (e.g.,~\cite{Biermann1987,Rachen1993,Biermann1998,Dermer2009,Eichmann2018}). Indeed, radio galaxies are among those source populations that are considered capable of accelerating charged nuclei to energies up to~${\sim}10^{20}$~eV, and hence, they fulfill the famous Hillas criterion~\citep{Hillas1984} as well as energetics criteria~\cite{NaganoWatson, Bhattacharjee2000} --- as do their beamed counterparts, blazars. In this latter case, a few powerful sources are considered as the origin of detected UHECRs, and a correspondingly anisotropic cosmic ray (CR) arrival distribution is expected (e.g.,~\cite{Rachen2019}). Indeed, recently the Pierre Auger Observatory (PAO) collaboration reported a~${\sim}5\sigma$ dipole signal above 8~EeV~\citep{Auger2017}.

Another search strategy for UHECR sources considers instead a sufficiently numerous class of (comparatively) low-luminosity sources more isotropically distributed in the nearby universe, such as starburst galaxies (see, e.g.,~\cite{AugerStarburst, Attallah2018}) or radio galaxies. The crucial influence of diffuse CR transport in the intergalactic magnetic field on the expected arrival distribution has been examined, e.g.,\ in~\citep{Eichmann2018, Wittkowski2018, Dundovic2019}. Studies of either scenario have typically focused on CR transport between the source distribution and Earth, thereby providing constraints on the required spectra and composition of UHECRs that have escaped the sources (e.g.,~\cite{Taylor2015, Das2019}). Only a few provide a study of acceleration and energy losses of CRs in the putative CR source environments (e.g.,~\cite{Allard2010,tavecchio2019}), which is required to answer questions regarding the energization and survival probability of charged particles in a particular source class.

Radio galaxies belong to the radio-loud jetted class of Active Galactic Nuclei~(AGN) and have been historically divided into Fanaroff-Riley~(FR)~1 and FR~2 radio galaxies depending on their morphology in the radio band~\cite{Fanaroff1974}. While FR~2s display prominent hot spots at the end of their powerful radio lobes and bright outer edges, FR~1 jets are more diffuse and weaker. Low power FR~1-type jets are usually linked with radiatively inefficient accretion flows, while radiatively efficient accreting objects produce FR 2-type jets.

Thanks to the ever-increasing sensitivity of large-area radio and optical surveys, a new population of very weak radio galaxies emerged by cross-matching the SDSS with the NVSS sample~\citep{Baldi2009} and has been named FR~0 RGs~\citep{Ghisellini2010}.

So far FR~0s have never been proposed as contributing to the observed UHECR flux\footnote{FR~0 Tol~1326-379 had previously been proposed as a neutrino emitter~\citep{tavecchio2019} due to its possible association with a Fermi-LAT source, which implies the presence of CRs. However, a recent refined LAT-analysis~\cite{Ajello2020} now questions this association, and hence this scenario.}. Their core radio properties turned out to be similar to typical FR~1 sources (see, e.g.,~\cite{Willott1999, Merloni2007, Birzan2008, Cavagnolo2010, Croston2018} for radio properties of FR~1/2), as well as their optical classification as low-excitation RGs~(LERGs); however, their extended radio emission shows a pronounced deficit (by a factor on the order of 100) of radio power (e.g.,~\citet{Baldi2009}). Indeed, in FIRST and VLA radio images, they appear rather compact down to a scale of~$\leq 1$~kpc~\citep{Baldi2019}; only on the pc-scale VLBI images could radio jets be resolved in most cases~\citep{Cheng2018}, and Doppler boostings in the range 1.7~to~6 were inferred. Their Eddington-ratios,~$L_{\mathrm{bol}}/L_{\mathrm{edd}}$, of~${\sim}10^{-3}\ldots 10^{-5}$ indicate radiatively very inefficient accretion taking place in FR~0s, compared to FR~1/2 (see,~\cite{Heckman2014, Croston2018}). Their spectral information in the X-ray band suggests a circum-nuclear environment depleted of a dusty torus~\citep{torresigrandi2018}. Their host galaxies are classified as (slightly less) massive ellipticals~($\lesssim 10^8\dots 10^9~M_\odot $), i.e., contain a rather old stellar population with a higher than solar metallicity.

FR~0s seem to prefer a lower galaxy density environment than that of FR~1s~\cite{Capetti2020}. Additionally, their intrinsic space density is about a factor of $\sim$5 times higher than that of FR~1s in the local universe,~$n_{\rm FR0} \sim 5n_{\rm FR1}$~\cite{Baldi2009, Baldi2010}, which seems to feed into the expectations of the cosmic down-sizing scenario (e.g.,~\cite{Rigby2015}).

 A sample of FR~0 RGs (``FR0\emph{CAT}"), limited to a small fraction of the sky, has been published in~\cite{Baldi2018}, which is used in the present study.\footnote{The sample is probably not complete at low radio fluxes (see, e.g.,~\cite{tavecchio2018}). However, this does not influence the prerequisites of this work as long as the missing sources' SEDs are not significantly different from the observed ones.} The population of FR~0s qualifies as a contributor to the observed UHECR flux from an energetics perspective. The required per source CR-power of~$U_{\mathrm{obs,UHECR}}/n_{\mathrm{FR0}}~{\sim}10^{40.5}\;\mathrm{erg}\,\mathrm{s}^{-1}$ (with $U_{\mathrm{obs,UHECR}}$ being the measured energy density of the UHECRs at Earth; see, e.g.,~\cite{Gaisser2013}) lies well below their typically available jet power of~${\sim}10^{42\ldots 43.5}\;\mathrm{erg}\,\mathrm{s}^{-1}$ (as estimated from their radio power following~\cite{Heckman2014}). Hence, FR~0s seem sufficiently numerous within the cosmic-ray horizon to account for an appreciable part of the observed UHECR flux measured on Earth.

In this work, we scrutinize this class of RGs as a contributor to the observed UHECR flux with a particular focus on their environment. For this purpose, we first~(\autoref{sec:background}) study various acceleration mechanisms for CR nuclei --- Fermi-I/II, gradual shear, and relativistic blast wave acceleration --- in tandem with all relevant particle losses in the radiative environment of AGN jets --- Bethe-Heitler pair production, photo-disintegration, photo-meson production synchrotron losses, and diffusive particle escape. This is then applied to the average radiative environments of FR~0 RGs in~\autoref{sec:sourceclasses}. Here, we assess the potential of these sources to accelerate nuclei to the highest energies observed so far; while at the same time surviving intact in this environment. We conclude this work by summarizing our results in~\autoref{sec:summary}. 

\section{Particle Energy Losses and Gains in Jets}
\label{sec:background}

To assess whether a class of astrophysical objects can be considered a cosmic-ray accelerator --- the total energetics (as discussed in~\autoref{sec:intro}) and the source environment must be evaluated. Models of the relevant processes that influence the maximal acceleration capability of a source are summarized here. These processes can be divided into three different competing groups:~1)~Particle acceleration,~2)~Energy losses of CRs, and~3)~Particle transport out of the region of interest. Whichever process happens on the shortest time scale will dominate the cosmic-rays' energy gains or losses. Comparing the three associated time scales~($\tau_\mathrm{acc},~\tau_\mathrm{loss},~\tau_\mathrm{esc}$) will allow the assessment of the maximum energies that can be reached in the sources. Accordingly, the first part of \autoref{sec:background} deals with established particle acceleration models for jetted AGN. In the second part, the loss time scales are discussed, where the focus is on losses due to electromagnetic pair production~(Bethe-Heitler), photo-disintegration, and photo-meson production. Finally, a parameter scan of the loss length is provided for reasonable target field energies,~$\epsilon$, and nuclei energies,~$E_\mathrm{N}$. The corresponding look-up plots will allow quick evaluation of approximate values for the acceleration chances by comparison with the derived acceleration time scales.

In this paper, we refrain from detailed modeling of cosmic-ray energy spectra and secondary spectral energy distributions~(SEDs) in the source region but rather attempt to answer whether FR~0 can potentially reach the required energies to contribute to the UHECR flux.

\subsection{Acceleration Time Scales}
\label{ssec:acceleration}

In this section, we will briefly discuss possible acceleration scenarios for the core region of FR~0 radio galaxies. We focus on the time scales and refer interested readers to the original publications for the proposed acceleration mechanisms' technical details. In general, the acceleration time scale is given by
\begin{align}
    \tau_\mathrm{acc}(E) = \eta(E)\,\frac{\lambda(E)}{c} \quad \label{eq:AccTime}, 
\end{align}
where $\eta\geq 1$ is a model-dependent scaling factor, and $\lambda$ is the scattering length. In general, this scaling factor's dependencies can complex, e.g., the scattering center speed (see \autoref{ssec:Fermi1}) or the shock, which is neglected in this work. The shortest acceleration time scale is given by $\eta=1$ and $\lambda=r_\mathrm{Larmor}$ (see, e.g., \cite{Aharonian2000}).   

\subsubsection{Fermi 2nd Order}
\label{ssec:Fermi2}
More than 70 years ago, Fermi proposed a model for cosmic ray acceleration in turbulent magnetized clouds~\citep{fermi1949}. The stochastic nature of this process allows for energy gains and losses due to the randomly moving scattering centers. This fact makes the second-order process relatively slow. On the other hand, it is easily realized since no large-scale shock structure is needed, only a turbulent magnetic field.

Second-order Fermi acceleration can be described by a corresponding (momentum diffusion) transport equation (see, e.g.,~\citet{Rieger2007}). From this equation, the acceleration time scale can be derived to be
\begin{align}
    \tau_\mathrm{acc} = p^3 \left( \frac{\partial}{\partial p}(p^2\,\kappa_p)\right)^{-1} \label{eq:Fermi2} \quad ,
\end{align}
where the momentum diffusion coefficient is given by~$\kappa_p$, and~$p$ is the absolute value of the momentum. This time scale now depends on the specific diffusion model that is applicable in the source region. 

In the turbulent cascading regime, where a power law describes the power spectrum, the following relation between the spatial and momentum diffusion coefficient holds~(see, e.g.,~\citet{Drury2017, sigl2017}):
\begin{align}
    \kappa_p\, \kappa_x \propto \frac{v_\mathrm{A}^2p^2}{\alpha(4-\alpha)(4-\alpha^2)} \quad , \label{eq:spatial-momentum-ratio}
\end{align}
where the Alfvén speed is~$v_\mathrm{A}=B/\sqrt{4\pi\,\rho}$, with~$\rho$ being the plasma density, and $B$ the magnetic field strength, while~$\alpha$ is the power-law index of the spatial diffusion coefficient~$D_x(p)\propto p^\alpha$. The relevant time scale can then be calculated by inserting~\autoref{eq:spatial-momentum-ratio} into~\autoref{eq:Fermi2} and assuming a proportionality constant of~4/3 \citep{sigl2017}, which yields:
\begin{align}
    \tau_\mathrm{acc} = \frac{3\alpha(4-\alpha^2)}{4}\, \frac{1}{v_\mathrm{A}^2} \kappa_x(p) \quad .
\end{align}

\subsubsection{Fermi 1st Order}
\label{ssec:Fermi1}
The concept of Fermi acceleration was later extended to include a more efficient acceleration mechanism, which can be seen as the predecessor of diffusive shock acceleration~\citep{fermi1954}, and is often the preferred acceleration model. An updated description of diffusive shock acceleration can be found, e.g., in~\cite{BELL2013}.

The first-order Fermi acceleration time-scale is determined by the conditions on both sides of the shock. When the shock is expanding with velocity~$u_s$ in the laboratory frame, the time scale is given by (see, e.g.,~\cite{lagage_cesarsky1983b,drury1983}): 
\begin{align}
    \tau_\mathrm{acc}(p) = \frac{3}{u_1-u_2}\int_0^{p} \frac{\mathrm{d}p'}{p'} \left(\frac{\kappa_1(p')}{u_1} + \frac{\kappa_2(p')}{u_2} \right) \label{eq:Fermi1} \quad .
\end{align}
Here,~$u_1=-u_{s}$~and~$u_1=s\, u_2$ are the upstream and downstream velocity components parallel to the shock normal measured in the shock frame, respectively, and $s$ is the compression ratio. The spatial diffusion coefficients describing the transport on either side of the shock are denoted with~$\kappa_i$.

Given that the FR 0 source class environment is not well known, we choose to consider the most optimistic scenario. This includes the assumptions of a strong non-relativistic shock with~$s=4$ and Bohm diffusion, which lead to the shortest acceleration time scales. This ansatz then yields: 
\begin{align}
    \tau_\mathrm{Fermi-Bohm} = \frac{20}{3} \frac{E}{u_s^2\,q\,B} \approx 7.4 \times 10^{7} \, Z^{-1}  \left(\frac{E}{\mathrm{EeV}}\right)\, \left(\frac{B}{\mathrm{G}}\right)^{-1} \, \left(\frac{u_s}{\mathrm{0.1\,c}}\right)^{-2} \;\mathrm{s} \; ,
\end{align}
where $q=Ze$ is the particle charge, and $Z$ is the charge number. Here, the ultra-relativistic approximation for particle speeds ($\gamma_N\gg1, v\approx c$) is applied throughout this paper, which implies $E\approx pc$.

Comparing the time scales of first- and second-order Fermi acceleration, it becomes clear that the Fermi-II process is slower by approximately a factor of~${\sim}(v_\mathrm{A}/u_s)^2$ since the Alfvén speed is usually much lower than typical shock speeds.

In general, the acceleration efficiency can hence be expressed by $\eta_\mathrm{Fermi-I}=20/(3\alpha)\,\beta_\mathrm{s}^{-2}$, where $\alpha$ is the diffusion coefficient power law index and $\beta_\mathrm{s}~=~u_\mathrm{s}/c$.

\subsubsection{Gradual Shear Acceleration}
\label{ssec:ShearAcc}

Gradual shear acceleration can be interpreted as a stochastic acceleration with scattering centers moving in an ordered direction. Therefore, the acceleration time scale can be calculated based on \autoref{eq:Fermi2}, when the corresponding effective diffusion coefficient is known. In~\citet{Rieger2006}, this is quantified for flows with a cylindrical profile as
\begin{align}
    \kappa_{p, \mathrm{shear}} = \frac{1}{15}\left(\frac{\partial u_z}{\partial r}\right)^2\tau_0\,p^{2+\alpha},
\end{align}
which yields an acceleration time scale
\begin{align}
    \tau_\mathrm{acc}(p, r) = \frac{1}{4+\alpha} \frac{1}{\tilde{\Gamma}\tau_0\,p^\alpha} \quad \text{with} \quad \tilde{\Gamma} = \frac{1}{15}\left(\frac{\partial u_z(r)}{\partial r}\right)^2\,\Gamma_\mathrm{j}^4 \quad . \label{eq:tau_shear}
\end{align}
Here,~$\partial u_z/ \partial r$ is the gradient of the flow profile,~$\tau_0$ is the energy-independent scattering time, and~$\Gamma_\mathrm{j}$ is the jet's Lorentz factor. The scattering time is related to the mean free path~$\lambda$ of the transport via~$\lambda=\tau_0\,p^\alpha\,c$. This time scale corresponds to an acceleration efficiency of $\eta_\mathrm{shear}=(c/\lambda)^2/(\tilde{\Gamma}(4+\alpha))$. More details, and the fully relativistic derivation, including other flow profiles, can be found in~\citet{Rieger2004}. \citeauthor{Rieger2004} assume that the particle's Larmor-radius is smaller than the width of the shear layer $r_\mathrm{Larmor}< \Delta r$, introducing an upper limit on the applicable energy range. When the gyroradius of the CR is larger, the shear layer can be approximated by a discontinuity; numerical results for the acceleration rate, in that case, can be found, e.g., in \cite{Ostrowski1990}.

Assuming a linear flow profile~($\partial u_z/\partial r=\Delta u/\Delta r$), where~$u_z$ is the flow speed along the jet axis, and Bohm diffusion for the scattering time, \autoref{eq:tau_shear} becomes
\begin{align}
    \tau_\mathrm{acc}^\mathrm{Bohm} = \frac{3 (\Delta r)^2 c}{\Gamma^4_\mathrm{j}\,(\Delta u)^2 r_\mathrm{g}} = 3\frac{1}{\Gamma_\mathrm{j}^4}\left(\frac{\Delta r}{\Delta u}\right)^2\frac{B\,q\,c^2}{E} \quad . \label{eq:tauBohmShear}
\end{align}
One interesting fact to notice is the energy dependency of shear acceleration: The acceleration time scale decreases with increasing CR energy --- the inverse of Fermi-type acceleration. This change in energy dependence behavior opens the possibility that the dominant acceleration mechanism changes with energy if the environment allows. Depending on the exact jet properties ($\Delta u$, $\Delta r$, and $\Gamma_\mathrm{j}$), this could lead to unphysically fast acceleration, corresponding to $\eta(E) <1$ at high energies. Therefore, this description of gradual shear acceleration can only be applied up to a maximum energy $E_\mathrm{max}^\mathrm{shear}$, making sure that $r_\mathrm{Larmor}(E_\mathrm{max}^\mathrm{shear})<\Delta r$, and $\eta(E_\mathrm{max}^\mathrm{shear})>1$. Comparing results from \citeauthor{Rieger2004} to \cite{Ostrowski1990, Ostrowski1998} suggests a transition of the acceleration time-scale energy-dependence, from $\propto E^{-1}$ to $\propto E^{+1}$, around $r_\mathrm{Larmor}\approx\Delta r$. Since the acceleration process in the transition is poorly known, we will apply the most optimistic scenario there: $\eta=1$ for energies above the threshold $E>E_\mathrm{max}^\mathrm{shear}$. As shown in \autoref{ssec:seds}, larger values for $\eta\approx 10$, as estimated in \cite{Ostrowski1998}, would not significantly change the resulting maximum energy.

In the following calculations it will be assumed that the central jet speed is given by~$u_z(r=0)=c\,\sqrt{1-\Gamma_\mathrm{j}^{-2}}$ and vanishes at the edge of the jet,~$u_z(r=\Delta r)=0$.

\subsubsection{Relativistic Blast Wave}
\label{ssec:Espresso}

The relativistic blast wave scenario~\citep{Gallant1999}, sometimes referred to as \emph{Espresso shot} acceleration~\citep{Caprioli2015_Espresso, Caprioli2018_Espresso}, differs from the other scenarios as it incorporates a drastic change of the energy gain per encounter over time. While the CRs' first crossing of the ultra-relativistic shock front can give an energy boost on the order of~$\Gamma_\mathrm{b}^2$~($\Gamma_\mathrm{b}$ is the Lorentz factor of the blast wave), all other crossings will give only an increase of a factor~$\sim$2. 

The downstream residence time~$t_\mathrm{d}$ approximately gives the time needed for the first boosting. In all subsequent cycles, the time scale is given by the sum of the up- and downstream residence time, which is mainly dominated by the upstream time scale~$t_\mathrm{u}$~\citep{Gallant1999}:
\begin{align}
    t_\mathrm{d} \approx \frac{pv_\mathrm{A}}{qc\Gamma_\mathrm{b} B}, \quad t_\mathrm{u} \approx \frac{p}{q\Gamma_\mathrm{b} B}\, \mathrm{max}\left(1, \frac{r_\mathrm{g}}{\Gamma_\mathrm{b} l_\mathrm{c}}\right)\quad .
\end{align}
Here,~$l_\mathrm{c}$ is the correlation length of the turbulent magnetic field, and~$r_\mathrm{g}$ is the gyro-radius. 

Therefore, this acceleration scenario is especially interesting when an already pre-accelerated particle distribution is available at the source site. This population can then be significantly boosted in energy on a very short time scale. However, it is not much more efficient as a sole acceleration mechanism than ordinary Fermi-I-type acceleration. Furthermore, the boosting factors are relatively low for FR~0 jets. Therefore, this mechanism is not further investigated in this work.

\subsection{Loss Time Scales}
\label{ssec:losses}

In this section, the principle loss processes relevant to the source region are discussed. Since abundant literature already exists on this topic (see, e.g.,~\cite{1993A&A...269...67M, 2001APh....15..121M} in application to AGN models or~\cite{sophia, 2008PhRvD..78c4013K, 2010ApJ...721..630H} for a general description), we refrain from detailed discussions and only present the necessary information to reproduce our results. 

The most relevant loss processes in the radiatively dominated source region are: 1) Electromagnetic Bethe-Heitler pair production, 2) Photo-disintegration of the primary nucleus, 3) Photo-meson production, and 4) Synchrotron losses when the magnetic field is relatively high (${\sim}1\,\mathrm{G}$). In this section, the losses are discussed for mono-energetic target fields to estimate the resulting time scales independently of the target photon field shape. In \autoref{sec:sourceclasses}, the loss time scales are evaluated for a concrete target photon field based on the average SEDs of FR~0 radio galaxies.

The energy loss length for photo-meson production~(\autoref{sssec:PPP}) and photo-disintegration~(\autoref{sssec:PDI}) is calculated based on the interaction rate~$\tau_{\mathrm{N}\gamma}^{-1}$, weighted with the mean energy loss per interaction. For Bethe-Heitler pair production~\citep{BH_pairproduction} and synchrotron losses, a direct (semi-) analytical description is possible and given in \autoref{sssec:BHloss} and \autoref{sssec:Synch}, respectively.

In general, the interaction rate for particle-photon processes is described by
\begin{align}
    \tau_{\mathrm{N}\gamma}^{-1}=c\int \mathrm{d}\Omega \int \mathrm{d}\epsilon n_\gamma (\epsilon, \Omega_\gamma)(1-\beta_N \cos(\theta))\sigma_{\mathrm{N}\gamma}(s) \quad .
\end{align}
Here,~$\sqrt{s}$ is the center-of-mass energy for the interaction between a photon with energy~$\epsilon$ and a nucleus with mass~$m_\mathrm{N}$ and energy~$E$:~$s=m_\mathrm{N}^2c^4+2\epsilon E(1-\beta_\mathrm{N} \cos(\theta))=m_\mathrm{N}^2c^4 + 2m_\mathrm{N}c^2\,\epsilon_\mathrm{r}$, where~$\epsilon_\mathrm{r}$ is the photon energy in the nucleus rest frame. The interaction angle is denoted with~$\theta$, and~$\beta_\mathrm{N}=v_\mathrm{N}/c$ is the normalized nucleus velocity. Quantities observed in the rest frame of the nucleus are denoted with a subscripted~'$\mathrm{r}$'.

Assuming an isotropic\footnote{The anisotropy introduced by boosting can shift the threshold energy~$\epsilon_\mathrm{r, thr}$ but does not change the results at high~$\epsilon_\mathrm{r, thr}$, e.g., in the multi-pion production regime~(see Reimer et~al.~2021 in prep.). Since our results, especially at the maximum energies, are not dominated by the thresholds' processes, we neglect the anisotropy's influence.} photon distribution~$n_\gamma (\epsilon, \Omega_\gamma)=n_\gamma(\epsilon)/4\pi$ allows integration over the azimuthal component, which results in
\begin{align}
   \tau_{\mathrm{N}\gamma}^{-1}=\frac{c}{2\,\gamma^2_\mathrm{N}\,\beta_\mathrm{N}} \int_0^\infty \mathrm{d} \epsilon \frac{n_\gamma(\epsilon)}{\epsilon^2} \int_{\epsilon_\mathrm{r, thr}} \mathrm{d}\epsilon_\mathrm{r}\,\epsilon_\mathrm{r}\sigma_{\mathrm{N}\gamma}(\epsilon_\mathrm{r}) \quad , \label{eq:Interactionrate}
\end{align}
where $\epsilon_\mathrm{r, thr}$ is the process threshold, which will be discussed in the following sections. For a monoenergetic photon field, $n_\gamma(\epsilon)=n_0\delta(\epsilon-\epsilon_0)$, one gets
\begin{align}
    \tau_{N\gamma}^{-1}=\frac{c n_0}{2\beta_\mathrm{N}} \frac{1}{\gamma_\mathrm{N}^2\,\epsilon_0^2} \int_{\epsilon_\mathrm{r, thr}}^{\gamma_\mathrm{N}\,\epsilon_0\,(1+\beta_\mathrm{N})} \mathrm{d}\epsilon_\mathrm{r}\,\epsilon_\mathrm{r}\sigma_{\mathrm{N}\gamma}(\epsilon_\mathrm{r}) \quad .
\end{align}

The energy loss length for photo-disintegration and photo-pion production is calculated as:
\begin{align}
    L_\mathrm{loss}(E)=\tau_{\mathrm{N}, \gamma}\, c \, \frac{E}{\langle \Delta E\rangle} \quad, 
\end{align}
where the interaction rate $\tau^{-1}$ is defined according to \autoref{eq:Interactionrate}, and $\langle \Delta E \rangle$ is the mean energy loss per interaction. For Bethe-Heitler pair production, only \autoref{eq:BHLL} is used.

\subsubsection{Photo-disintegration}
\label{sssec:PDI}
\begin{figure}[htbp]
    \centering
    \begin{minipage}{.49\linewidth}
    \includegraphics[width=\linewidth]{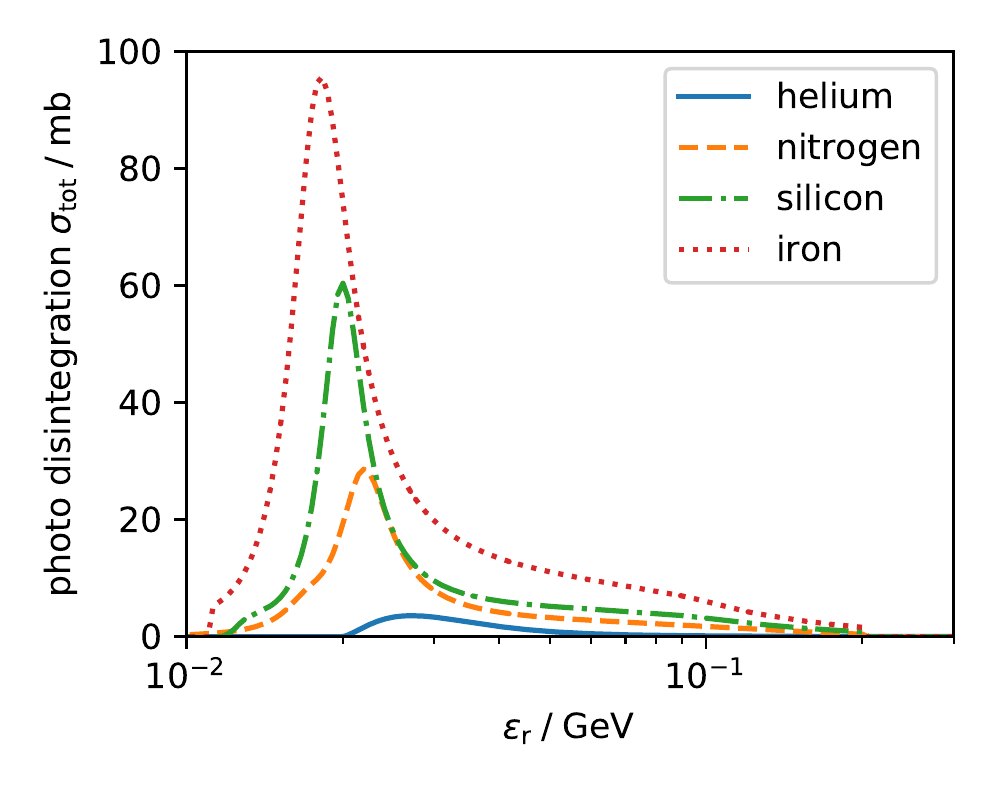}
    \end{minipage}
    \begin{minipage}{.49\linewidth}
    \includegraphics[width=\linewidth]{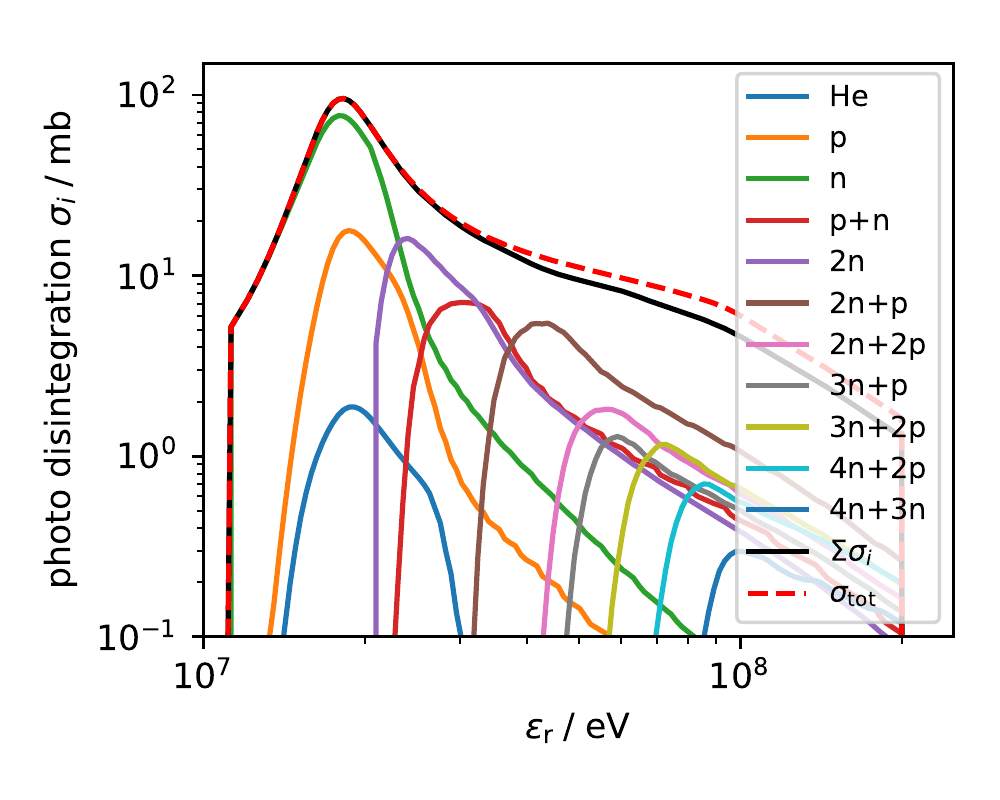}
    \end{minipage}
    \caption{The left figure shows the total photo-disintegration cross-sections $\sigma_\mathrm{tot}$ for the tracer elements --- the maximum increases and shifts to lower photon energies $\epsilon_\mathrm{r}$ for increasing masses of the nuclei. The right figure shows the exclusive cross-section for iron. The legend gives the spallation products.}
    \label{fig:xs-PDI}
\end{figure}

In this work, we refer to the photo-disintegration regime when the target photon's nucleus rest-frame energy is in the range of~$10^{-2}\leq \epsilon_\mathrm{r}/\mathrm{GeV}\leq 0.14$, where the upper boundary is approximately the pion production threshold. The left plot in \autoref{fig:xs-PDI} shows the total cross-sections used for the four relevant tracer elements\footnote{The tracer elements --- protons, helium, nitrogen, silicon, and iron --- are usually used to model the cosmic-ray chemical composition since they approximate their mass group's interactions sufficiently~(see, e.g.,~\cite{GST2013} and references therein for this approach).}, helium, nitrogen, silicon, and iron. 

The cross-sections are taken from tabulated values, as published in the CRPropa data repository\footnote{https://github.com/CRPropa/CRPropa3-data}. The elements with $A>12$ are based on the TALYS~$1.8$ Monte-Carlo simulations of the photo-hadronic cross-section~\citep{TALYS}. The cross-section for lower mass numbers~$A\leq12$ are taken from various references, including~\citep{Rachen1996, Kossov2002, Varlamov1986, Kulchitskii1963}.\footnote{See~\cite{Nierstenhoefer, KAMPERT201341}, for an explanation and the data repository for the exact listings.} The energy loss length is calculated as a weighted average over all possible disintegration channels
\begin{align}
    \langle\Delta E\rangle(E) = \sum_i b_i(E)\,\Delta E_i (E) \quad .
\end{align}
Here,~$b_i$ is the branching ratio, which is the normalized exclusive interaction rate of the specific process:~$b_i=\tau_i(E)/\sum_j \tau_j(E)$. The relative energy loss is proportional to the mass ratio of the fragment(s) and primary particle~$\Delta E/E~=~m_\mathrm{sec}/m_\mathrm{N}$, where $\Delta E$ is the energy loss and $m_\mathrm{sec}$ is the mass of the fragment(s). As an example, we show the exclusive cross-sections for iron in the right plot of~\autoref{fig:xs-PDI}.

\subsubsection{Photo-Meson Production}
\label{sssec:PPP}
\begin{figure}[htbp]
    \centering
    \includegraphics[width=.5\linewidth]{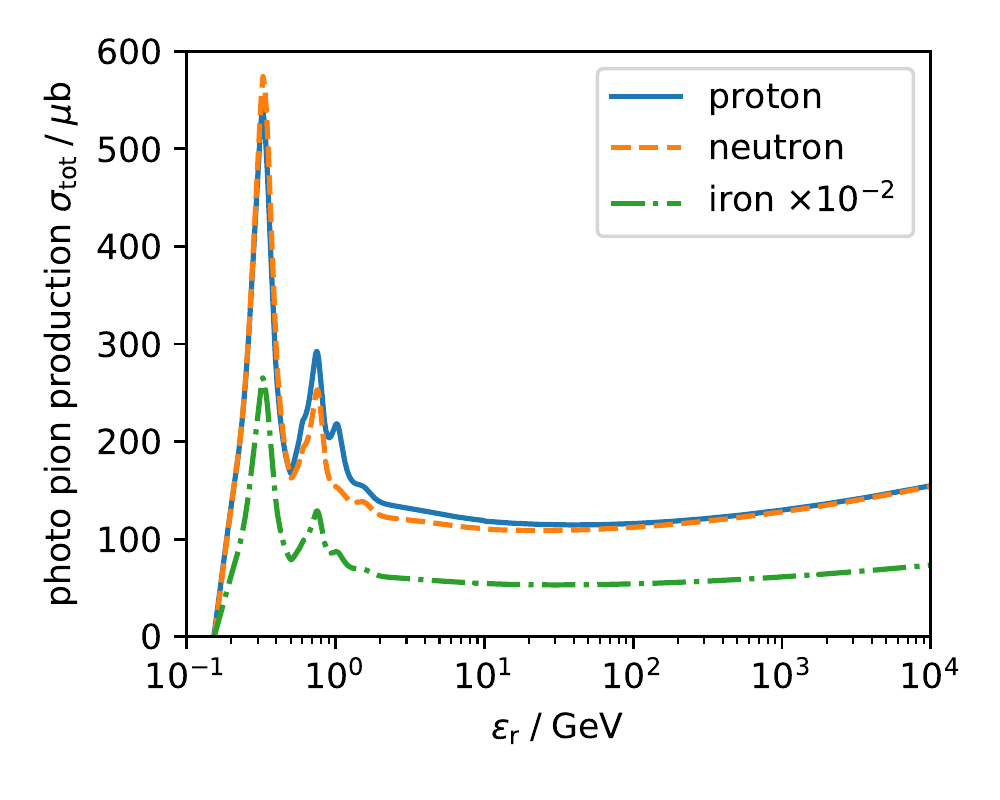}
    \caption{The photo-pion production cross-section for protons, neutrons, and iron is shown. In this work, we used a simple superposition model to construct the cross-section of heavier elements. The cross-section data comes from the CRPropa~\citep{crpropa30} implementation of the SOPHIA code~\citep{sophia}.}
    \label{fig:xs-PPP}
\end{figure}

For interaction energies above~$\epsilon_\mathrm{r, thr}\approx 140$~MeV photo-meson production sets in. The cross-section, see~\autoref{fig:xs-PPP}, is first dominated by several hadronic resonances before multi-pion-production dominates at higher center-of-mass energies. The neutron cross-section deviates from that of protons, mainly in the secondary resonance region. In this work, the cross-section for more massive particles is approximated using the superposition model:~$\sigma_\mathrm{N}=A^\alpha (N_\mathrm{p}\sigma_{\mathrm{p\gamma}}~+~N_\mathrm{n}\sigma_{\mathrm{n\gamma}})$. The variable~$A^\alpha$ describes a shielding factor, determining which portion of the nucleons are shielded and not available for interactions. The exponent~$\alpha$ can vary from unity to~$2/3$, depending on the mass number~$A$ of the nucleus.

In this simple superposition model, exactly one nucleon takes part in the interaction, making the energy loss calculation a simple scaling with the mass number:~$\Delta E=E/A$. The energy loss for protons and neutrons is approximated as occurring in the~$\Delta$-resonance regime, leading to a decrease by a factor of~$\Delta E_{p/n}=E(1-m_{p/n}/m_\Delta)\approx 0.24 E$.

\subsubsection{Bethe-Heitler Pair Production}
\label{sssec:BHloss}
The calculation of the electron-positron pair production loss length is based on the semi-analytical approach described in~\citet{Chodorowski1992}. The assumptions made there --- $\gamma_\mathrm{N}\gg1$ and $\epsilon_\mathrm{r}\gg m_ec^2$ --- are usually very well justified for UHECR. Assuming isotropy of the target field, the energy loss is then described by:
\begin{align}
    -\frac{\mathrm{d}\gamma_\mathrm{N}}{\mathrm{d}t} = \alpha_\mathrm{S}\,r_0^2\,c\, Z^2 \frac{m_e}{m_\mathrm{N}}\,\int_2^\infty \mathrm{d}\kappa\,n_\gamma\left(\frac{\kappa}{2\gamma_\mathrm{N}}\right)\frac{\phi(\kappa)}{\kappa^2} \quad \label{eq:BHLL} .
\end{align}
Here,~$\alpha_\mathrm{S}\approx 1/137$ is Sommerfeld's fine-structure constant,~$r_0$ is the classical electron radius, and~$Z$ is the nucleus charge number. The nuclear rest frame photon energy is parameterized in units of the electron momentum as~$\kappa=2\gamma_\mathrm{N}\epsilon/(m_e c)$. The function~$\phi(\kappa)$ is a semi-analytical description of the cross-section integral~(see Eq. 3.12 of the original paper~\cite{Chodorowski1992}). 

\subsubsection{Synchrotron Radiation}
\label{sssec:Synch}
The average energy loss of an isotropic nuclei ensemble due to synchrotron radiation can be written as:
\begin{align}
    -\left\langle \frac{\mathrm{d}E}{\mathrm{d}t}\right\rangle = \frac{4}{3}\sigma_\mathrm{T}c\,u_\mathrm{mag}\,\left(\frac{v}{c}\right)^2\gamma^2\,\left(\frac{m_e}{m_\mathrm{N}}\right)^2\,Z^4 \quad . \label{eq:SynchLoss}
\end{align}
Here, $\sigma_\mathrm{T}$ is the Thomson cross-section, and $u_\mathrm{mag}$ is the magnetic energy density. Furthermore, $v$ is the nucleus speed, $\gamma$ its Lorentz factor, $Z$ the charge number, $m_e$ is the electron mass, and $m_\mathrm{N}$ the nucleus mass. 

\subsubsection{Combined Loss Length}
\label{sssec:comb-loss}
\begin{figure}[htbp]
    \centering
    \includegraphics[width=.65\textwidth]{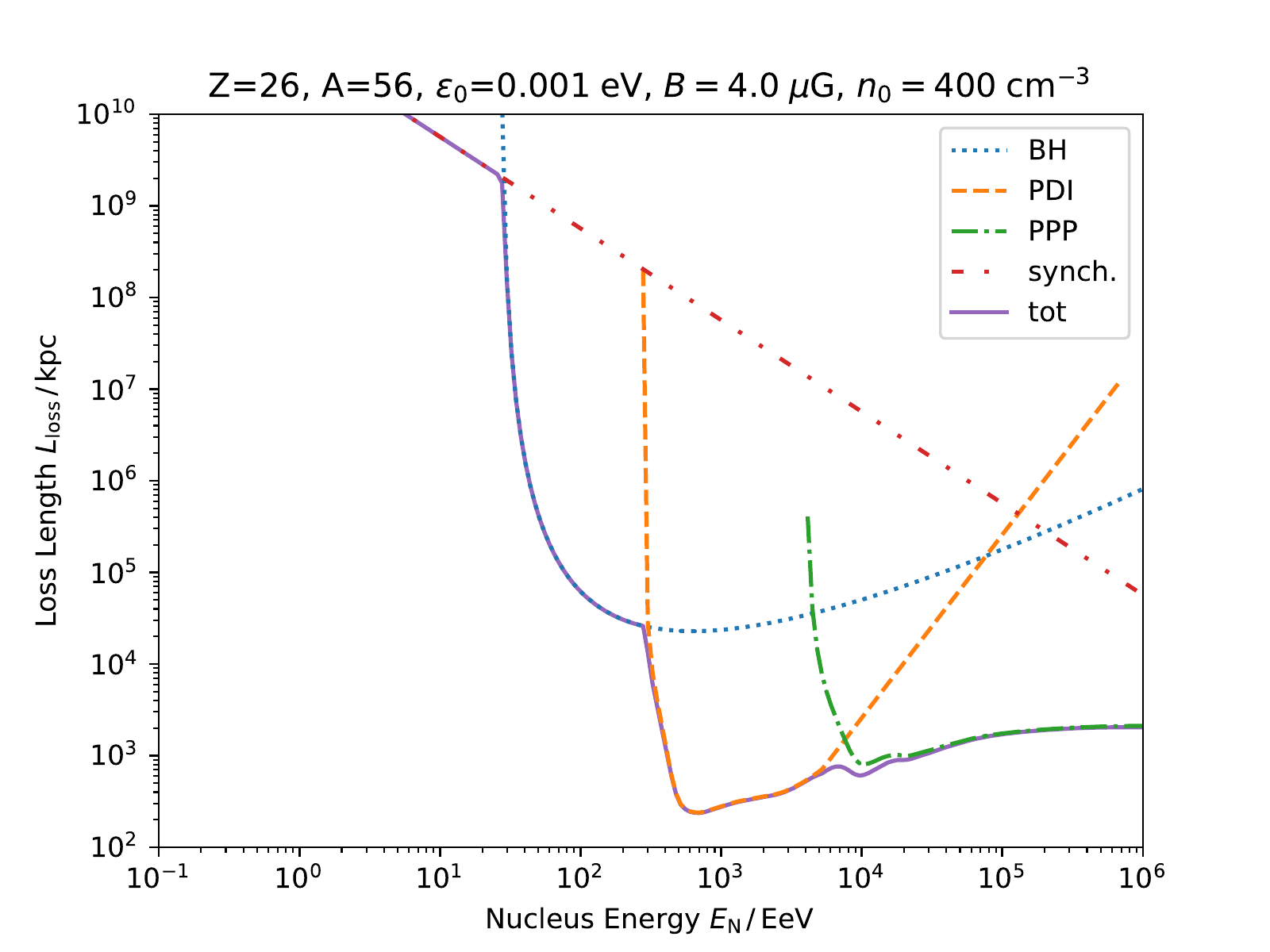}
    \caption{Combined losses for the tracer elements are shown. The blue dotted line~(BH) indicates Bethe-Heitler pair production, which is combined with photo-disintegration~(PDI, orange-dashed), photo-pion-production~(PPP, green-dash-dotted), and synchrotron radiation (synch., red-dash-dot-dotted) to give the total loss length scale~$L_\mathrm{loss}$~(violet-solid).}
    \label{fig:loss}
\end{figure}

The total energy loss length is the combination of all four processes given by~$L_\mathrm{loss, tot}^{-1}=\sum_i L_{\mathrm{loss}, i}^{-1}$. \Autoref{fig:loss} shows the individual loss lengths for the four loss processes and the total loss length depending on the nucleus energy~$E_\mathrm{N}$ for iron; all other elements can be found in~\autoref{fig:LL_monochromatic}. Here, an exemplary mono-energetic target photon field with an energy of~$\epsilon_0=1\;\mathrm{meV}$ and a number density of~$n_0=400\;\mathrm{cm^{-3}}$, close to the CMB monochromatic approximation, was chosen. The magnetic field strength ($B=4\;\mu\mathrm{G}$) was chosen such that equipartition between the magnetic and the target photon field energies is reached. Depending on the nucleus energy, different processes dominate, with photo-disintegration leading to the highest losses at a given energy above its threshold. The position of the minimum changes, of course, with changing energy of the target photon field.

\begin{figure}[htbp]
    \centering
    \includegraphics[width=.65\textwidth]{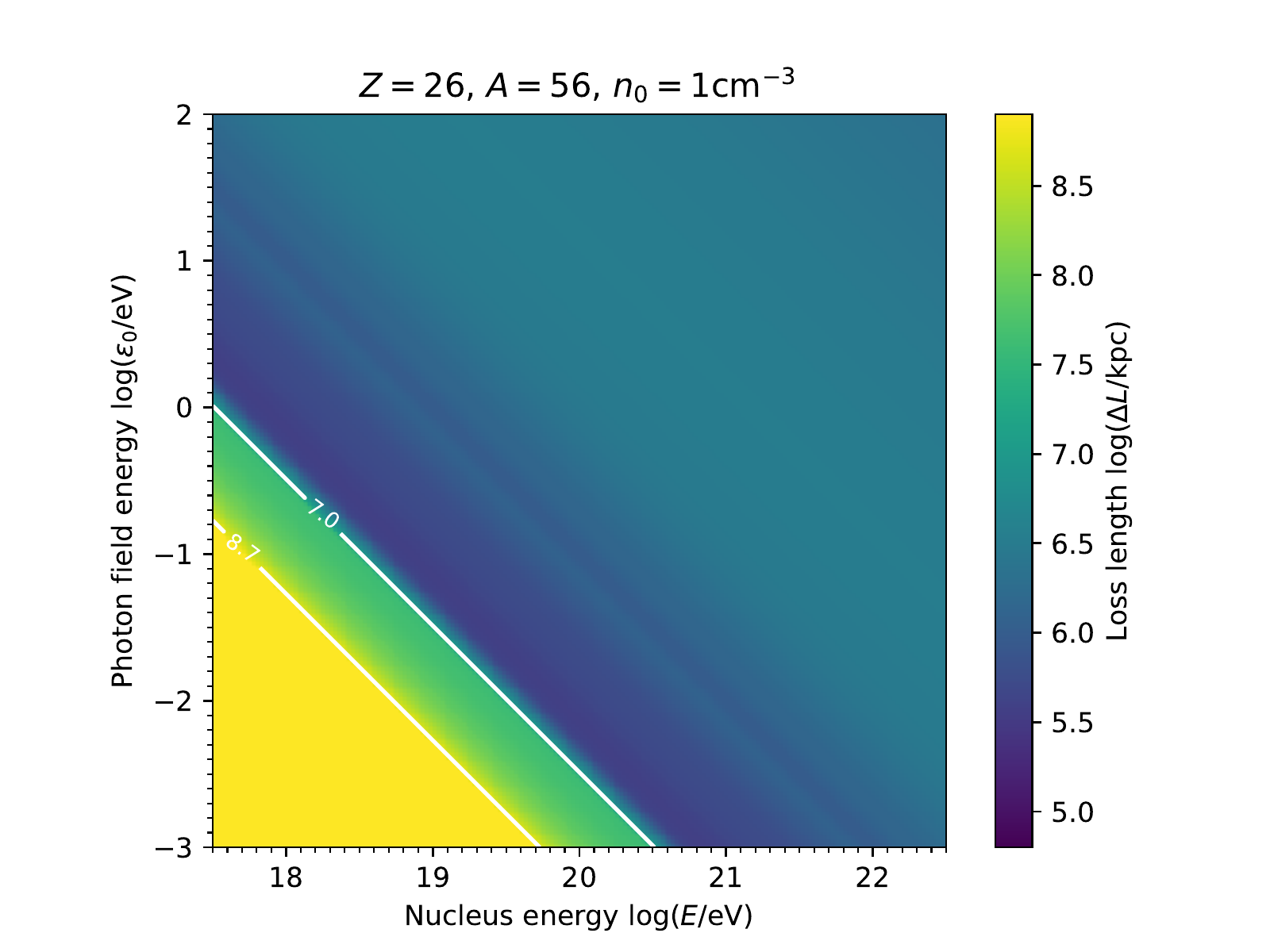}
    \caption{The combined loss length for iron~(see~\autoref{fig:2dscan} for the other tracer elements) is shown. The loss length is color-coded, where bright yellow corresponds to large and dark blue to short loss lengths. The nuclear rest frame photon energy,~$\epsilon_\mathrm{r}$, increases along the main diagonal. Note that the results have to be scaled to the concrete total target field density~$n_0$.}
    \label{fig:loss-contour}
\end{figure}

As an initial step, a parameter scan for reasonable target field energies was performed in the range~$10^{-3}\leq\epsilon_0/\mathrm{eV}\leq10^2$. These scans are shown in~\autoref{fig:loss-contour} and~\autoref{fig:2dscan} for a monochromatic target field number density of~$n_0=~1\;\mathrm{cm}^{-3}$ as an example. Note, that synchrotron radiation has not been included in these plots. However, the simple inverse scaling with energy for any given nucleus allows adding this loss to the combined loss rate of the other processes.

The gradient is found along the major diagonal, which corresponds to an increasing nuclear rest frame photon energy~$\epsilon_\mathrm{r}$.
The target field energies in the optical regime~($\sim$eV) have the largest influence on nuclei with energies around the ankle~(${\sim}10^{18}$ eV). For higher nuclei energies, the most important target field energies are in the range of the CMB. Furthermore, at least for a monochromatic target field, only a relatively narrow range in nuclei energies will suffer from the highest losses. A broader target photon field, such as the SEDs discussed in~\autoref{sec:sourceclasses}, will smooth the relatively peaked loss rate. 

\subsection{Escape Time Scale}
\label{ssec:escape}

Cosmic-ray escape from the acceleration region is not yet completely understood. Ideally, a transport equation (including advection, streaming, and diffusion) should be solved --- though this is beyond this paper's scope. With a velocity profile along the jet axis, perpendicular escape is only possible by diffusion ($\tau_\perp=R^2/(2\kappa)$). In the parallel direction, escape could be dominated by advection ($\tau_\parallel=L/u$).

However, for the modeled jet structures (see~\autoref{sec:sourceclasses}), perpendicular diffusive escape is the more relevant loss process, when the jet is longer than $L\gtrsim 10$~pc.\footnote{For Bohm a turbulence spectrum advection might dominate the escape process at energies $\lesssim 0.1$~EeV. However, at these energies Fermi-I acceleration is much faster than any escape process, leaving only a diffusion dominated escape scenario at relevant energies.} Additionally, at the highest energies, where gradual shear acceleration dominates the energy gains (see \autoref{ssec:CRsources}), perpendicular diffusion is the most likely escape process (see, e.g., \cite{rieger_review, rieger2019}). Therefore, we focus here only on diffusive escape with a time scale of
\begin{align}
    \tau_\mathrm{esc} &= \frac{R^2}{2\,\kappa} \quad , \label{eq:tau_esc}
\end{align}
where~$R$ is the size of the acceleration region, and~$\kappa$ is the diffusion coefficient. Assuming again Bohm diffusion, which gives a lower limit on the escape time scale,~\autoref{eq:tau_esc} yields
\begin{align}
    \tau_\mathrm{esc}^\mathrm{Bohm} = \frac{3}{2}\frac{R^2\,Bq}{E} \quad .
\end{align}

\subsection{Maximum Energy}
\label{ssec:comparison}
The maximum energy for any multiplicative acceleration process is given by the Hillas criterion, which yields for the maximum energy:
\begin{align}
    E_\mathrm{max}^\mathrm{Hillas} = q\,B\,R\,\beta c \approx 10^{21} \, Z\beta \left(\frac{B}{\mathrm{G}}\right) \left(\frac{R}{\mathrm{pc}}\right) \;\mathrm{eV} \quad . \label{eq:EmaxHillas}
\end{align}
Here,~$R$ is the size and~$B$ the magnetic field strength of the acceleration region. The speed of the scattering centers is~$\beta c$.

However, in most cases, the maximum energy will be smaller than defined by the Hillas criterion. Neglecting losses --- which is a valid assumption in some cases~(see~\autoref{sec:sourceclasses}) --- the maximum energy can be calculated by comparing the escape time scale with the acceleration time scale. 

For Fermi-I acceleration --- assuming a strong shock and the same diffusion coefficient~$\kappa$ on both sides of the shock~(see~\autoref{eq:Fermi1}) --- and an escape process driven by the same diffusion process~(see~\autoref{eq:tau_esc}), the maximum energy becomes

\begin{align}
    \tau_\mathrm{Fermi-I}&=\frac{20}{\alpha}\frac{\kappa}{u_s^2} = \frac{R^2}{2\kappa}=\tau_\mathrm{esc} \quad ,
\end{align}
which yields
\begin{align}
    E_\mathrm{max} &= Z\,E_0 \left(\frac{R\,u_s\sqrt{\alpha}}{2\sqrt{10}\,\kappa_0}\right)^{1/\alpha} \label{eq:EmaxFermi}
\end{align}
for the maximal particle energy. Here,~$\kappa_0$ is the diffusion coefficient normalization with~$\kappa(E_0)=\kappa_0$. In the case of Bohm diffusion \autoref{eq:EmaxFermi} can be approximated by
\begin{align}
    E_\mathrm{max}^\mathrm{Bohm} &= 4.4\times 10^{19} \, Z\left(\frac{R}{\mathrm{pc}}\right)\left(\frac{B}{\mathrm{G}}\right)\left(\frac{u_s}{0.1c}\right)\;\mathrm{eV} \quad .
\end{align}

Since shear acceleration has the same energy dependence as the escape process, a general condition for efficient acceleration is given by ~$\Gamma_\mathrm{j}\gtrsim 1.6$, where $\tau_\mathrm{acc}^\mathrm{shear} < \tau_\mathrm{esc}$.

The gradual shear acceleration breaking point is given by $\eta(E_\mathrm{max}^\mathrm{shear})=1$. For Bohm diffusion and a linear shear profile (see~\autoref{ssec:ShearAcc}) the maximal shear acceleration energy is~$E_\mathrm{max}^\mathrm{shear}=\sqrt{3/(\Gamma^2-1)}\,(qBc)/\Gamma\,\Delta r$. Above this energy the acceleration process cannot be described by~\autoref{eq:tau_shear}, therefore we assume the most optimistic case ($\tau_\mathrm{acc}=\lambda/c$) for higher energies.

Comparing the acceleration length scale with the scales of losses and escape leads to a convenient definition for the maximum energy~$E_\mathrm{max}^\mathrm{Acc}$ by
\begin{align}
    \sum_i \tau^{-1}_{\mathrm{acc}, i}\left(E_\mathrm{max}^\mathrm{Acc}\right) = \tau^{-1}_\mathrm{esc}\left(E_\mathrm{max}^\mathrm{Acc}\right) + \sum_j \tau^{-1}_\mathrm{{loss}, j}\left(E_\mathrm{max}^\mathrm{Acc}\right) \quad . \label{eq:defAcc}
\end{align}
Here,~$i$ and~$j$ are wildcards for all possible acceleration and loss processes, respectively. 

Based on~\autoref{eq:defAcc} we define the acceleration probability:
\begin{align}
    P_\mathrm{acc}(E) = \frac{\sum_i {L_{\mathrm{acc}, i}}^{-1}(E)}{\sum_i {L_{\mathrm{acc}, i}}^{-1}(E) + {L_{\mathrm{esc}, i}}^{-1}(E) + \sum_j {L_{\mathrm{loss}, i}}^{-1}(E)} \label{eq:Pacc} 
\end{align}
which describes the chances for acceleration, with~$P_\mathrm{acc}\left(E_\mathrm{max}^\mathrm{Acc}\right) = 0.5$.

\section{FR~0 source environment}
\label{sec:sourceclasses}
In this section we aim to find a reasonable description of the target-photon field expected for CR-interactions in the environment of FR~0s. For this purpose we derive the average source SED in the relevant target field energy range. This includes internal~(jet) and external photon fields, where both particle acceleration and energy losses take place. We note that a complete and self-consistent description of the source physics is beyond the scope of this work which focuses only on the nuclei and their acceleration chances, which are examined in \autoref{ssec:CRsources}.
\subsection{Average source SEDs}
\label{ssec:seds}

The FR~0s sample used here consists of the FR0\emph{CAT}~\cite{Baldi2018} (in its revised version, containing~104 sources, more details can be found in the appendix of~\cite{Baldi2019}) to which were added 10 of the FR~0s studied in~\cite{torresigrandi2018}\footnote{Nine FR~0s in this study were already included in the original FR0\emph{CAT}.} for their X-rays properties. In total, there are then~114 FR~0s for which we collected SED data from the NASA Extra-galactic Database\footnote{\url{ https://ned.ipac.caltech.edu/}} and SSDC Sky Explorer\footnote{\url{https://tools.ssdc.asi.it/}}. 
The selection criteria for the~104 sources included in FR0\emph{CAT} are the following \citep{Baldi2018}:
\begin{itemize}
    \item Redshift $z \leq 0.05$.
    \item FIRST flux $\geq 5$ mJy and no extended emission.
    \item Observed major-axis in FIRST images $\leq 6.7"$, at redshift $z=0.05$, this corresponds to a radius of $\sim$2.5~kpc.
    \item Radio sources with a maximum offset of 2" from the optical center.
\end{itemize}
Similar criteria (less constrained to enlarge the sample) were applied to select the other ten sources~\citep{torresigrandi2018}:
\begin{itemize}
    \item Redshift $z \leq 0.15$.
    \item FIRST flux $\geq 30$ mJy (to ensure X-ray emission) and no extended emission.
    \item Radio size $\leq 10$ kpc.
\end{itemize}
\begin{figure}
    \centering
    \includegraphics[width=.7\textwidth]{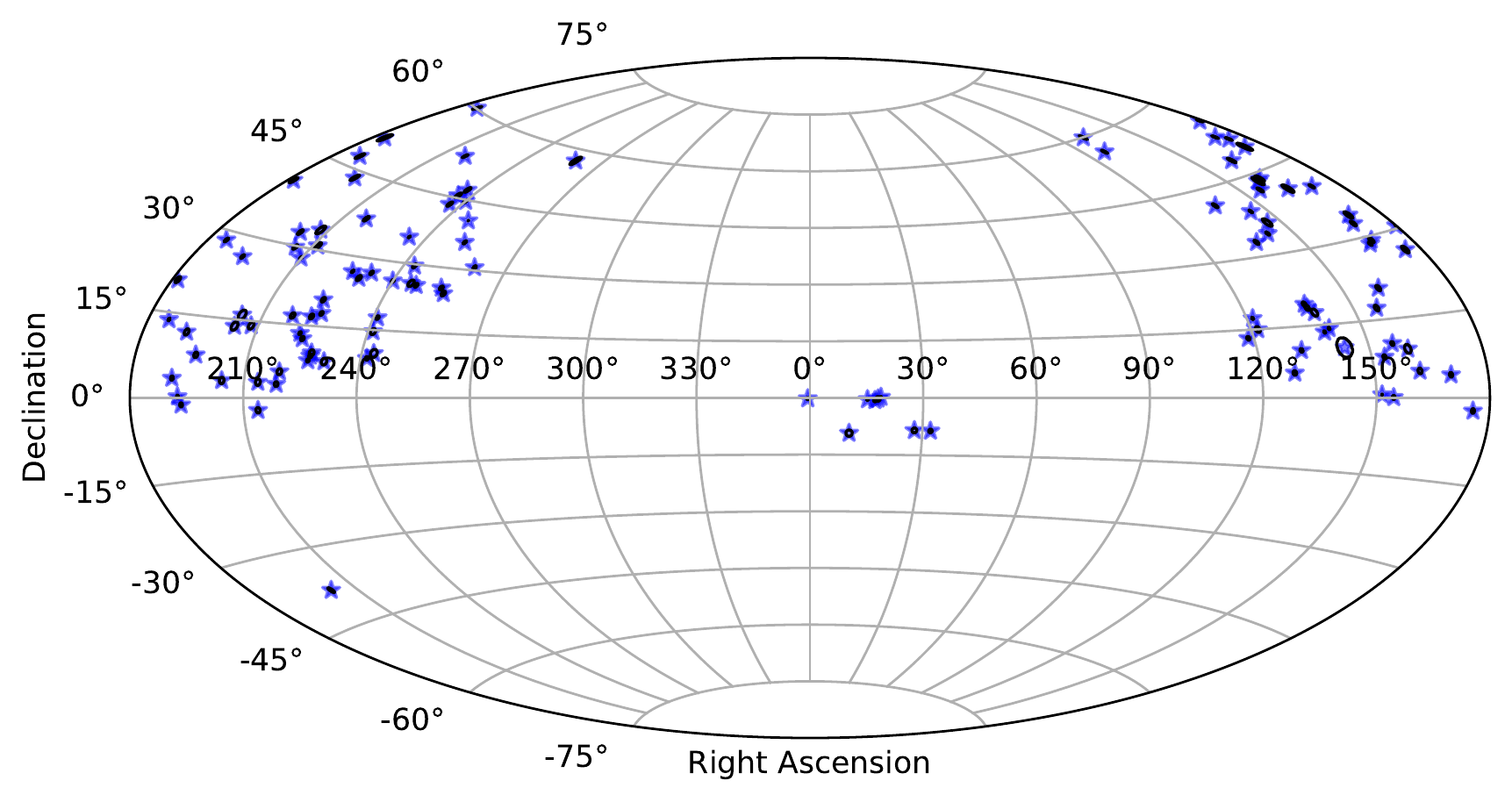}
    \caption{Distribution of the FR~0 sample over the sky, in equatorial coordinates.}
    \label{fig:coordinates}
\end{figure}
While collecting data for the~114 sources, cuts were made on some of the available data points.
First, spectral data points without corrections from other bright sources in the instruments' beam were excluded. Further, a comparison between the sources' position ellipses and the IRAS observations' targeted positions led us to not consider these IRAS data points, as the ellipses did not overlap for the sources concerned.

Due to the low spatial resolution of the UV images provided by Galaxy Evolution Explorer~(\emph{Galex})~\cite{galex}, we follow~\cite{Grandi2016} and treat these data points as upper limits. 

As mentioned before, the case of Tol~1326-379 (which was claimed to be the first FR~0 associated with a $\gamma$-ray counterpart) is uncertain since it is no longer associated with a $\gamma$-ray emission in the last version of the Fermi catalog~4LAC~\cite{Ajello2020}. For this reason, and because they are not relevant as target photon fields for this work, the $\gamma$-ray data obtained for Tol~1326-379 are not included in the global SED.

Based on all the source sample data described above, we construct an average SED, from which we derive the average target photon field in the jet frame. We note that the FIR to UV band is likely dominated by the host galaxy, whereas the rest of the multi-wavelength data is associated with the jet. Hence, two different photon fields, one external and one internal to the jet, need to be considered. From now on, the unprimed quantities refer to the ones in the galaxy frame. In this frame, each source's luminosity is~$L_{\nu} = (4\pi d_\mathrm{L}^2 )/(1 +~z)F_{\nu, \mathrm{obs}}~= 4\pi d_\mathrm{L}^2 F_{\nu}$, where $d_\mathrm{L}$ is the luminosity distance, and $z$ is the redshift. The mean value for the luminosity distance of all the sources in our sample is~$d_\mathrm{L} \approx~5.6\times~10^{26}$cm and~$z \approx 0.04$.

\subsubsection{Jet component}
\label{sssec:jet}

The Synchrotron-Self-Compton~(SSC) model can be considered a \emph{minimal} model to describe the jet component as synchrotron radiation and Inverse Compton scattering inevitably occurs in any magnetized region that hosts relativistic particles. This motivates us to use a simple steady-state SSC model, in the following, to describe the typical jet component of FR~0s at low energies. The apparent positive slope in the X-ray SED of FR~0s may indicate the high-energy component's onset. Different parameters were investigated, focusing on those that impact the maximum energy particles may reach~(see~\autoref{ssec:comparison}). Specifically, the product of the emission region size and the magnetic field strength $r_\mathrm{em}'B'$ was maximized in order to optimize the maximal energy limited by the Hillas criterion. All other parameters have been tuned in order not to overestimate the observed fluxes. In doing so, it was considered that some of the parameters, e.g., the maximal electron Lorentz factor $\gamma_\mathrm{max}'$, might be limited; the theoretical upper limit for $\gamma_{\max}'$ can be estimated by comparing acceleration and synchrotron losses for electrons.

Here we present six different models, computed using the \emph{SSC/EC Simulator} by ASI\footnote{\url{http://www.isdc.unige.ch/sedtool/PROD/SED.html}}. These models are able to represent the synchrotron component of the jet and do not violate the X-ray data. The models use a broken power-law for the electron differential density~$n'(\gamma')$, described with the parameters~$\gamma'_\mathrm{min}$,~$\gamma'_\mathrm{max}$,~$\gamma'_\mathrm{cut}$,~$p_1$ and~$p_2$, such that:
\[ n'(\gamma') \propto \left\{
  \begin{array}{rl}
     \gamma'^{-p_1} & \mbox{if } \gamma'_\mathrm{min} \leq \gamma' < \gamma'_\mathrm{cut} \\
     \gamma'^{-p_2} & \mbox{if } \gamma'_\mathrm{cut} \leq \gamma' \leq \gamma'_\mathrm{max} \\
  \end{array}
\right.\]
where~$\gamma$' is the electrons' Lorentz factor. The parameters for each model can be found in~\autoref{tab:SSCparams}.

\begin{table}
	\centering
    \caption{Parameters of the SSC models. The following parameters are the same for all models:~$\gamma'_\mathrm{min}=100$,~$p_1=2$,~$p_2=3$, and~$\theta = 20 ^\circ$. The luminosity is given in logarithmic units~$\log_{10}(L_i/(\mathrm{erg}\,\mathrm{s}^{-1}))$, where a luminosity ratio between protons and electrons of~$\xi=10$ is assumed.}
    \label{tab:SSCparams}
    \begin{tabular}{lrrrrrrrrrrr}
        \toprule
        & \multicolumn{1}{l}{r$_\mathrm{em}'/\mathrm{cm}$} & \multicolumn{1}{l}{$B'/\mathrm{G}$} &  \multicolumn{1}{l}{$\Gamma$} & \multicolumn{1}{l}{$\gamma_\mathrm{cut}'$} &  \multicolumn{1}{l}{$\gamma_\mathrm{max}'$} &  \multicolumn{1}{l}{$u_e'/(\mathrm{erg}\,\mathrm{cm}^{-3})$} &  \multicolumn{1}{l}{$u_\gamma'/(\mathrm{erg}\,\mathrm{cm}^{-3})$} &   \multicolumn{1}{l}{$L_\mathrm{equ}$} & \multicolumn{1}{l}{$L_\mathrm{CR}$}\\
        \midrule
        
        Model A &  $1\times 10^{18}$        & 1    & 2 &$2\times 10^3$ & $5\times 10^3$   &  $1.29\times 10^{-7}$ &  $7.32\times 10^{-6}$ & 46.8 & 46.5 \\
        
        Model B &  $5\times 10^{17}$ & 0.1  & 2 & $1\times 10^4$ & $2\times 10^4$ & $ 3.05 \times 10^{-5}$ &  $3.09\times 10^{-5}$ & 44.2 & 44.16 \\
                
        Model C\tnote{*} &  $5\times 10^{16}$ & 0.2  & 2 & $1\times 10^4$ & $3\times 10^4$ &   $3.81 \times 10^{-2}$ &  $1.80\times 10^{-2}$  & \multicolumn{1}{c}{--\tnote{**}} & 44.92 \\
        
        Model D & $1.7\times 10^{17}$ & 0.06 & 1.2 & $1\times 10^4$ & $3\times 10^4$ &  $2.67 \times 10^{-3}$ &  $2.26\times 10^{-3}$  & \multicolumn{1}{c}{--\tnote{**}} & 43.94 \\
        
        Model E & $5\times 10^{17}$ & 0.04 & 2 & $1\times 10^4$ & $3\times 10^4$ &  $1.90 \times 10^{-5}$ &  $3.39\times 10^{-6}$  & 43.39 & 43.72 \\
                
        Model F & $5\times 10^{17}$ & 0.04 & 1.26 & $1\times 10^4$ & $3\times 10^4$  &   $6.09 \times 10^{-5}$ &  $1.11\times 10^{-5}$ & 43.02 & 43.75 \\
                
        \bottomrule
    \end{tabular}
    
    \begin{tablenotes}
        \item[*] For \emph{model~C} the second index is set to~$p_2=4.8$.
        
        \item[**] The equipartition scenario is only shown for~$u'_B\gtrsim u'_e$.
    \end{tablenotes}
\end{table}

\begin{figure}[htbp]
    \centering
    \includegraphics[width =.7\textwidth]{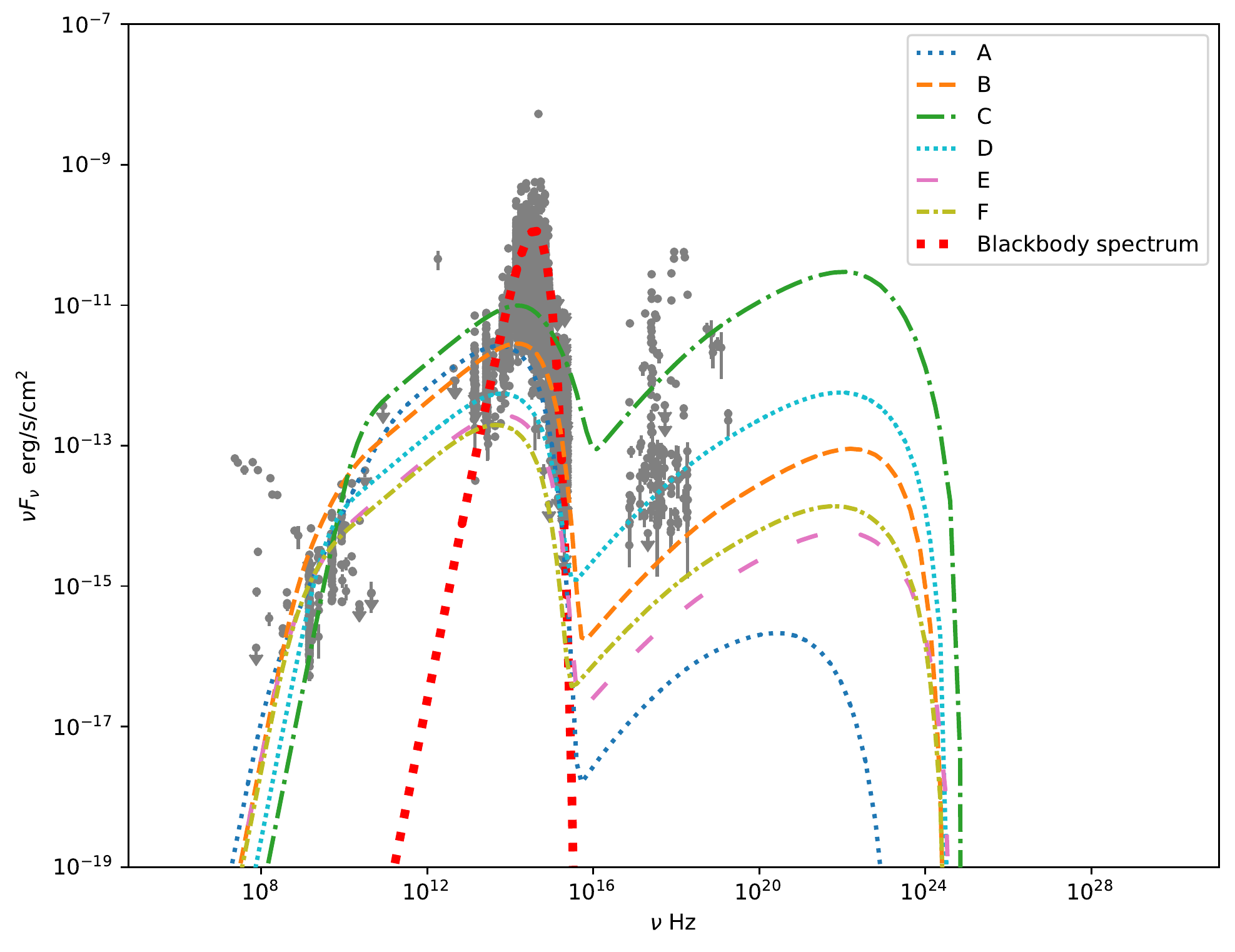}
    \caption{The combined flux spectrum of the~114 FR~0s (black dots) with the different models is shown. The models are mainly distinguished by their prediction of the high energy flux, which increases with decreasing emission region size,~r$_\mathrm{em}'$.} 
    \label{fig:average_SED}
\end{figure}

In \autoref{fig:average_SED}, the model SEDs are shown together with the data. All models agree well on the description of the low energy data~$\nu\lesssim 10^{16}$~Hz. Given the different sizes of the emission region --- and therefore different photon-energy densities --- the second bump is quite different for each of the models. Especially, models A, B, E, and F need an additional (possibly hadronic) component to explain the observed flux above~${\sim}10^{16}$~Hz.

A Lorentz transformation is used to evaluate the jet radiation field in the comoving frame. For an internal photon field, if the emission is isotropic in the comoving frame~(denoted with primed quantities), the jet frame specific energy density is~\citep{Dermer2009b}:

\begin{align}
     u'_{\nu'} = \frac{3d_\mathrm{L}^2 F_{\nu}}{\delta_D^3 r_\mathrm{em}'^2 c} 
\end{align}
where~$F_\nu$ is the spectral flux density,~$\delta_D = (\Gamma_\mathrm{j}(1 - \beta \mu))^{-1}$ is the Doppler factor, $\mu$ is the cosine of the angle between the line of sight and jet direction, $\Gamma_\mathrm{j}$ is the jet's Lorentz factor,~$\beta = (1 - \Gamma_\mathrm{j}^{-2})^{1/2}$ is the normalized velocity, and~$r_\mathrm{em}'$ is the emission region comoving size. 

For all models, the energy densities of the three major constituents are listed in~\autoref{tab:SSCparams}:~1) The 
magnetic field energy density (where a homogeneous turbulent field strength is assumed),~2) The total radiation energy density~$u_\gamma'$, and lastly,~3) The CR energy densities~$u_\mathrm{CR}'=u_e' + u_p'$, with contributions from electrons~$u_e'$ and protons~$u_p'$.

We estimate the protons' contribution in two ways: 1) Assuming equipartition of the magnetic field energy density and the cosmic-ray energy density~$u_B'=u_\mathrm{CR}'$ (which makes sense only if the former is larger than the electron energy density~$u_B'>u_e'$) and,~2)~Assuming a constant energy density ratio of protons and electrons~$u_p'=\xi u_e'$. Additionally, assuming charge neutrality for the non-thermal proton and electron distribution and the same power-law index~$\alpha\approx 2.2$ for their spectra, the luminosity ratio becomes~$\xi\approx 10-100$. The ratio~$\xi$ can change significantly when both populations follow different spectral slopes (see, e.g., \cite{Persic2014, Merten2017}). However, in this work, we fix the ratio to~$\xi=10$ for our calculations.

The total jet luminosity is given by
\begin{align}
    L_\mathrm{tot} = 2\pi\,c\,r'^2_\mathrm{em}\,\Gamma^2_\mathrm{j}\sum_i u_i' \quad , 
\end{align}
where~$u_i'$ are all relevant energy densities, and is calculated for all models (see~\autoref{tab:SSCparams}).

Comparing the derived values with the jet's average mechanical power,~$L_\mathrm{jet}\lesssim 10^{43.5}\;\mathrm{erg}\,\mathrm{s}^{-1}$ (cf.\ \autoref{sec:intro}), shows that only models~E and~F comply with that boundary. However, the underlying correlation is not well restricted~\citep{Heckman2014}; therefore, it is plausible that some of the FR~0s will have enough jet power to support models~B,~C, and~D, and only \emph{model~A} will be neglected for the rest of this work.

\subsubsection{Host galaxy component}
\label{sssec:host}

Following the treatment of~\cite{Stawarz2006}, the spectral and radial dependencies are separated to express the energy density provided by the host galaxy inside a given region of radius R:~$u(E,R) = u(E)\, u(R)$. A Sersic profile can adequately reproduce the radial dependence of the surface brightness from elliptical galaxies~\citep{Sersic1968}:
\begin{align}
    I(R) = I_0\exp{(-b\eta^{1/m})} \quad ,
\end{align}
where~$\eta = R/R_e$ and~$R_e$ is the half-light radius of the elliptical galaxy that encloses half of the total luminosity. For elliptical galaxies,~$R_e \sim 1$~kpc, and with~$m~=~4$, we recover the De Vaucouleur's profile~\citep{DeVaucouleurs1948}. The advantage of this formulation is that it can be solved analytically~\citep{Ciotti1999} to give the projected monochromatic luminosity inside a given radius~R:
\begin{align}
    L(R) = I_0\,R_e^2\,\frac{8\pi }{b^{8}}\gamma(8, b\eta^{1/4}) \label{lumi_radius}
\end{align}
with~$b \approx 7.669$.
Here~$\gamma(\alpha, x)$ is the incomplete gamma function.
The total luminosity can then be calculated as
\begin{align}
    L_\mathrm{tot} = I_0 R_e^2 \frac{2\pi m}{b^{2m}} \Gamma(2m),
\end{align}
which gives its normalization
\begin{align}
    I_0 = \frac{L_\mathrm{tot} b^{2m} }{R_e^2 2\pi m \Gamma(2m)}
\end{align}
with the complete gamma function~$\Gamma(\alpha) = \gamma(\alpha, \infty)$.

For the spectral part, we used a diluted black-body spectrum to reproduce the FIR to UV component, for which the normalized energy density is:
\begin{align}
    u(E) = \epsilon_\mathrm{dil} \frac{8\pi}{h^3 c^3}\frac{E^3}{e^{\frac{E}{k_\mathrm{B} T}} - 1} \nonumber \quad ,
\end{align}
with~$R = \dfrac{r_\mathrm{em}'}{\delta_D}$ being the size of the emitting region in the host galaxy frame. The luminosity is related to the energy density by
\begin{align}
    L(E) = 4\pi R^2 c u(E) \label{lumi_energy} \quad .
\end{align}
We then determine the dilution factor~$\epsilon_\mathrm{dil}$ by adjusting~$\int_0^{\infty}L(E,R) \mathrm{d}R = L_\mathrm{tot}\, u(E)$ to match the data with~$L(E,R) = u(E)\,L(R)$.

Finally the energy density for a region of size~R is then given by:
\begin{align}
    u(E,R) = \frac{L(E,R)}{4\pi R^2 c} = u(E)\,\frac{L(R)}{4\pi R^2 c} = u(E)\,u(R) \label{energy_density} \quad .
\end{align}

\begin{table}[htbp]
    \centering
    \caption{Parameters of the host galaxy model}
    \label{tab:BBparams}

    \begin{tabular}{lrrrrr}
        \toprule
        & \multicolumn{1}{l}{$L_\mathrm{tot}/(\mathrm{erg}\,\mathrm{s}^{-1})$} & \multicolumn{1}{l}{$T/\mathrm{K}$} & \multicolumn{1}{l}{$\epsilon_\mathrm{dil}/(\mathrm{cm}^3\,\mathrm{erg}^{-1})$} & \multicolumn{1}{l}{$R_e/\mathrm{kpc}$} & \multicolumn{1}{l}{$m$} \\
        \midrule

        blackbody &  $ 6.4\times 10^{44}$ & 4801 & 0.29 & 1 & 4 \\
        \bottomrule
        
        \end{tabular}
\end{table}

We use the invariance of~$u(E,\Omega)/E^3$ for the transformation of this external photon field energy density into the jet's frame: $u'(E',\mu', r_\mathrm{em}') = u(E, R)/(2\Gamma_\mathrm{j}^3(1 + \beta \mu')^3)$ for an isotropic photon field in the galaxy frame.
Replacing~$u(E,R)$ by~\autoref{energy_density} and integrating over~$\mu'$ one gets:
\begin{align}
    u'(E', r_\mathrm{em}') &=& \frac{8\pi k_\mathrm{B} T}{2 h^3 c^3 \Gamma_\mathrm{j} \beta} \epsilon_\mathrm{dil} \, E'^2 \, u(R) 
    \left[ \ln{\left(\frac{\exp{(x_\mathrm{max})} - 1}{\exp{(x_\mathrm{min})} - 1}\right)} - (x_\mathrm{max} - x_\mathrm{min})\right] 
\end{align} 
with~$ x_\mathrm{min} =\dfrac{\Gamma_\mathrm{j} E'}{k_\mathrm{B} T}(1 - \beta)$ and~$ x_\mathrm{max} = \dfrac{\Gamma_\mathrm{j} E'}{k_\mathrm{B} T}(1 + \beta)$.

\subsection{CR Acceleration in Loss Environment}
\label{ssec:CRsources}
Based on our estimates above for the typical target photon fields expected in FR~0 environments, it is now possible to evaluate their role as potential cosmic-ray sources. The target photon field's modeling fixes many parameters of the overall source region model, as explained above. It should be noted that some of the parameters are degenerate in the sense that other combinations of the fitting parameters might have led to equally good approximations of the observed photon fluxes. However, the FR 0 SEDs spread is much larger than the changes introduced by varying some of the degenerated parameters.

In the following, two scenarios will be tested based on Bohm and Kolmogorov diffusion. Bohm diffusion is the limiting case with the shortest time scale of all Fermi-I acceleration processes and does not need the introduction of parameters in addition to those given by the SSC model. However, at least on Galactic scales, observational evidence points towards a Kolmogorov diffusion~$\kappa(E)= \kappa_0\,(E/(Z E_0))^{1/3}$~(see, e.g.,~\cite{BECKERTJUS2020}). We fix the normalization by requiring that a particle with an energy according to the Hillas criterion $E_\mathrm{max}^\mathrm{Hillas}$ has a mean free path smaller than the emission region size.\footnote{This gives comparable values to the isotropic turbulence case described in \cite{Dundovic2020}, assuming that the maximum wavelength of the turbulence spectrum $l_\mathrm{max}=5l_\mathrm{c, iso}$ fits into the emission region $r_\mathrm{em}$} This will lead by construction to comparable time scales for the Bohm and Kolmogorov case and any difference will be mainly due to the different spectral indices. If another turbulence model or normalization were realized, the time scales will change significantly. However, the currently available observations are not able to discriminate such speculations from each other. Furthermore, we are assuming a wide shear layer which is given by the size of the emission region $r_\mathrm{em} = \Delta r$ and constant gradient $\partial u_z/\partial r = \beta_\mathrm{jet}c/\Delta r$ (see \autoref{eq:tauBohmShear}).

In addition, we will focus on Fermi-I and gradual shear acceleration for the rest of this work. These two cases are among the most promising processes considering FR~0s as sources of CRs. Classical Fermi-II acceleration can be neglected since it is not fast enough compared with the former two. Blast wave acceleration only provides a fast energy increase for the acceleration cycle's first encounter. For this reason, it provides a better mechanism for re-acceleration than for the full acceleration cycle. One of the neglected scenarios could, however, still play some role in the acceleration. 

Therefore, two acceleration scenarios are considered. For the first acceleration scenario, we consider pure diffusive shock acceleration, whereas the second scenario assumes a \emph{hybrid} approach that models the energy gain by a combination of Fermi-I and gradual shear acceleration.

\begin{figure}[htbp]
    \centering
    \includegraphics[width=.65\linewidth]{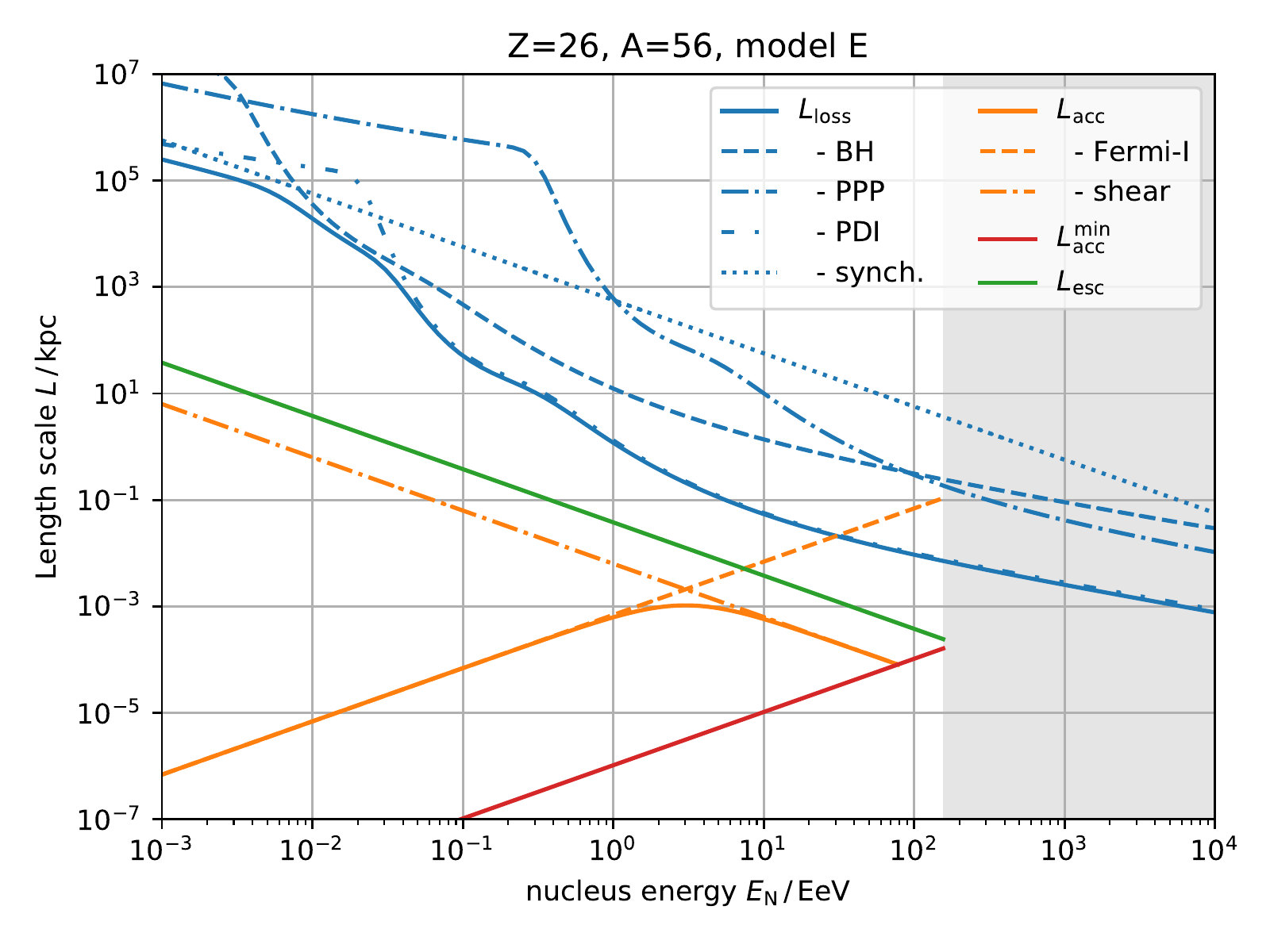}
    \caption{A summary of all relevant length scales for iron nuclei is shown. The parameters for the target photon field, magnetic field strength, and size of the emission region used here are that of \emph{model~E}. It is clear that the interplay between losses, escape, and acceleration can become complex. The total loss length~$L_\mathrm{loss}$~(blue line) combines pair production~(BH), photo-pion production~(PPP), photo-disintegration~(PDI), and synchrotron losses. The same is done for gradual shear and Fermi-I acceleration~(orange lines). The escape length scale is shown in green, and the minimum acceleration length, corresponding to $\eta=1$, is depicted in red. The gray shaded area shows the region above the Hillas energy calculated for $\beta=1$.}
    \label{fig:lengthsiron}
\end{figure}

\Autoref{fig:lengthsiron} shows all relevant length scales for iron in a single plot\footnote{Similar plots for all other tracer elements and target field models are shown in the appendix~\autoref{fig:AllLengths1} and~\autoref{fig:AllLengths2}. \Autoref{fig:AllLengths1_kolm} and~\autoref{fig:AllLengths2_kolm} show the same plots based on Kolmogorov diffusion.}. Exemplarily, the Bohm diffusion scenario is presented here using \emph{model~E}. The complexity of the processes involved makes it hard to draw quantitative conclusions about the maximum energy that might be reached in the underlying \emph{model~E} from a plot like this alone. Due to the rather broad-banded target photon field, all four loss processes are no longer sharply distinguished but smoothed into one another over several orders of magnitude in energy. As expected from the results presented in~\autoref{ssec:losses}, the most relevant process for heavy nuclei is photo-disintegration. In contrast to the monochromatic target fields discussed above, photo-disintegration has a non negligible --- sometimes dominant --- influence on the total loss length even at the highest nuclei energies~$E>1000\;\mathrm{EeV}$.

Gradual shear acceleration can become the dominant acceleration mechanism at high energies, as reported in, e.g., \cite{rieger_review}. The transition energy can be derived as
\begin{align}
    E_\mathrm{shear-Fermi} = Z\,E_0 \left(\sqrt{\frac{\alpha}{4(4+\alpha)}} \frac{1}{\sqrt{b_\mathrm{j}}\,\Gamma_\mathrm{j}^2} \frac{u_s\,r_\mathrm{em}}{\kappa_0}\right)^{1/\alpha} \quad .
\end{align}

\begin{figure}[htbp]
    \centering
    \includegraphics[width=.65\linewidth]{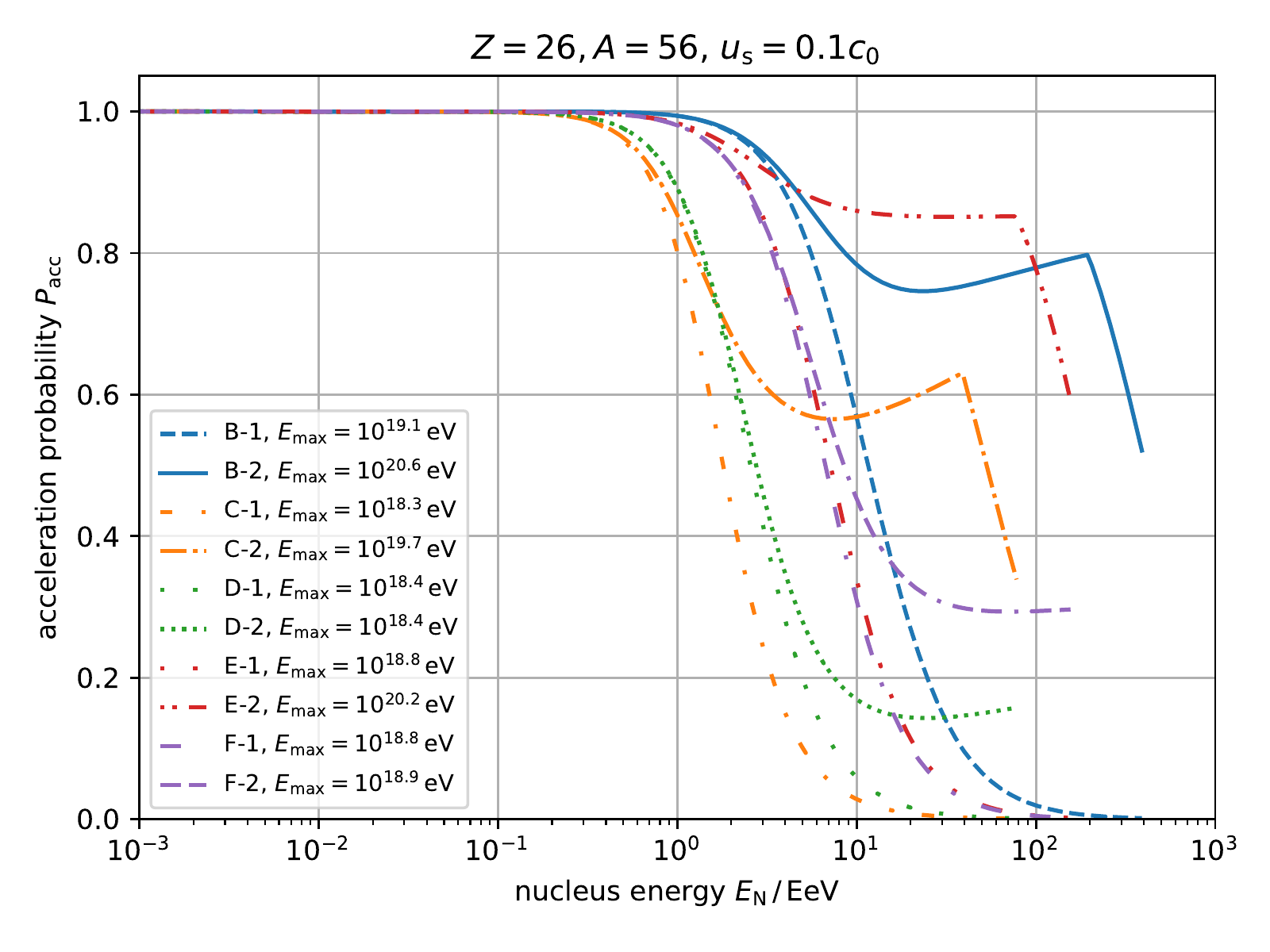}
    \caption{Acceleration probability for iron, based on all investigated models, is shown. The lines are cut-off at the Hillas criterion. The sharp break, e.g., in \emph{model B-2} occurs at the maximum shear acceleration energy $E_\mathrm{max}^\mathrm{shear}$. For most acceleration models, the jet frame's actual maximum energy is lower:~$E_\mathrm{max}^\mathrm{acc}=E(P_\mathrm{acc}=0.5)$.}
    \label{fig:P_acc_Fe}
\end{figure}

\Autoref{fig:P_acc_Fe} shows ten models for the acceleration probability of iron~(B-1 to F-2) for comparison. For the models based on pure Fermi-I acceleration, the probability will drop below~$P_\mathrm{acc}<0.5$ at some point. This statement holds independently of the chosen target field mode, as was already shown in~\autoref{ssec:acceleration}. The maximum energy might be even smaller when the losses become significant.

In contrast, the hybrid scenario could lead to a saturation of the acceleration probability at larger energies~$\lim_{E\to\infty} P_\mathrm{acc}(E)>0.5$, when $E_\mathrm{max}^\mathrm{shear}\geq E_\mathrm{max}^\mathrm{Hillas}$. This saturation is because the loss length depends only weakly on energy at the highest energies, and the escape and gradual shear acceleration rates have the same energy dependence~$\propto~E^{-\alpha}$ (and are stronger compared with the losses). When shear acceleration is cut off by the maximum acceleration rate ($E_\mathrm{max}^\mathrm{shear} < E_\mathrm{max}^\mathrm{Hillas}$) a steep decrease of the acceleration probability can be observed for $E>E_\mathrm{max}^\mathrm{shear}$. This is due to the fact that the sign of the power-law index $\alpha$ of energy dependence of the acceleration time scale flips at that energy.

The maximum energy in the jet frame for all the models has been derived by calculating the upper limit of the Hillas energy~$E_\mathrm{max}^\mathrm{Hillas}$ assuming $\beta=1$, and the maximum acceleration energy given by~$P_\mathrm{acc}\left(E_\mathrm{max}^\mathrm{acc}\right)=0.5$; the smaller of the two defines the maximum energy. The results for all models, and the five tracer elements, are given in~\autoref{tab:Emax} for the Bohm diffusion ansatz and in \autoref{tab:Emax_kolm} for Kolmogorov diffusion. From these five tracer elements, the average maximal rigidity~$\zeta$ was also calculated. The shock speed~$u_s$, has of course, some influence on the time scale of the Fermi-I acceleration. However, the effect on the maximum energy is almost negligible for~$0.03<u_s/c<0.1$. Therefore, all numbers shown in this work are calculated for~$u_s=0.1\,c$. 

\begin{table}[htbp]
    \centering
    \caption{Maximum Energy in the jet frame. Here the Bohm diffusion scenario is shown where the X-1 models are based on pure Fermi-I, and the X-2 models use the hybrid approach. The shock speed was fixed at~$u_s=0.1\,c$.}
    \label{tab:Emax}

        \begin{tabular}{lrrrrrr}
            \toprule
            model & \multicolumn{5}{c}{$\log_{10}(E_\mathrm{max}'/\mathrm{eV})$} & \multicolumn{1}{l}{$\langle \log_{10}(\zeta_\mathrm{max}'/ \mathrm{V})\rangle$\tnote{a}} \\
                   & \multicolumn{1}{l}{p} & \multicolumn{1}{l}{He} & \multicolumn{1}{l}{N} & \multicolumn{1}{l}{Si} & \multicolumn{1}{l}{Fe} & \\ 
            \midrule
            B-1 & $17.9$ & $18.0$ & $18.4$ & $18.8$ & $19.1$ & $17.70\pm0.11$\\
            C-1 & $17.1$ & $17.3$ & $17.6$ & $17.9$ & $18.3$ & $16.90\pm0.14$\\
            D-1 & $17.2$ & $17.4$ & $17.7$ & $18.1$ & $18.4$ & $17.02\pm0.12$\\
            E-1 & $17.4$ & $17.8$ & $18.3$ & $18.6$ & $18.8$ & $17.44\pm0.04$\\
            F-1 & $17.4$ & $17.7$ & $18.2$ & $18.5$ & $18.8$ & $17.38\pm0.02$\\
            \midrule
            B-2 & $19.2$ & $19.5$ & $20.0$ & $20.3$ & $20.6$ & $19.18\pm0.02$\\
            C-2 & $18.5$ & $18.7$ & $17.8$ & $19.4$ & $19.7$ & $18.08\pm0.57$\\
            D-2 & $17.2$ & $17.4$ & $17.8$ & $18.1$ & $18.4$ & $17.04\pm0.10$\\
            E-2 & $18.8$ & $19.1$ & $19.6$ & $19.9$ & $20.2$ & $18.78\pm0.02$\\
            F-2 & $17.6$ & $17.9$ & $18.3$ & $18.7$ & $18.9$ & $17.54\pm0.06$\\
            \bottomrule
        \end{tabular}
        \begin{tablenotes}
            \item[a] $\zeta=E/q$ is the rigidity of the particle in units of volt.
        \end{tablenotes}
\end{table}

Models based on the hybrid ansatz lead, on average, to an increase of the maximum rigidity ~$\langle \zeta_\mathrm{max} \rangle$ (which can be as large as a factor of $\sim$20). Furthermore, a higher charge number~$Z$ generally leads to a higher maximum energy as expected from the acceleration mechanism. However, when the target photon field density is high, and the FR~0 SED shows a larger second bump (e.g., in \emph{model~C}), a clear proportionality with~$Z$ cannot be found. This non-proportionality effect is more pronounced in the hybrid scenario than Fermi-type acceleration alone. In general, it can be observed that only in a hybrid scenario, energies above the ankle can be reached. Fermi-type acceleration alone is only efficient enough to reach energies below the ankle. 

Looking at the magnetic turbulence model's influence, it is interesting that pure Fermi-1 order acceleration in a Kolmogorov turbulence spectrum does not generate UHECR at all. In this case, the maximum energy is determined by balancing escape and acceleration and is, therefore, independent of the SED. However, gradual shear acceleration leads to the highest maximum energies that we have found within our model; see \emph{model~B-2} and \emph{model~E-2} in~\autoref{tab:Emax_kolm}.

In terms of the maximal energy reachable, \emph{model~B} and \emph{model~E} are the most promising ones. 

\section{Summary \& Conclusions}
\label{sec:summary}
In this work, the relatively recently discovered class of FR~0 radio galaxies has been discussed as a potential source class for UHECRs. FR~0s are much more numerous than more energetic radio galaxies such as FR~1s or FR~2s (see~\autoref{sec:intro}). In this case, each source does not need to provide a high CR flux, which can be advantageous when modeling the rather isotropic CR arrival directions. In doing so, a brief summary of the most relevant processes has been given in~\autoref{sec:background}, where different acceleration, energy loss, and escape models have been explained.

To answer the question of whether UHECRs can be accelerated, and survive, in the source region of FR~0s the available multi-wavelength data was compiled into an average SED to describe the expected target field for CR-photon interactions. This SED was decomposed into a jet component (based on six different SSC models) and a host galaxy component. The host galaxy was assumed to be a typical elliptical galaxy and represented by a de-Vaucouleurs'-radial-luminosity-profile combined with a diluted blackbody energy spectrum. Based on the Doppler and Lorentz factor constraints of the jetted regions the jet frame's expected photon target field has been derived. This modeling allowed a reasonable estimation of the time scales, of all relevant particle-photon interactions, to be expected in the environment of FR~0s, in tandem with acceleration and escape. 

We define the probability~$P_\mathrm{acc}$ as a measure for the acceleration potential to derive the expected maximal energy for all different target field models. Since the nature of the magnetic field is not known, two turbulence models~(Bohm or Kolmogorov) have been tested, as well as different shock speeds~$u_s$. From our calculations, we can draw the following main conclusions:
\begin{enumerate}
    \item Acceleration of UHECR, up to the highest measured energies, in FR~0s, is possible under certain conditions:~a)~Acceleration due to the \emph{hybrid} scenario, where HECR are predominantly accelerated by Fermi-I acceleration and the most energetic particles are driven by gradual shear acceleration. b)~A jet Lorentz factor~$\Gamma_\mathrm{j}\gtrsim 1.6$, so gradual shear acceleration happens on faster time scales than the escape (given a linear flow profile). c)~According to our models an emission and acceleration region size of~${\sim}5\times 10^{17}$ cm (as described in, e.g., \emph{model~E}) is the most promising size scale. 
    
    \item Less optimistic scenarios, based on pure Fermi-I acceleration with Bohm diffusion, still reach energies large enough to contribute to the CR flux, between the knee and the ankle. Models based on pure Fermi-I acceleration within Kolmogorov turbulence are not able to reach energies significantly above the knee.
    
    \item The influence of the realized turbulence model --- Bohm or Kolmogorov --- is found to be relatively weak. This is partly due to the similar normalization of the two turbulence models. In reality, the diffusion coefficient might be different and may be incapable of keeping the most energetic particles confined in the acceleration region.
    
    \item Within the hybrid scenario, the shock speed's~$u_s$ influence on the maximum energy is weak as well. However, it has an impact on the transition energy between the two acceleration modes.

    \item A scenario based on gradual shear acceleration alone is not realistic, as the acceleration time scales at the lowest energies are too large due to the small boosting factors of FR~0 jets. In the hybrid approach, this problem is overcome by Fermi-I acceleration but could also be realized by other mechanisms such as, e.g., magnetic reconnection~\cite{Comisso2019}. 

    \item Despite the fact that FR~0s are radiatively weak sources, radiative losses not only play an important role for the maximum CR energy of FR~0s but they will have a significant impact on the expected cosmic-ray spectrum at the source~(see~\autoref{fig:AllLengths1}, ~\autoref{fig:AllLengths2}, ~\autoref{fig:AllLengths1_kolm}, and ~\autoref{fig:AllLengths2_kolm}). However, we find that synchrotron losses of nuclei are largely negligible. This suggests that a simple unbroken power-law with a cut off is probably insufficient to describe the cosmic-ray energy distribution escaping the source region.
\end{enumerate}

We can summarize that FR~0s make an interesting candidate source class for UHECRs that deserves further investigations.

The code developed for this work is available upon request in a repository for reference. 

\section*{Acknowledgments}
This work acknowledges financial support from the Austrian Science Fund (FWF) under grant agreement number I 4144-N27 and from the Slovenian Research Agency - ARRS~(project no. N1-0111). MB has for this project received funding from the European Union’s Horizon 2020 research and innovation program under the Marie Sklodowska-Curie grant agreement No 847476. The views and opinions expressed herein do not necessarily reflect those of the European Commission. G.B.\ acknowledges financial support from the State Agency for Research of the Spanish MCIU through the "Center of Excellence Severo Ochoa" award to the Instituto de Astrofísica de Andalucía (SEV-2017-0709).

This research has made use of the NASA/IPAC Extragalactic Database (NED), which is funded by the National Aeronautics and Space Administration (NASA) and operated by the California Institute of Technology (CIT). Furthermore, part of this work is based on archival data, software, or online services provided by the Space Science Data Center (SSDC) via ASI.

This work benefited from the following software: CRPropa~\citep{crpropa30, crpropa31}, NumPy \citep{numpy}, Matplotlib \citep{matplotlib}, pandas \citep{pandas}, SOPHIA \cite{sophia}, jupyter notebooks \cite{ipython}, AGN SED tool \citep{Massaro2006, Tramacere2009, Tramacere2011}.

\bibliography{literature}{}
\bibliographystyle{elsarticle-num-names}

\clearpage
\appendix

\section{Loss, Escape, and Acceleration Lengths}

\begin{figure}[!ht]
    \centering
        \begin{minipage}{.49\textwidth}
        \includegraphics[width=\textwidth]{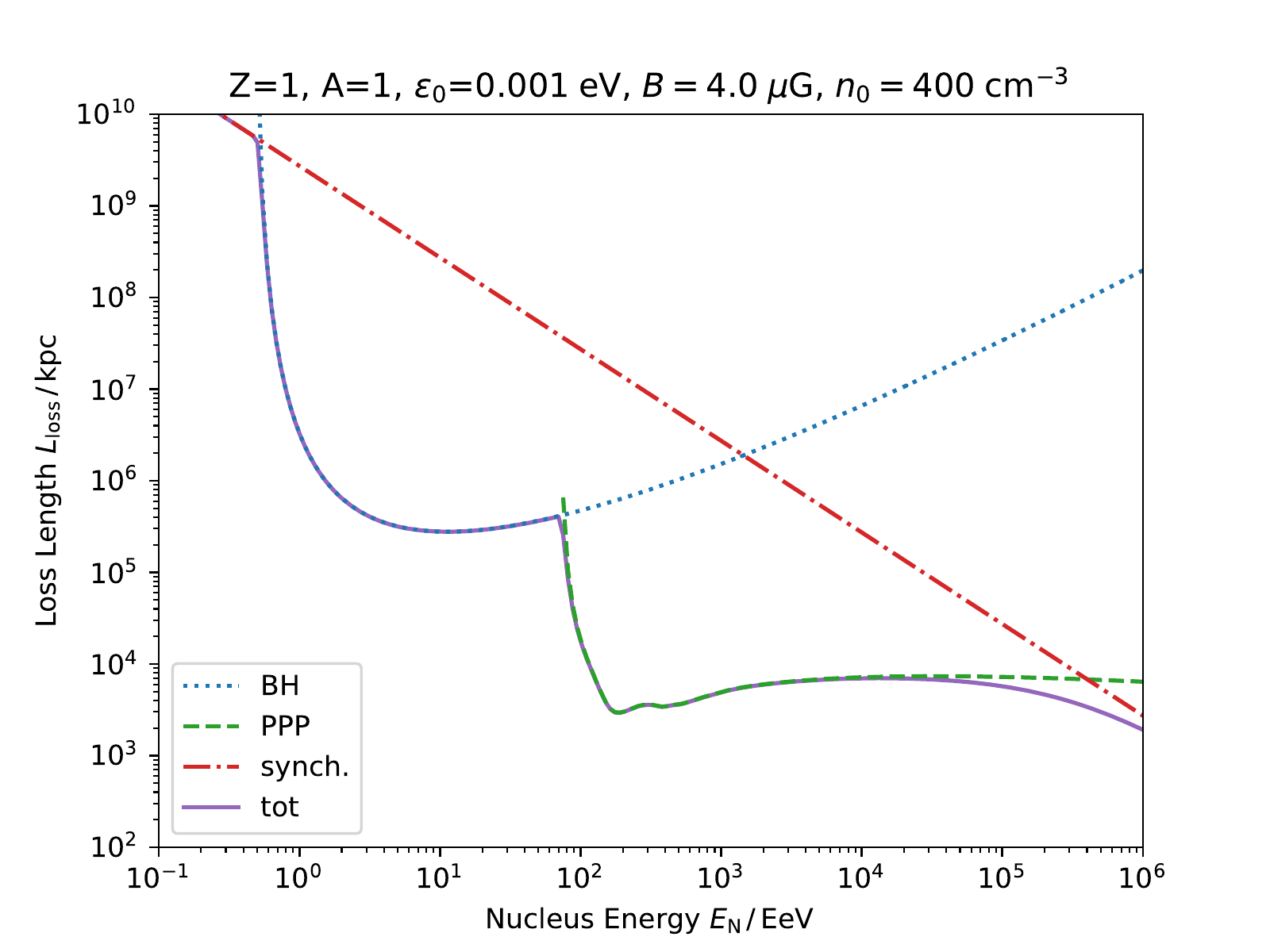}
        \includegraphics[width=\textwidth]{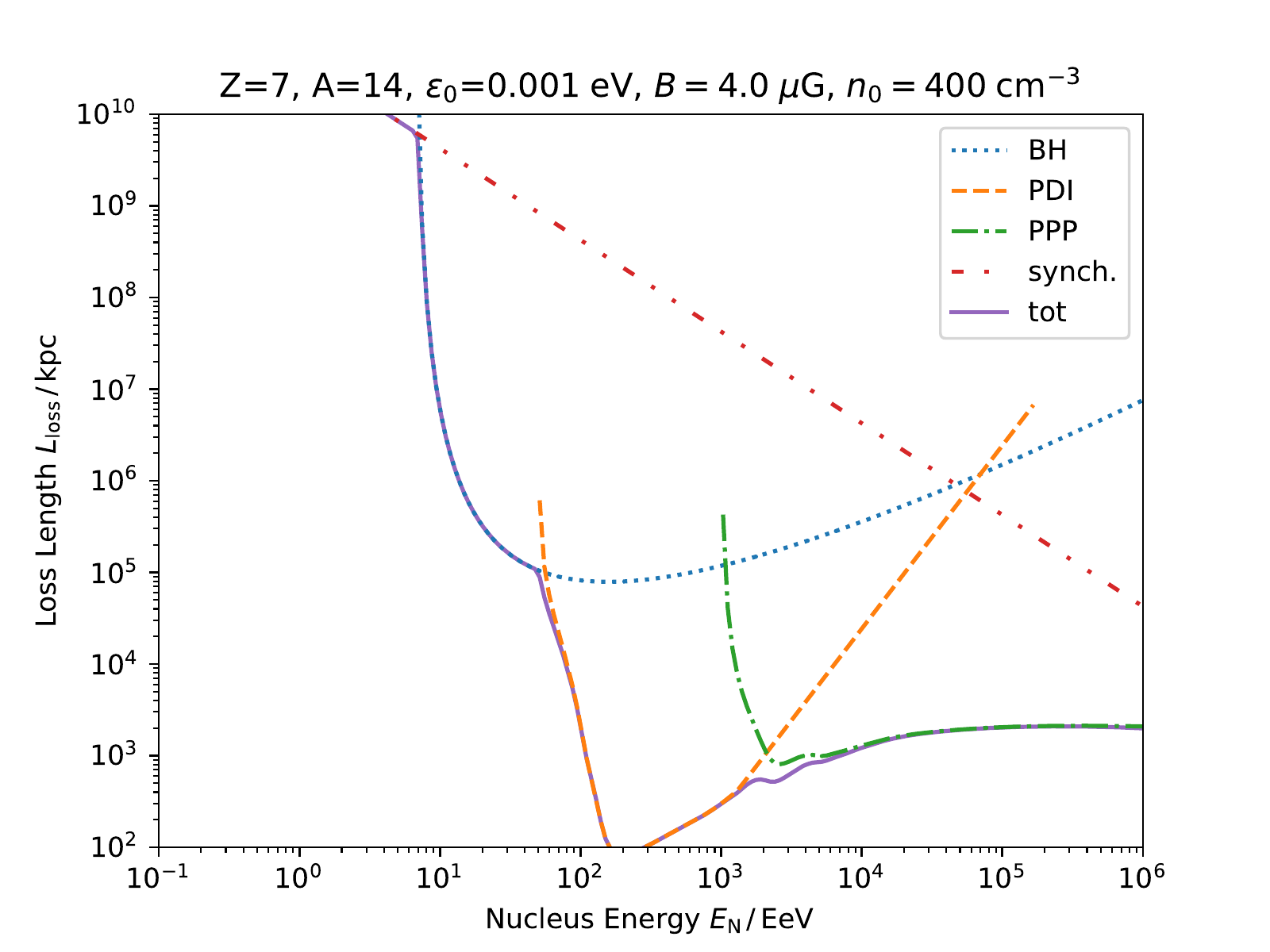}
    \end{minipage}%
    \begin{minipage}{.49\textwidth}
    \includegraphics[width=\textwidth]{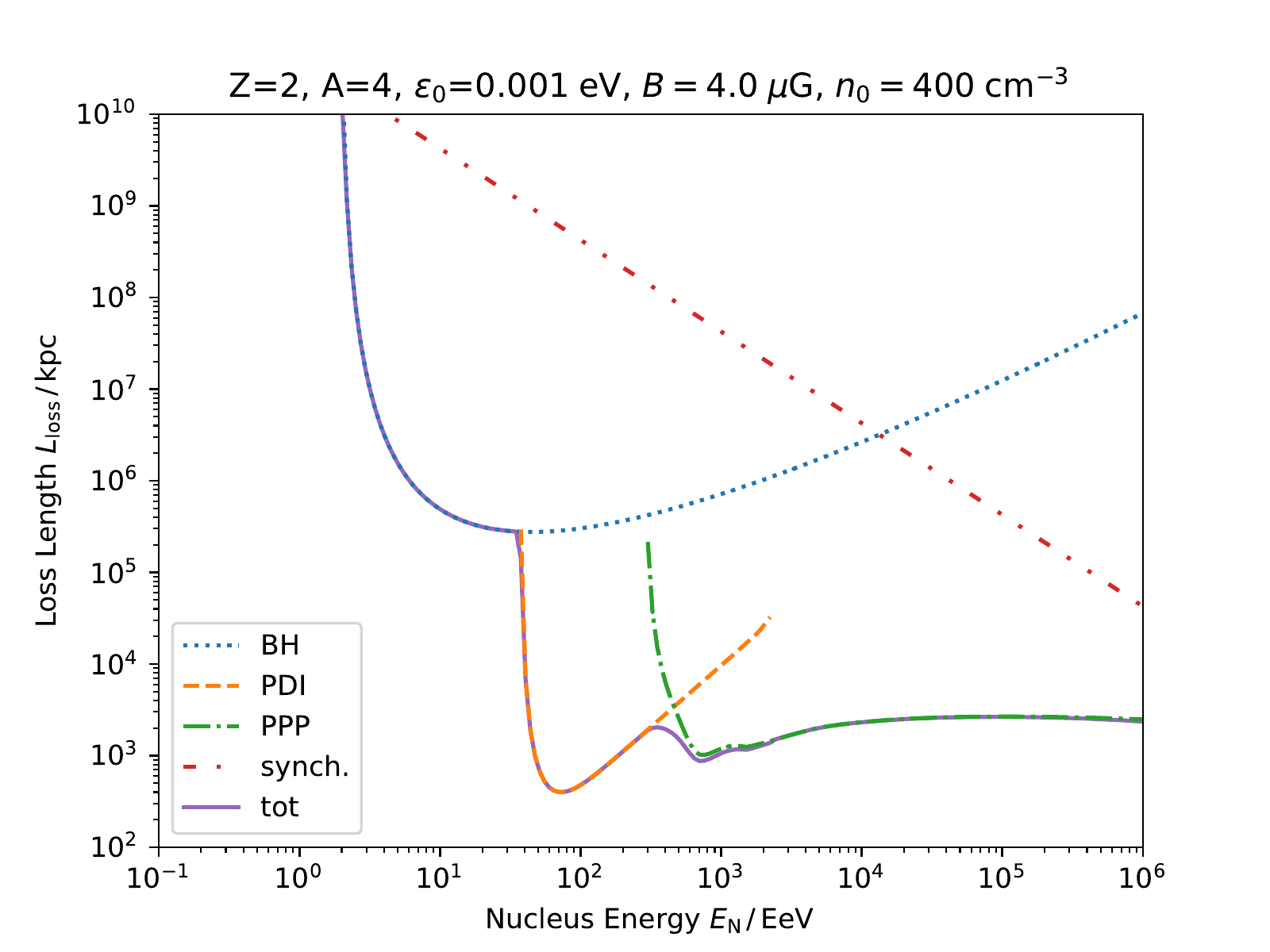}
    \includegraphics[width=\textwidth]{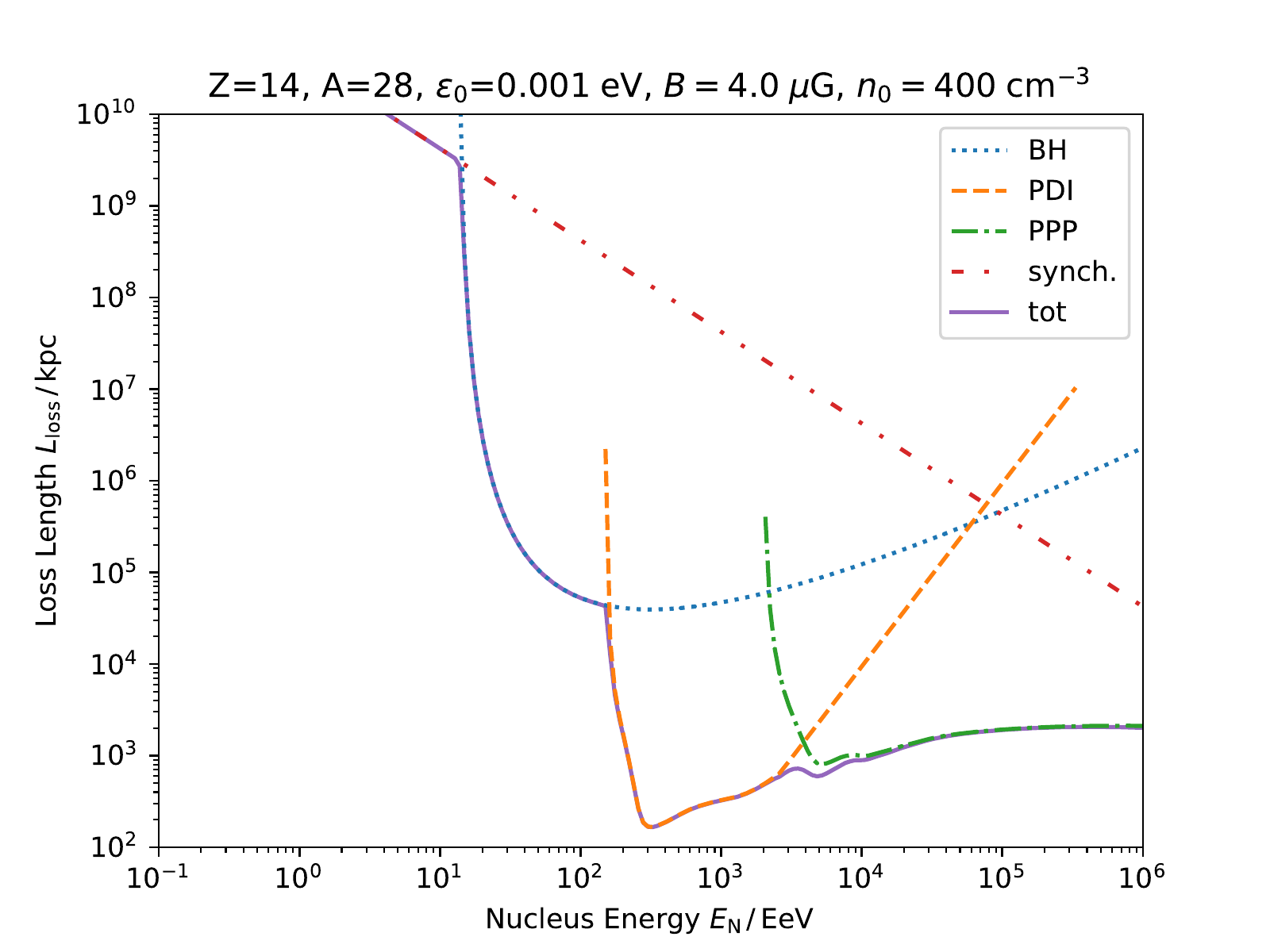}
    \end{minipage}
    \caption{Loss lengths for the tracer elements hydrogen, helium, nitrogen, and silicon is shown. It is clearly visible that pair production (blue dotted line) can almost be neglected for a CMB-like monoenergetic target field. For the chosen magnetic field strength, yielding equipartition between the magnetic and the target photon field $u_\mathrm{mag}=u_\mathrm{target}$, synchrotron losses (red dash-dash-dotted) can be neglected as well. The short loss length is due to photo-disintegration (orange dashed line), while at the very highest energies, the losses are dominated by photo-pion production (green dash-dotted line).}
    \label{fig:LL_monochromatic}
\end{figure}

\begin{figure}[htbp]
    \centering
    \begin{minipage}{.49\textwidth}
        \includegraphics[width=\textwidth]{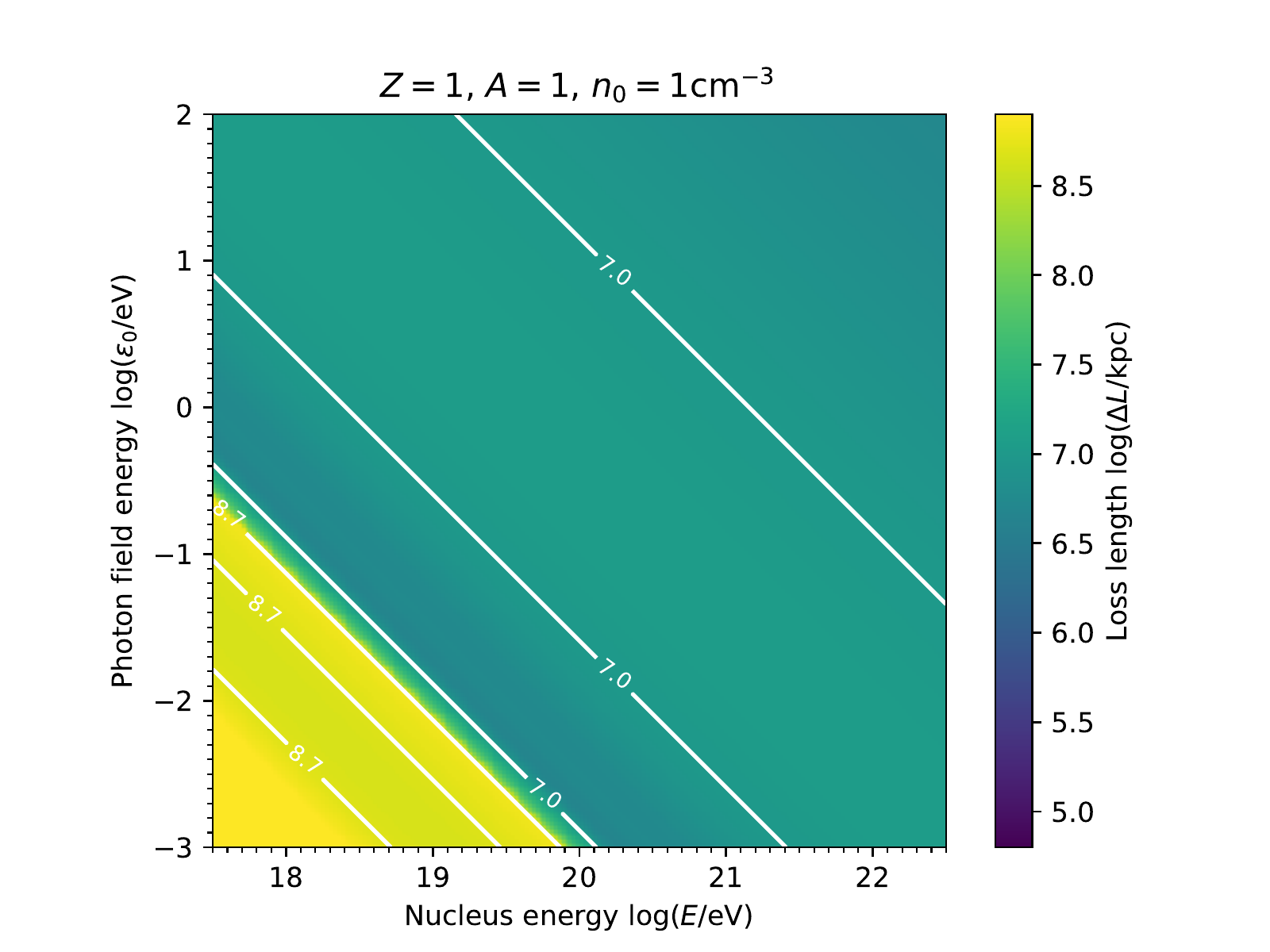}
        \includegraphics[width=\textwidth]{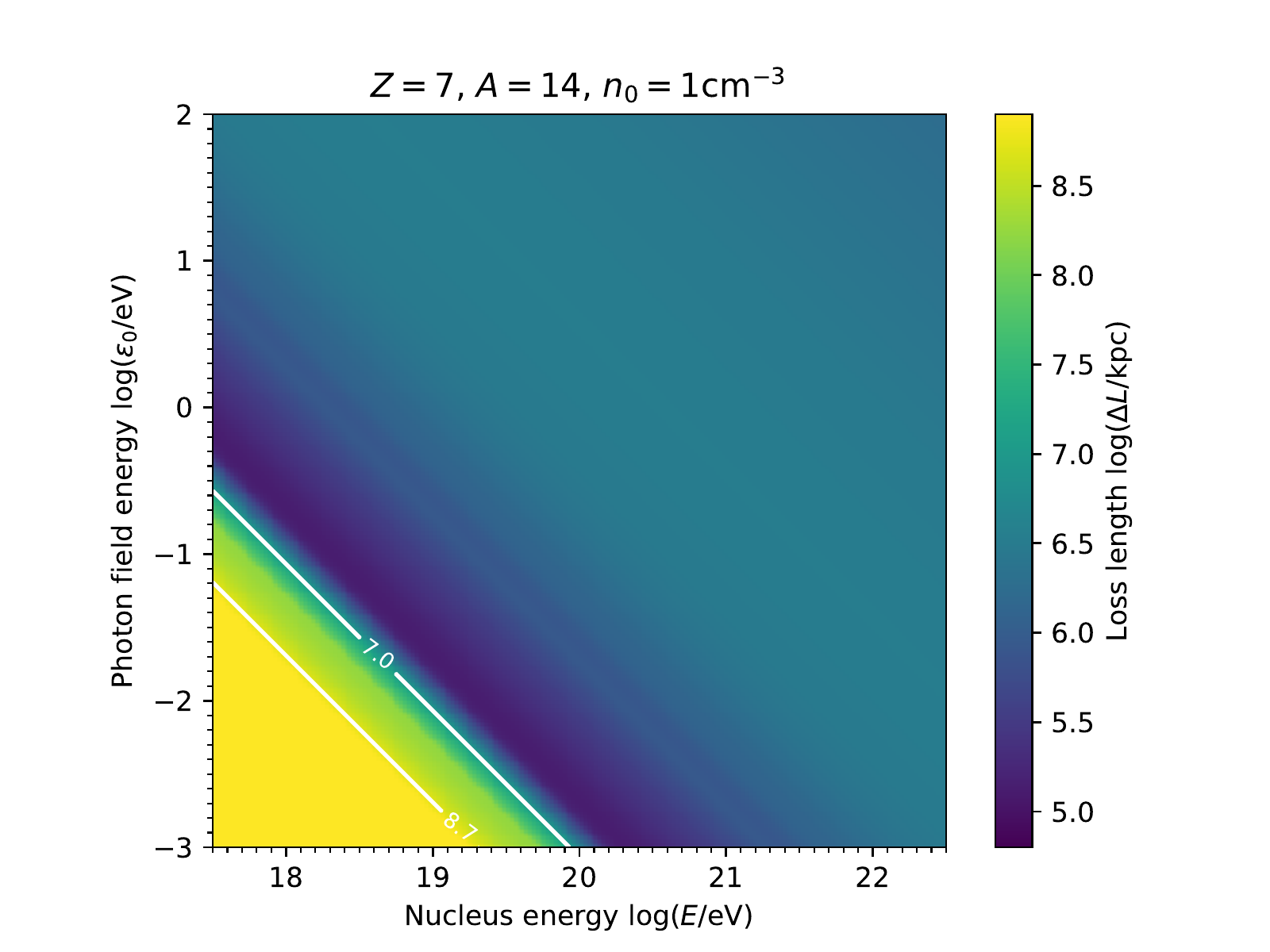}
    \end{minipage}
    \begin{minipage}{.49\textwidth}
        \includegraphics[width=\textwidth]{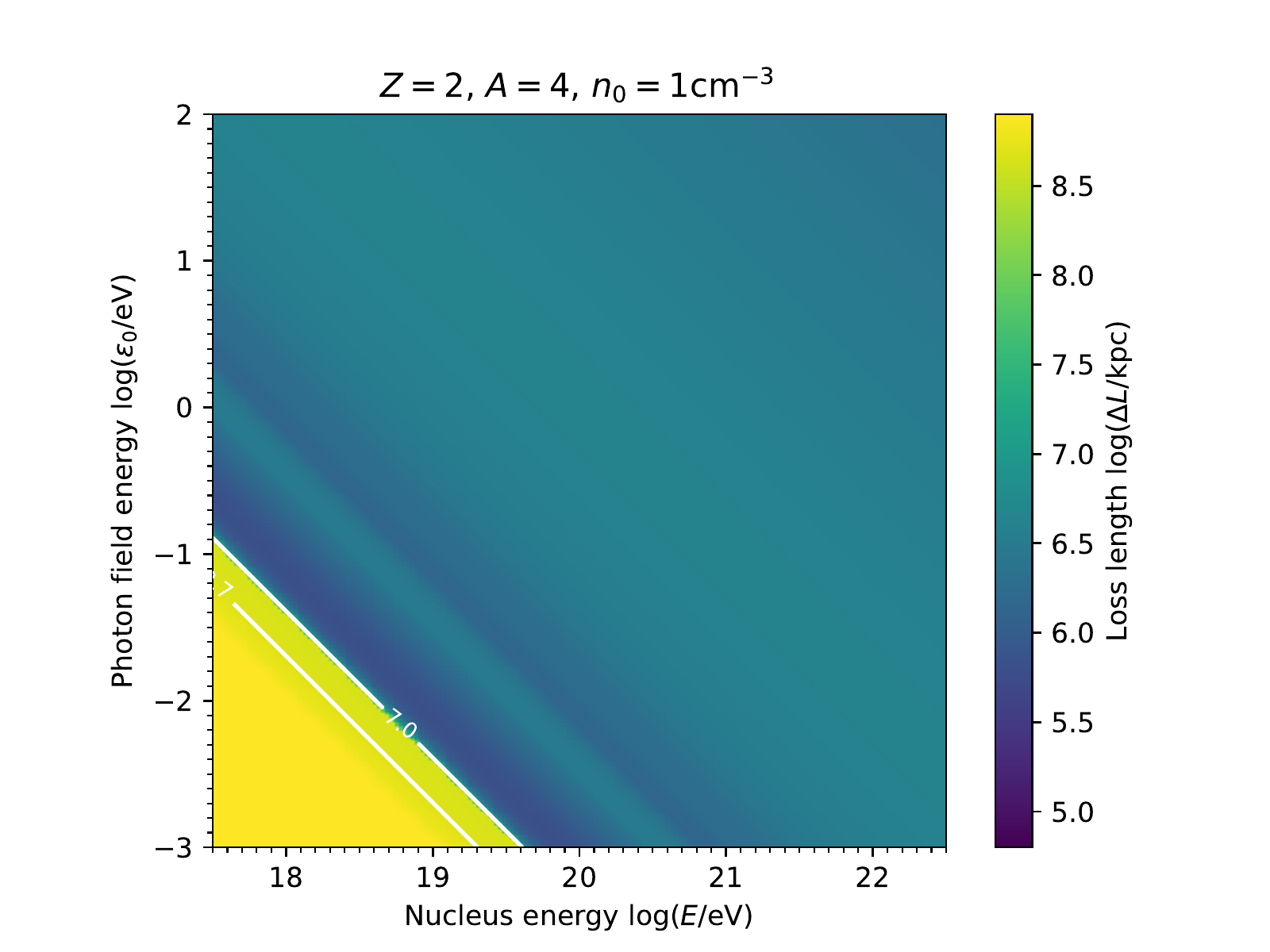}
        \includegraphics[width=\textwidth]{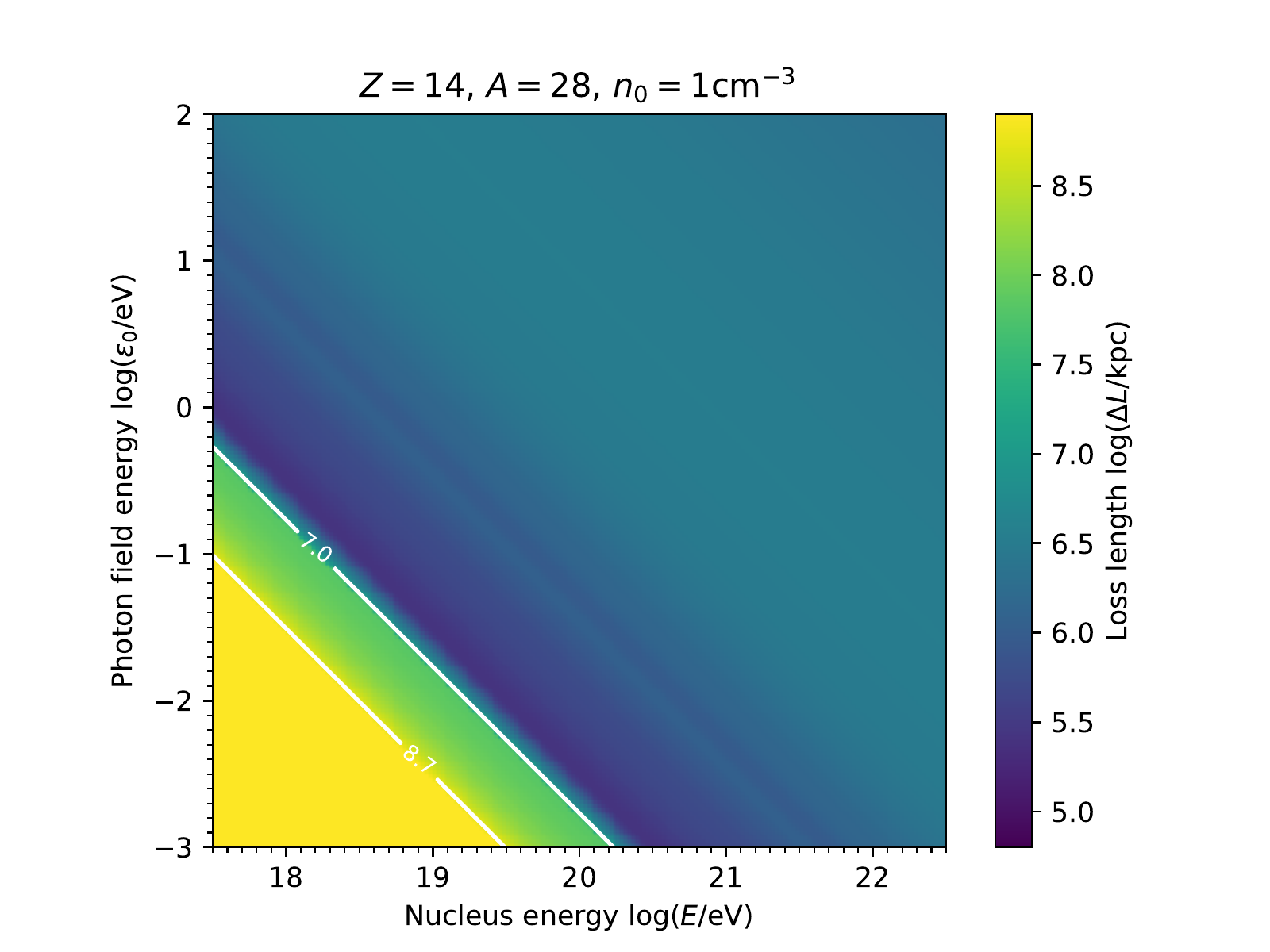}
    \end{minipage}
    \caption{Parameter scan for different monochromatic target fields (vertical axis) and varying nuclei energies $E_\mathrm{N}$ (horizontal axis) is shown. The logarithm of the loss length ($\log_{10}L$) is color-coded with the same scaling for all four plots. The nuclear rest frame photon energy $\epsilon_\mathrm{r}$ increases from the lower left to the upper right along the major diagonal. The number density was fixed to $n_0=1\;\mathrm{cm}^{-3}$, but the results can be easily re-scaled due to the linear dependence $L\propto n_0$.}
    \label{fig:2dscan}
\end{figure}

\begin{figure}[htbp]
\centering
    \begin{minipage}{.31\textwidth}
        \includegraphics[width=\textwidth]{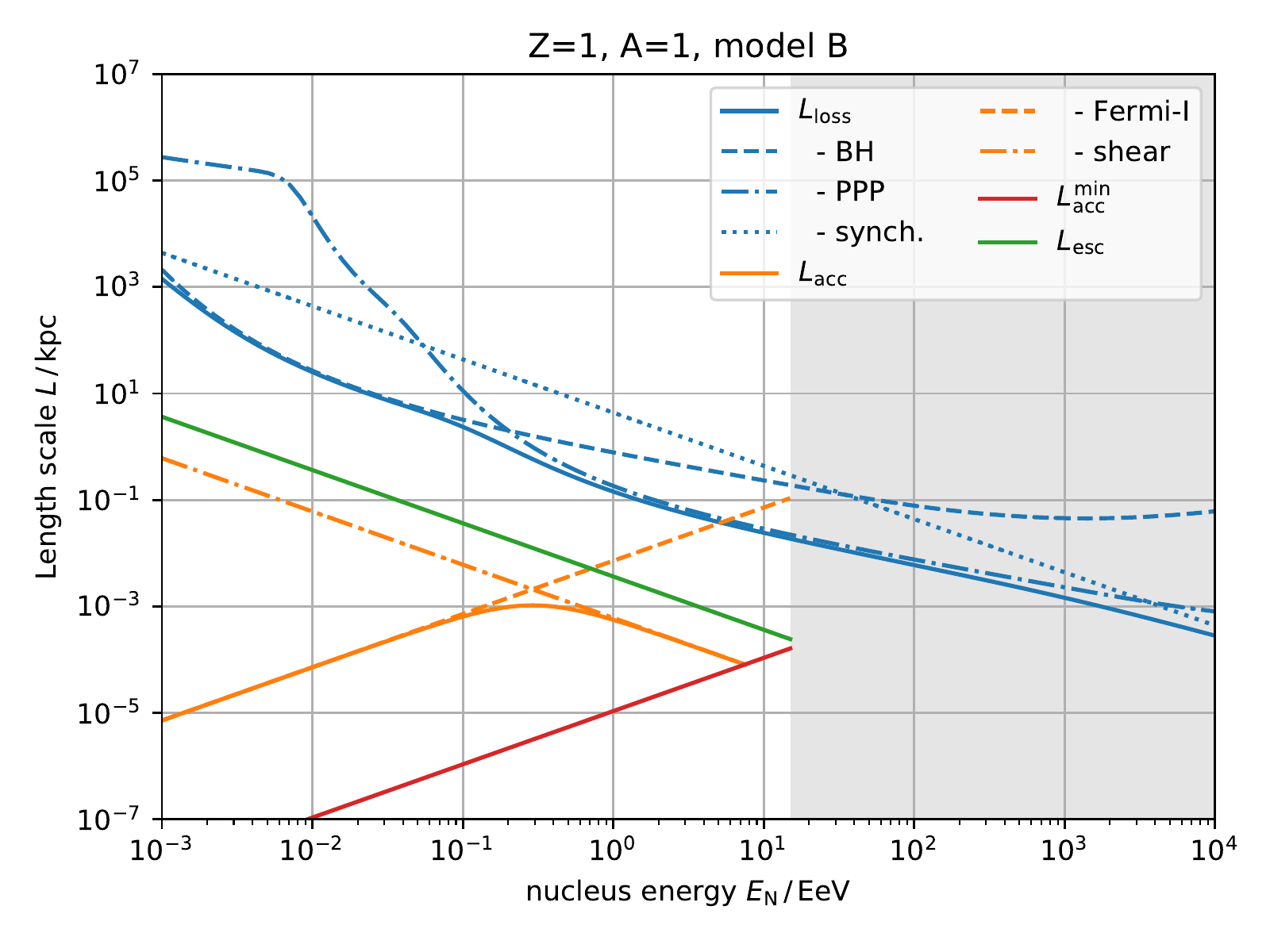}
        \includegraphics[width=\textwidth]{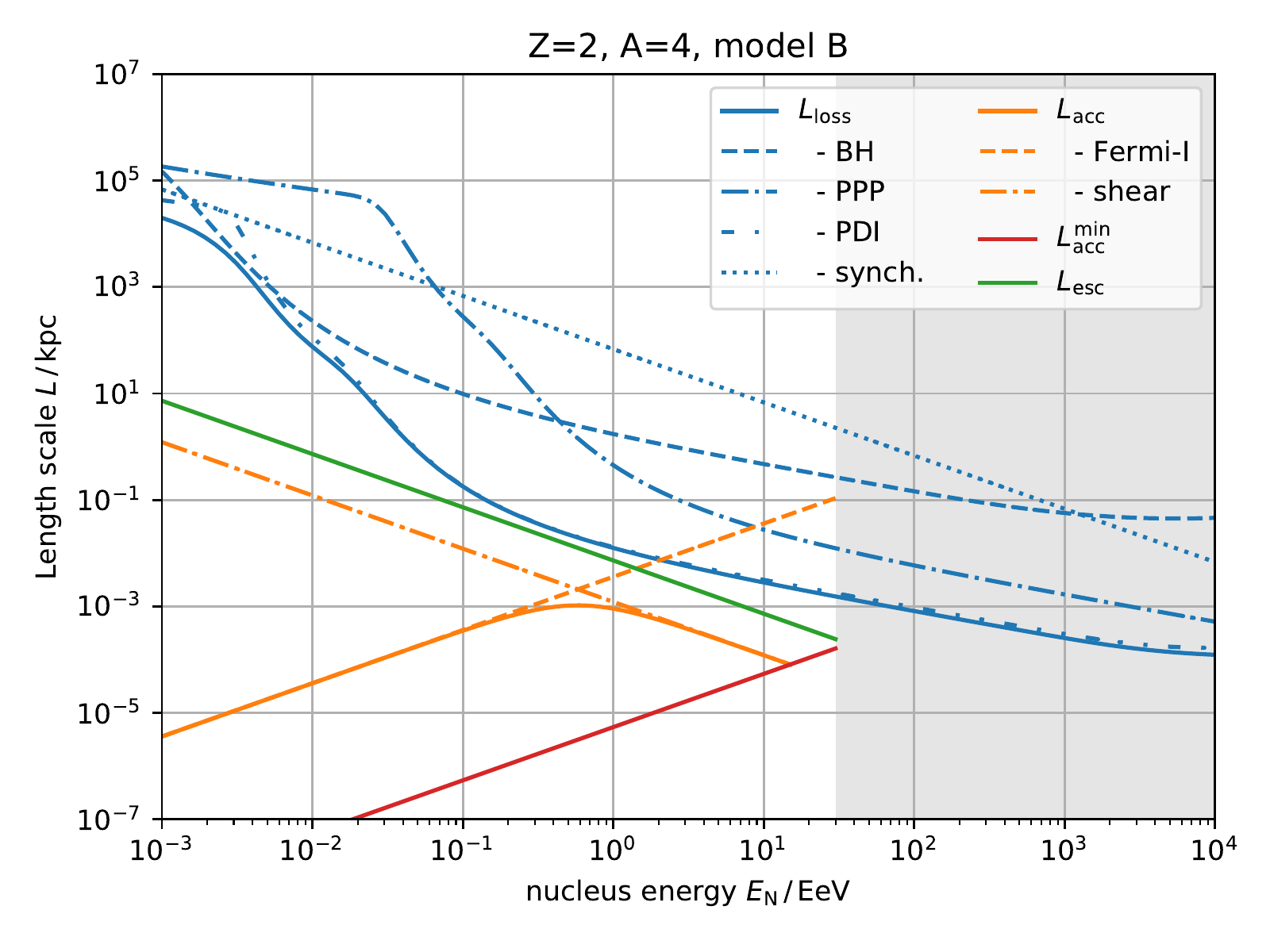}
        \includegraphics[width=\textwidth]{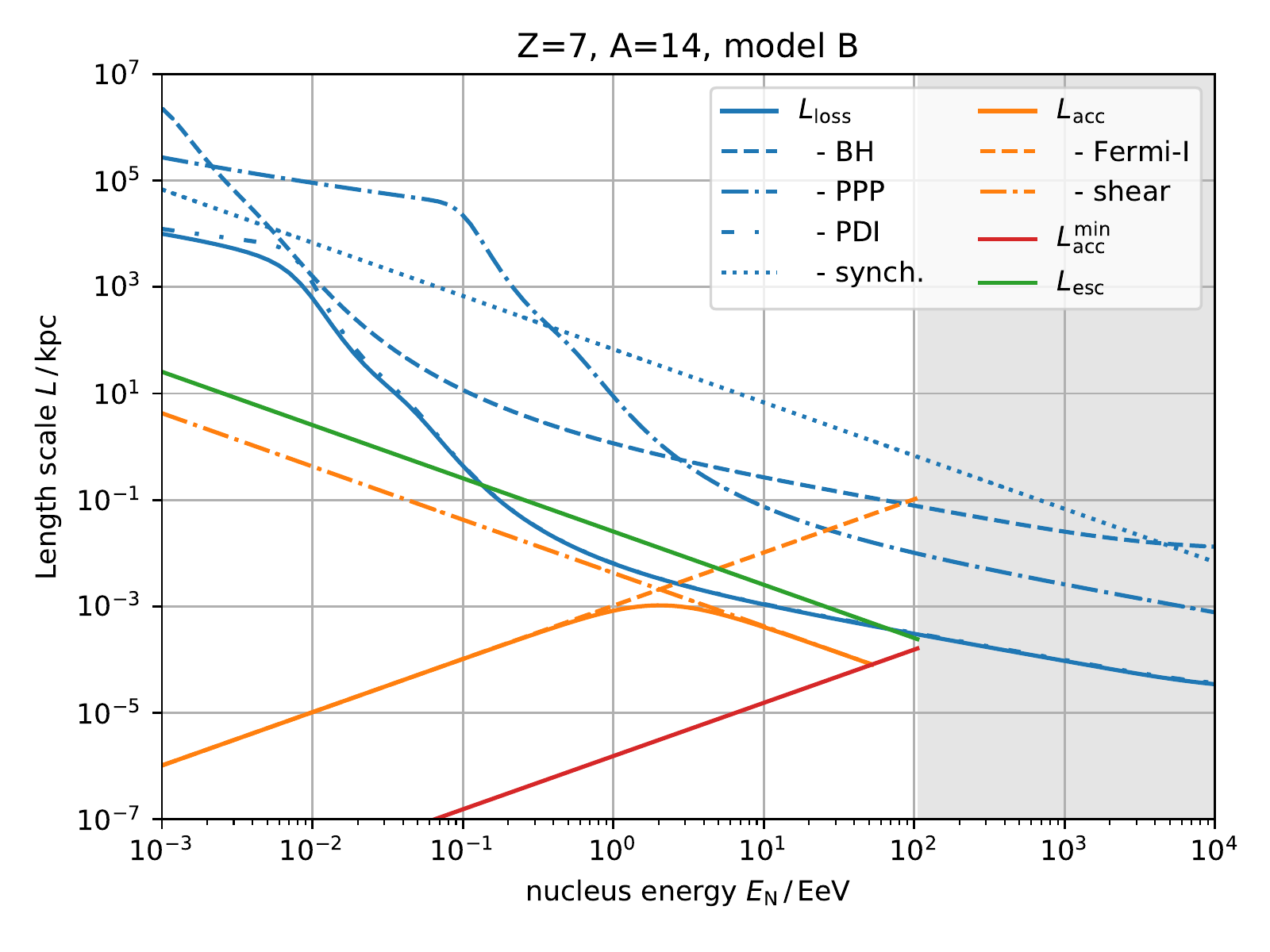}
        \includegraphics[width=\textwidth]{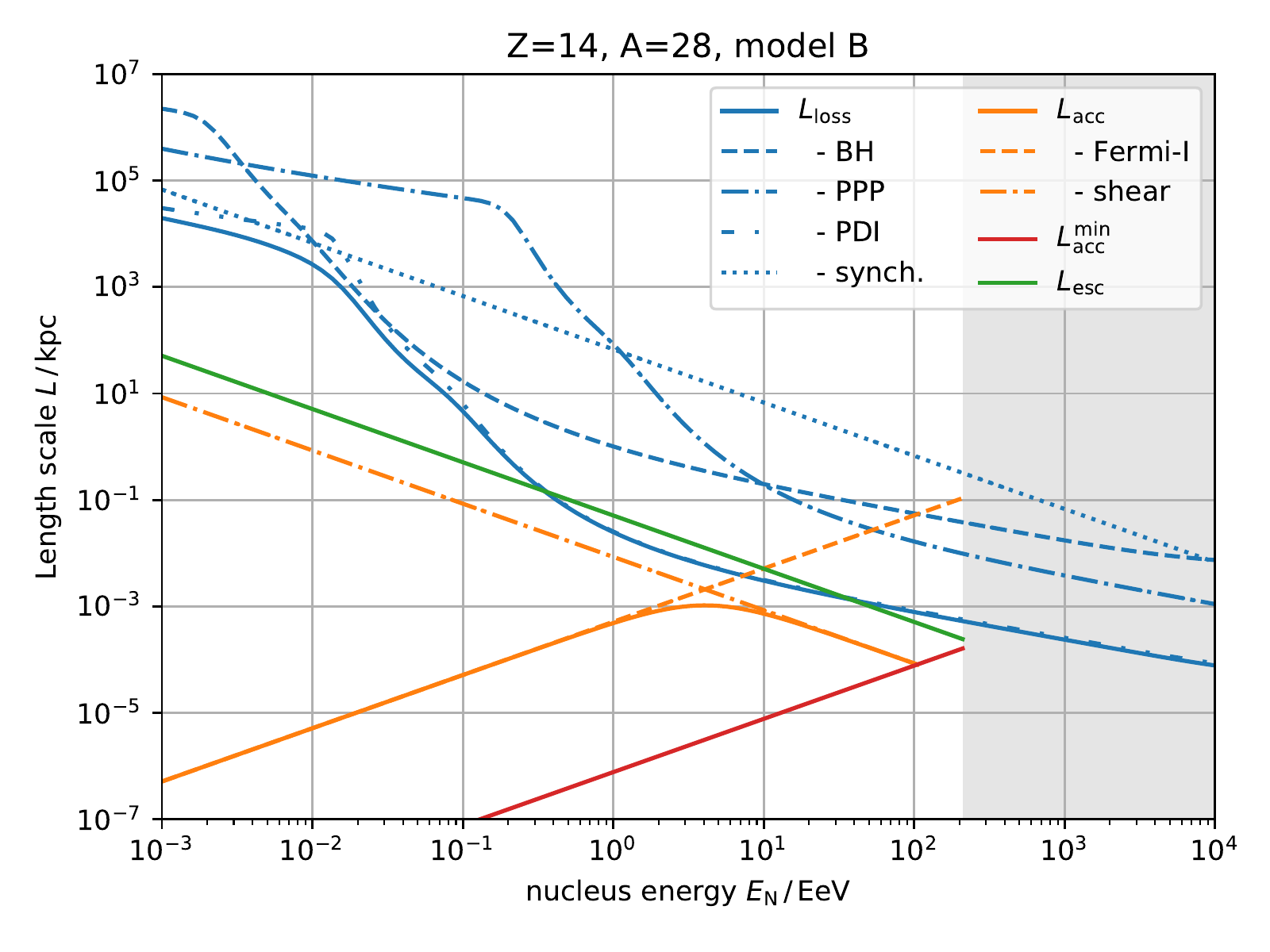}
        \includegraphics[width=\textwidth]{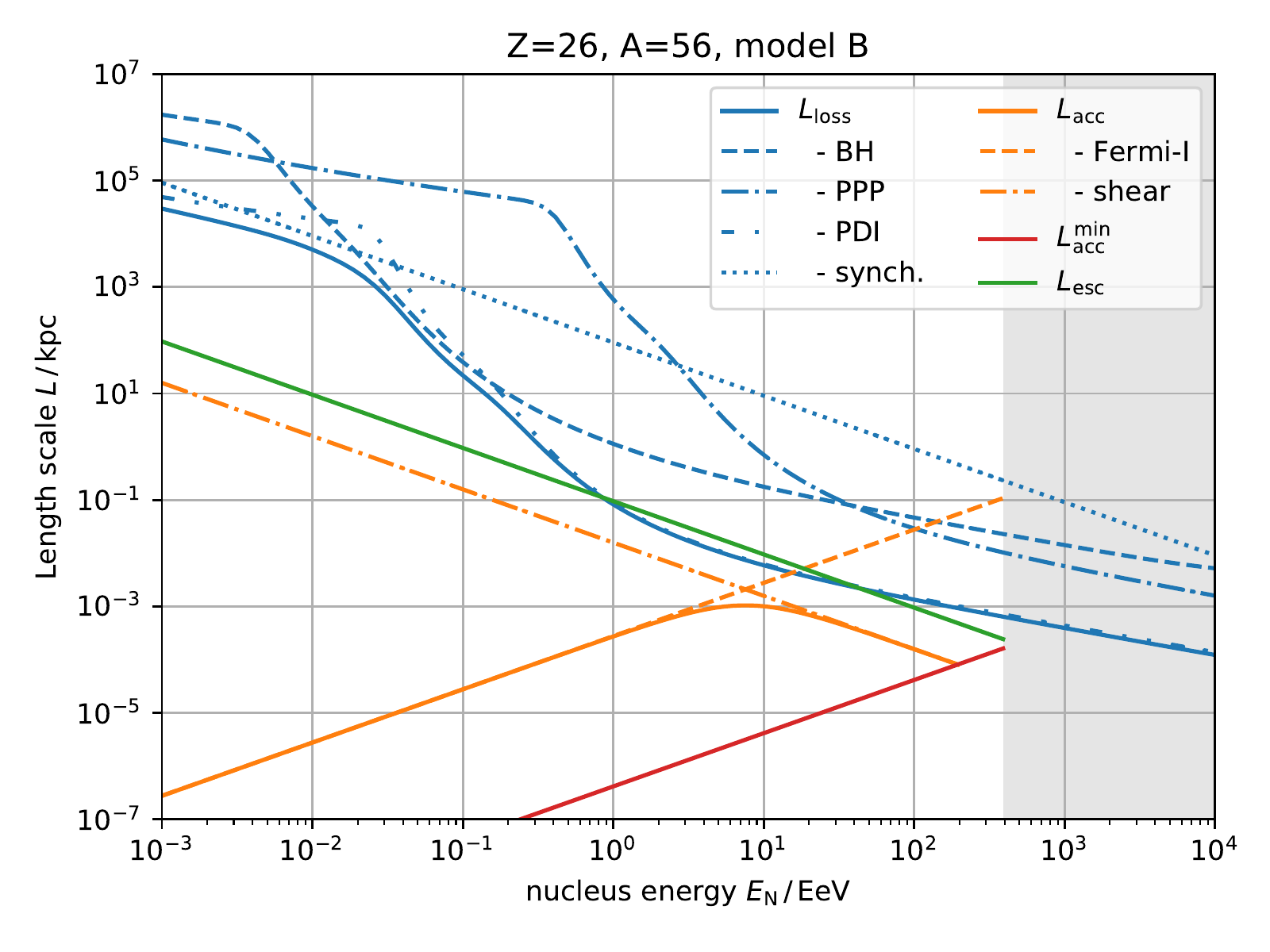}
    \end{minipage}
    \begin{minipage}{.31\textwidth}
        \includegraphics[width=\textwidth]{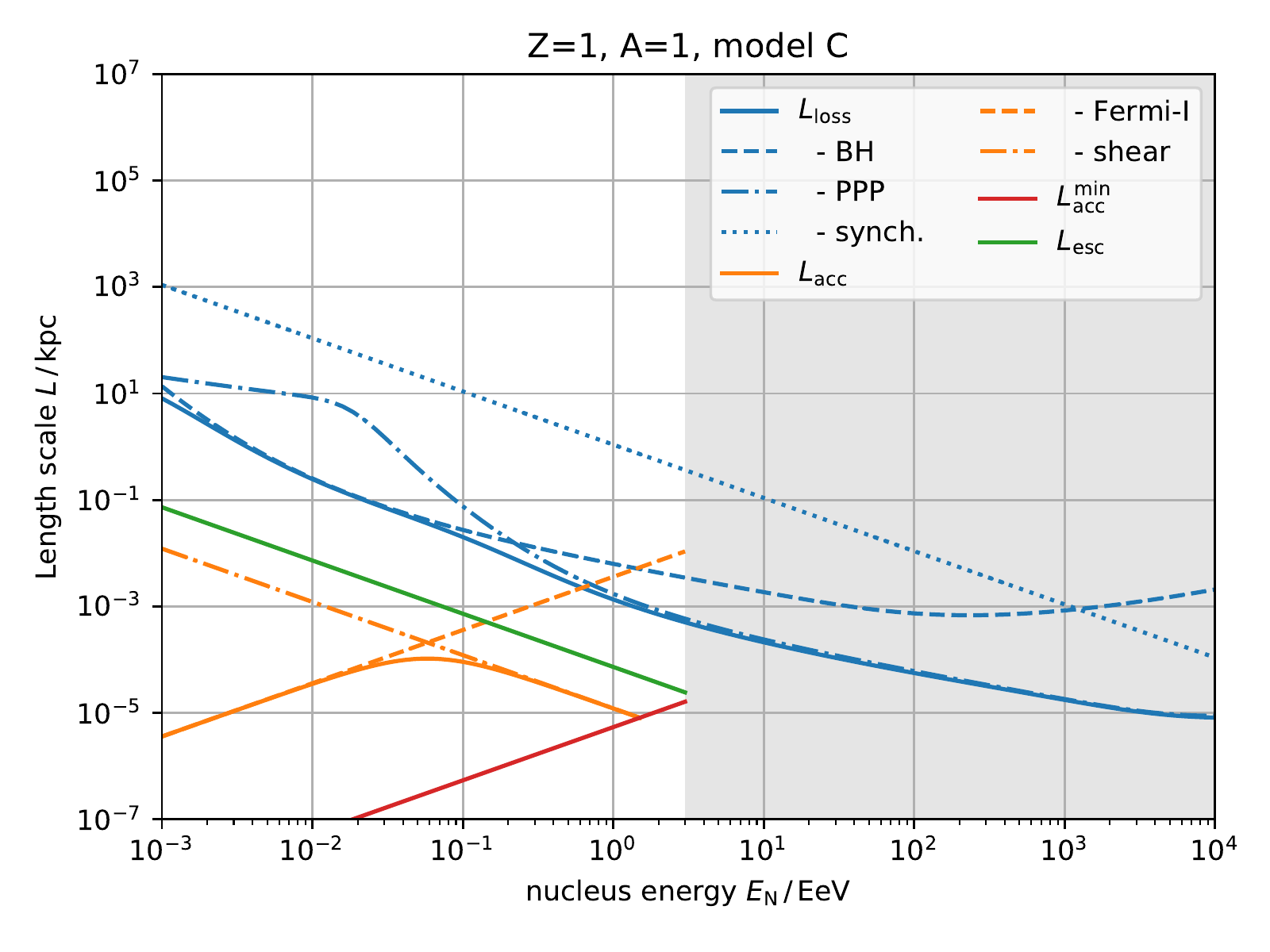}
        \includegraphics[width=\textwidth]{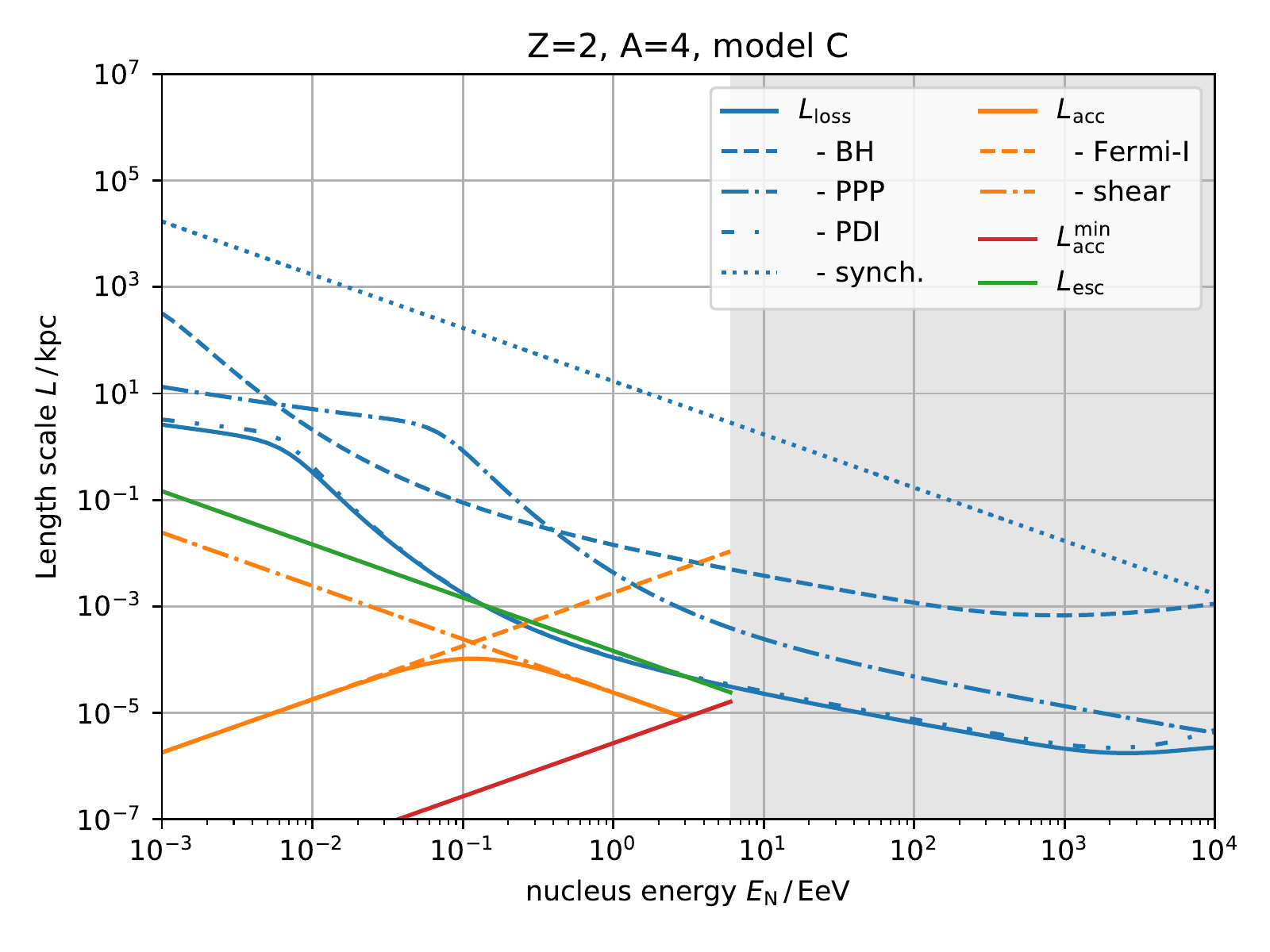}
        \includegraphics[width=\textwidth]{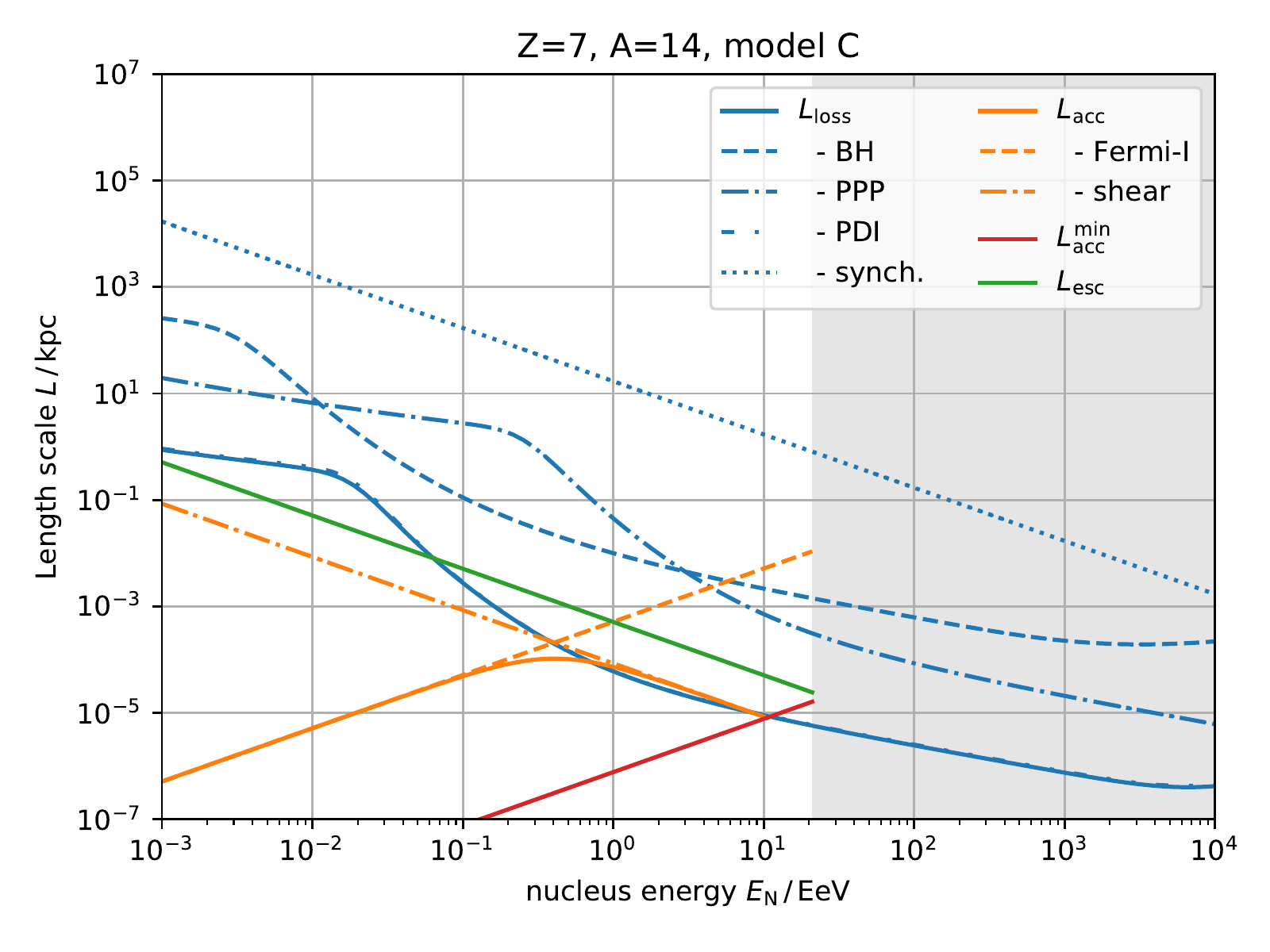}
        \includegraphics[width=\textwidth]{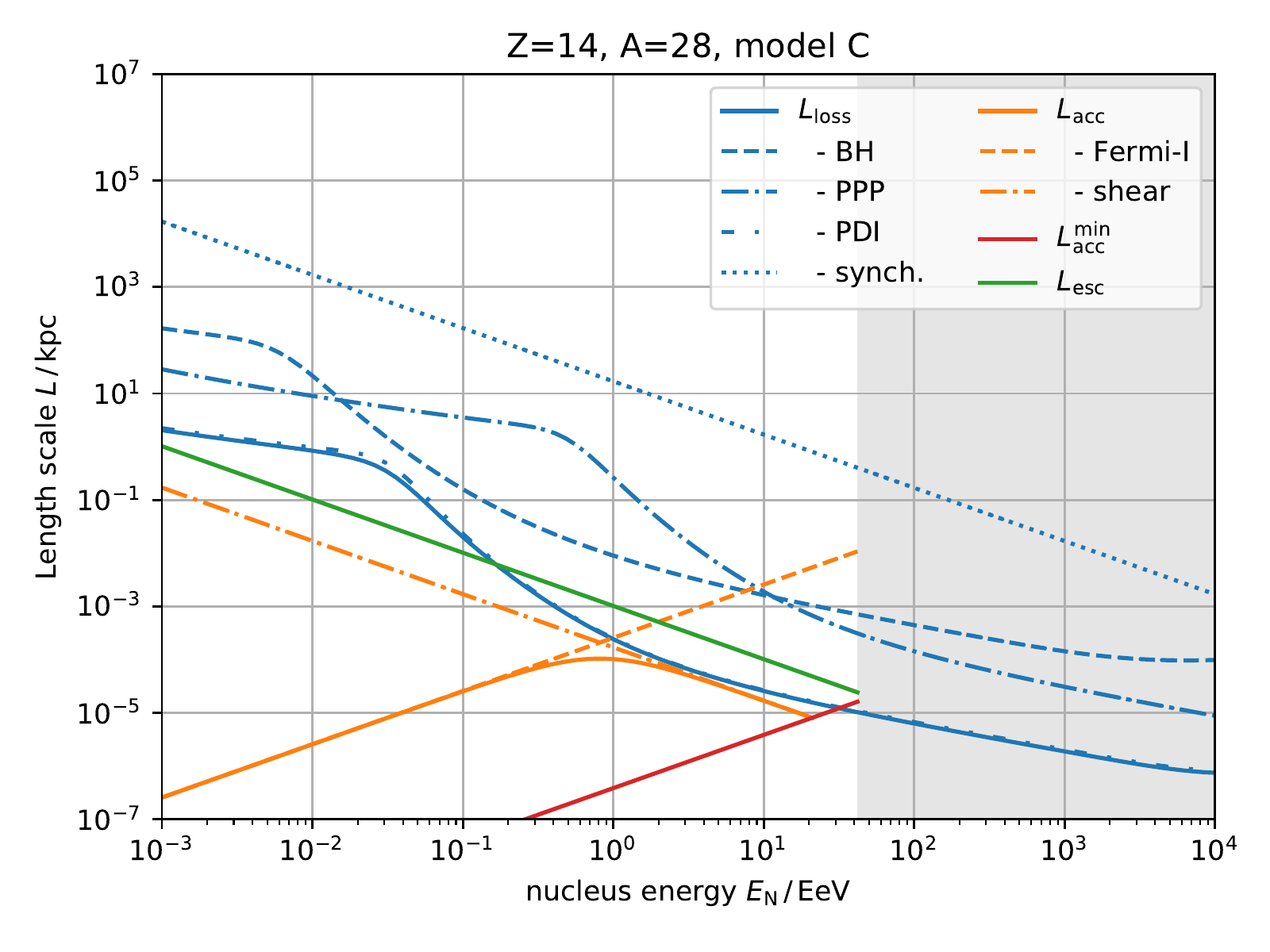}
        \includegraphics[width=\textwidth]{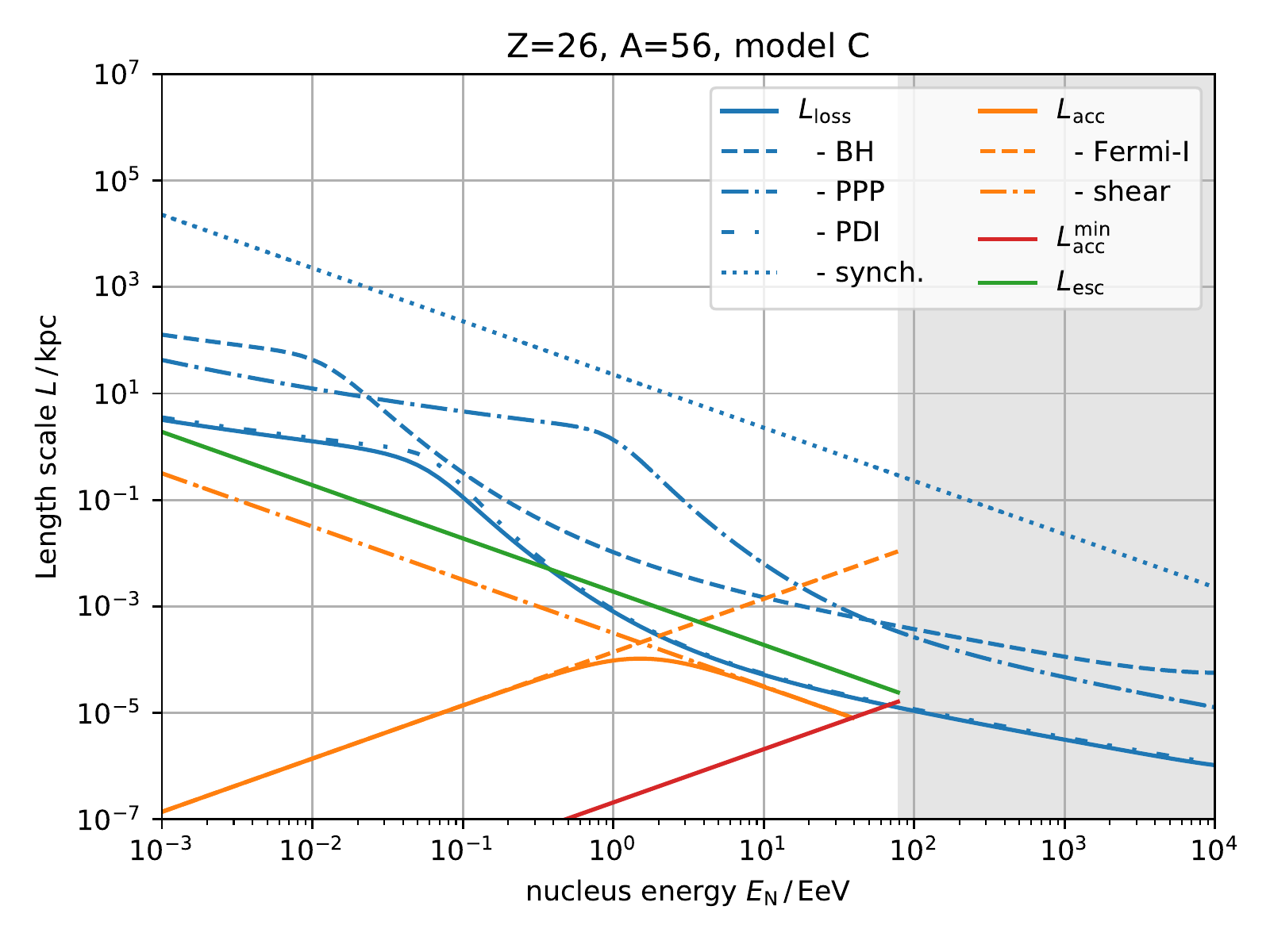}
    \end{minipage}
    \begin{minipage}{.31\textwidth}
        \includegraphics[width=\textwidth]{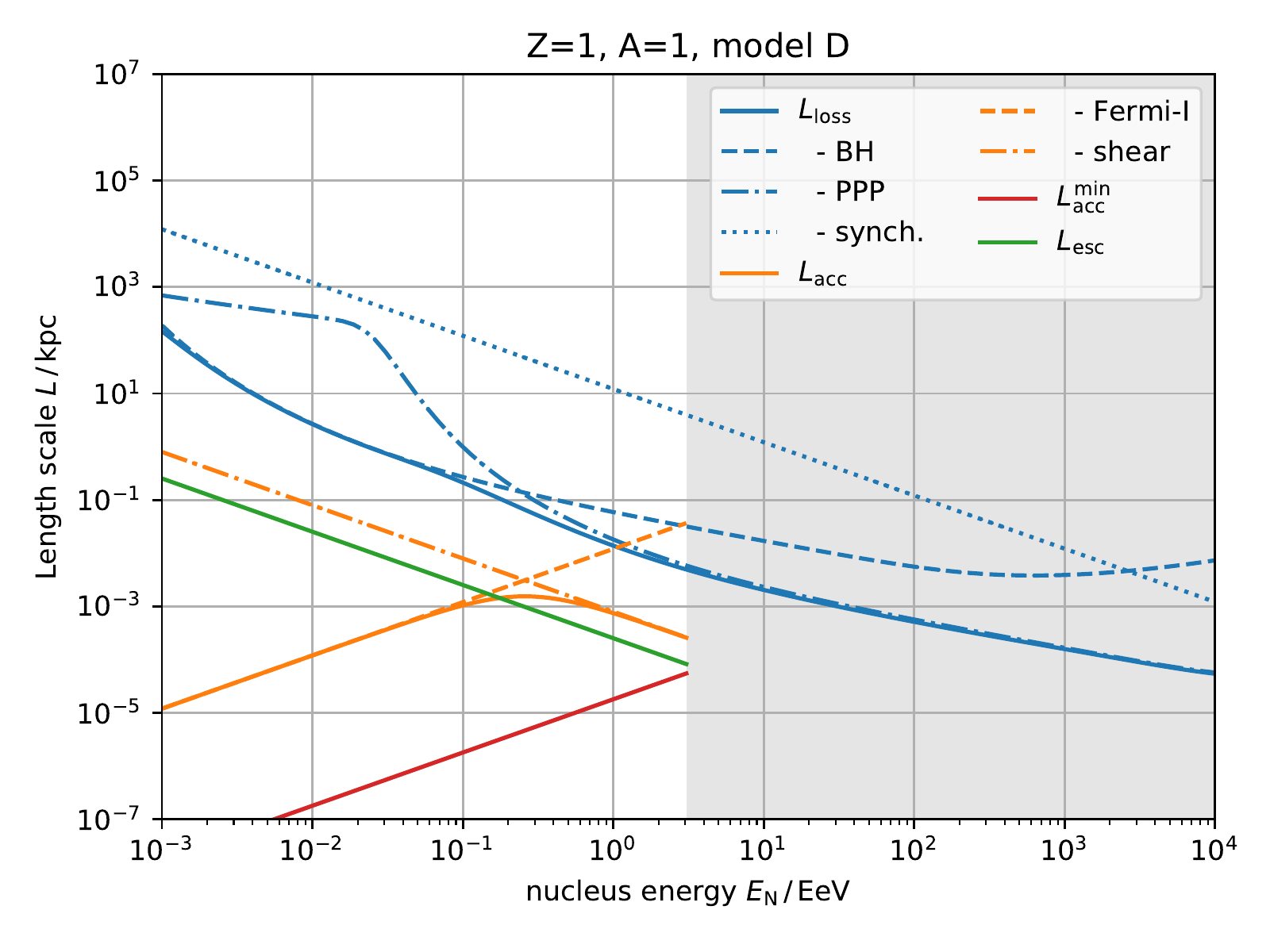}
        \includegraphics[width=\textwidth]{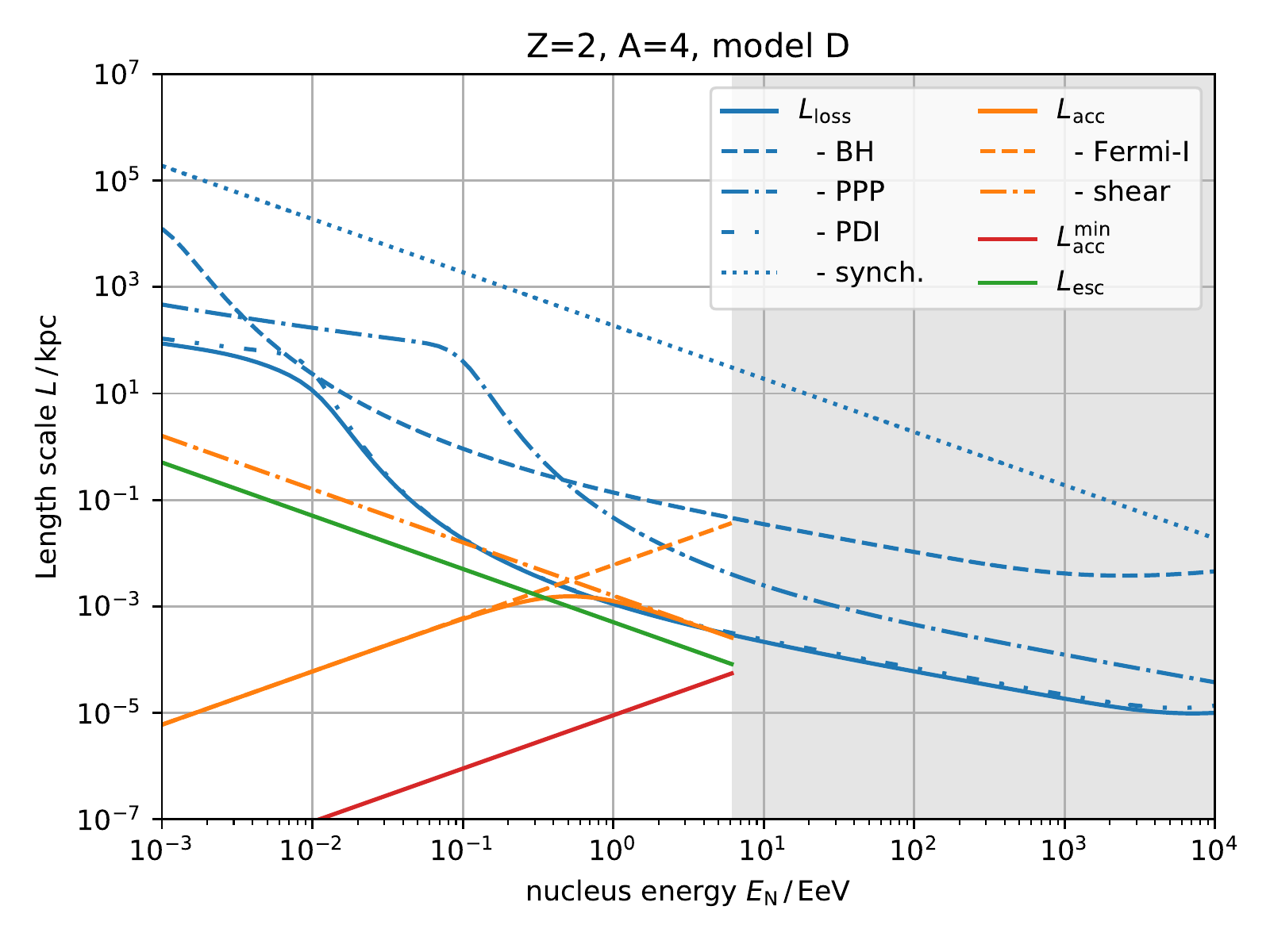}
        \includegraphics[width=\textwidth]{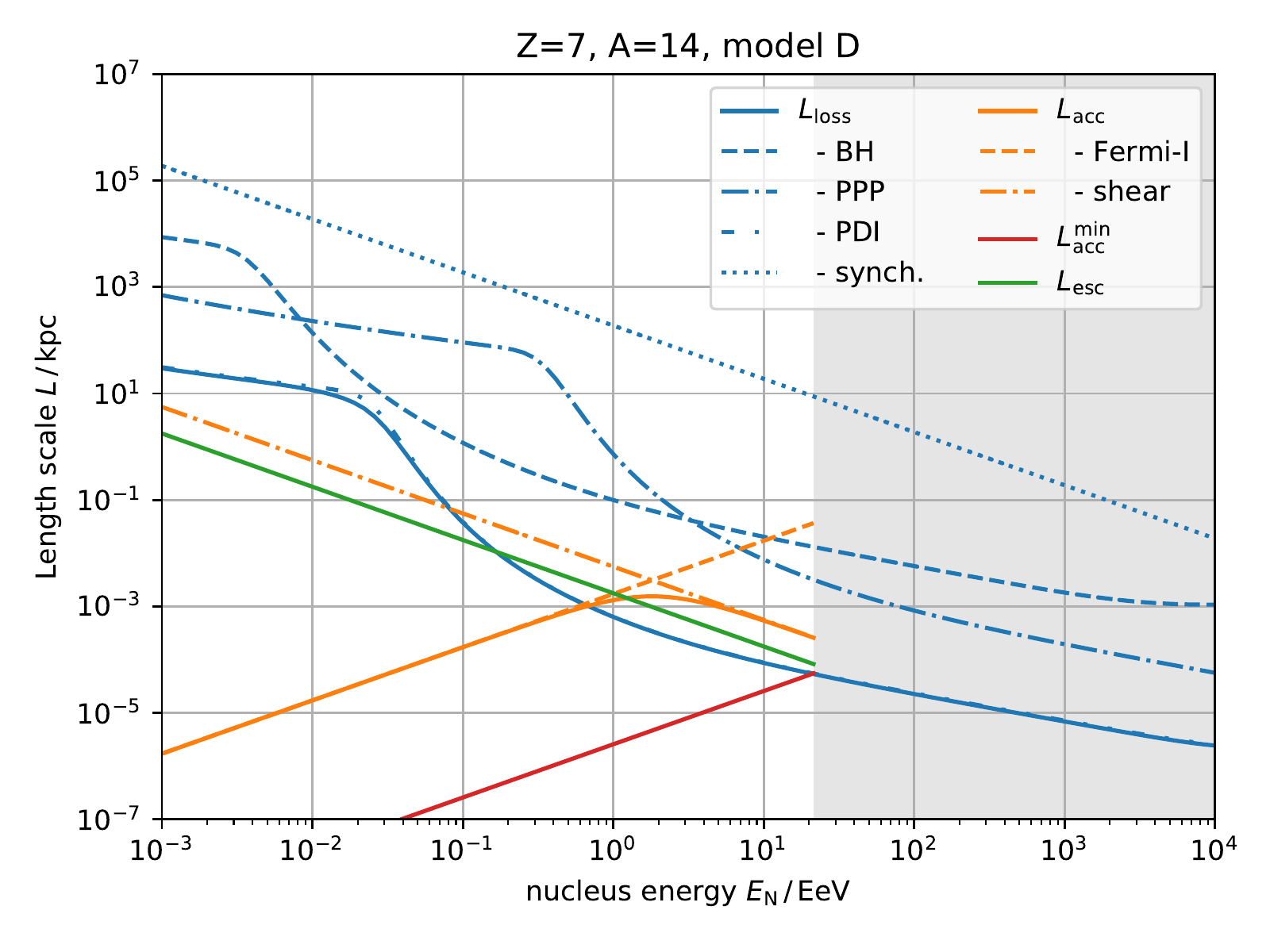}
        \includegraphics[width=\textwidth]{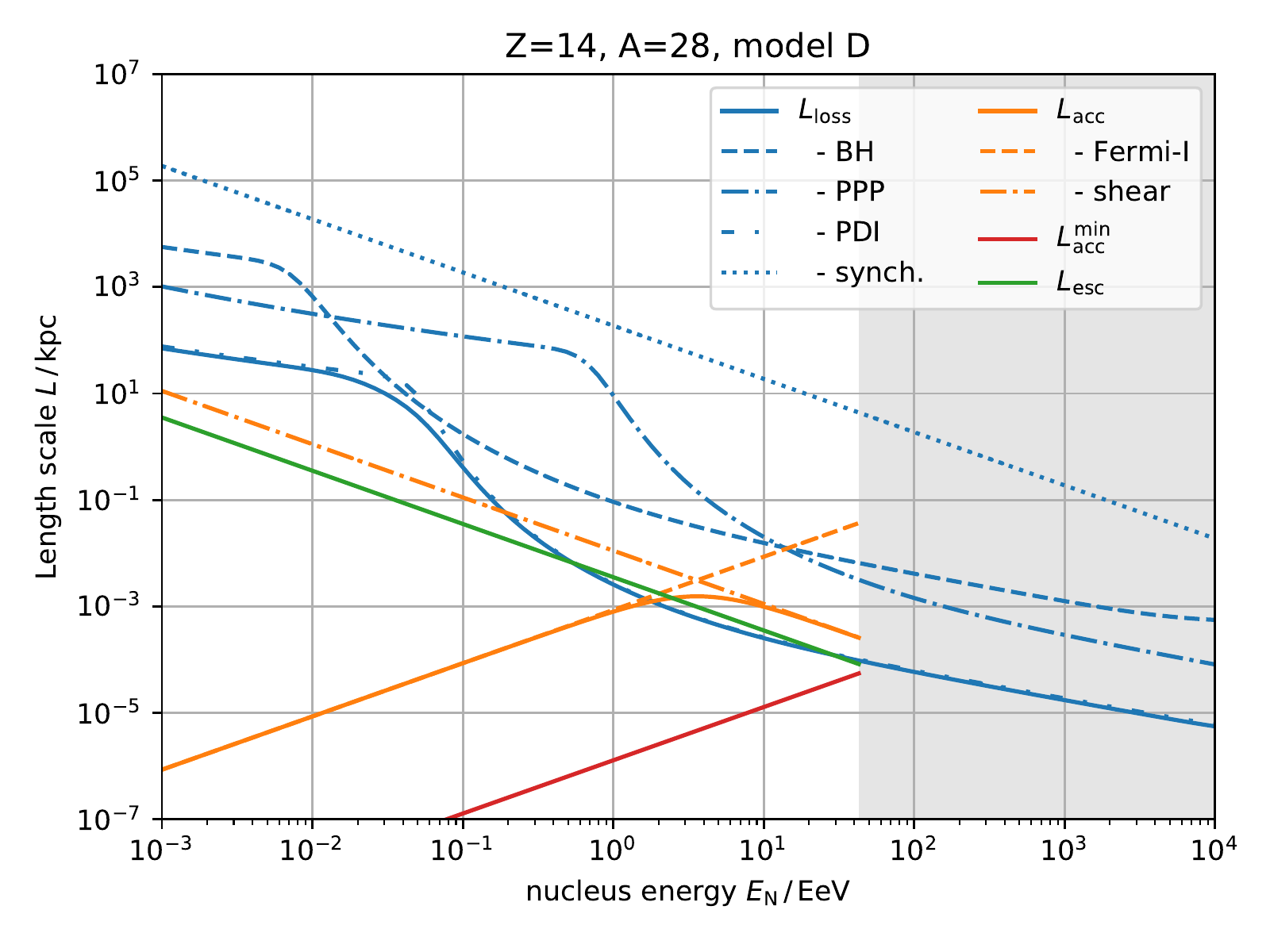}
        \includegraphics[width=\textwidth]{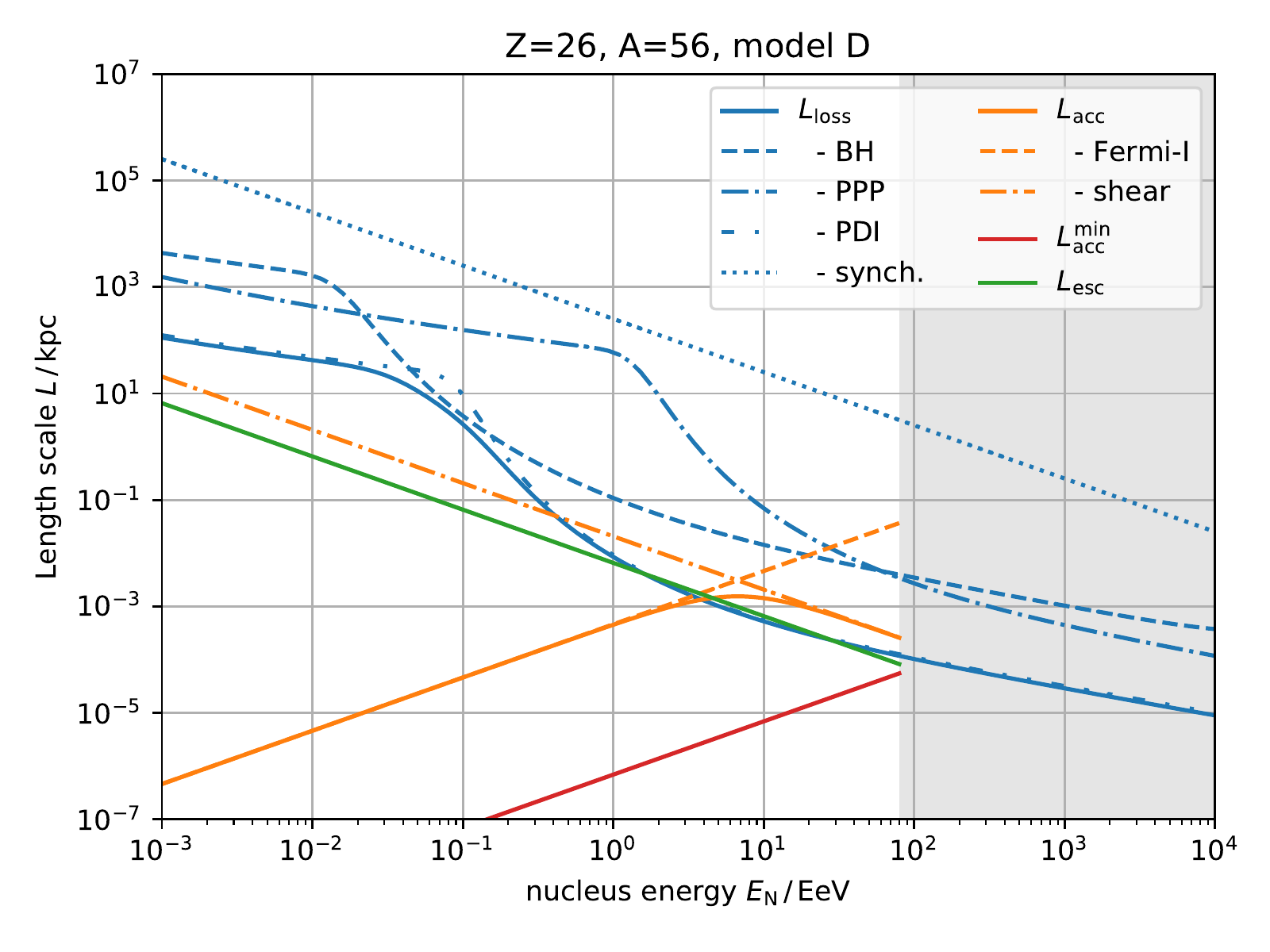}
    \end{minipage}
    \caption{Comparison of the lengths scales for all relevant processes based on Bohm diffusion is shown. From left to right the columns show models B, C, and D respectively. From top to bottom the tracer elements hydrogen, helium, nitrogen, silicon, and iron are shown.}
    \label{fig:AllLengths1}
\end{figure}

\begin{figure}[htbp]
\centering
    \begin{minipage}{.33\textwidth}
        \includegraphics[width=\textwidth]{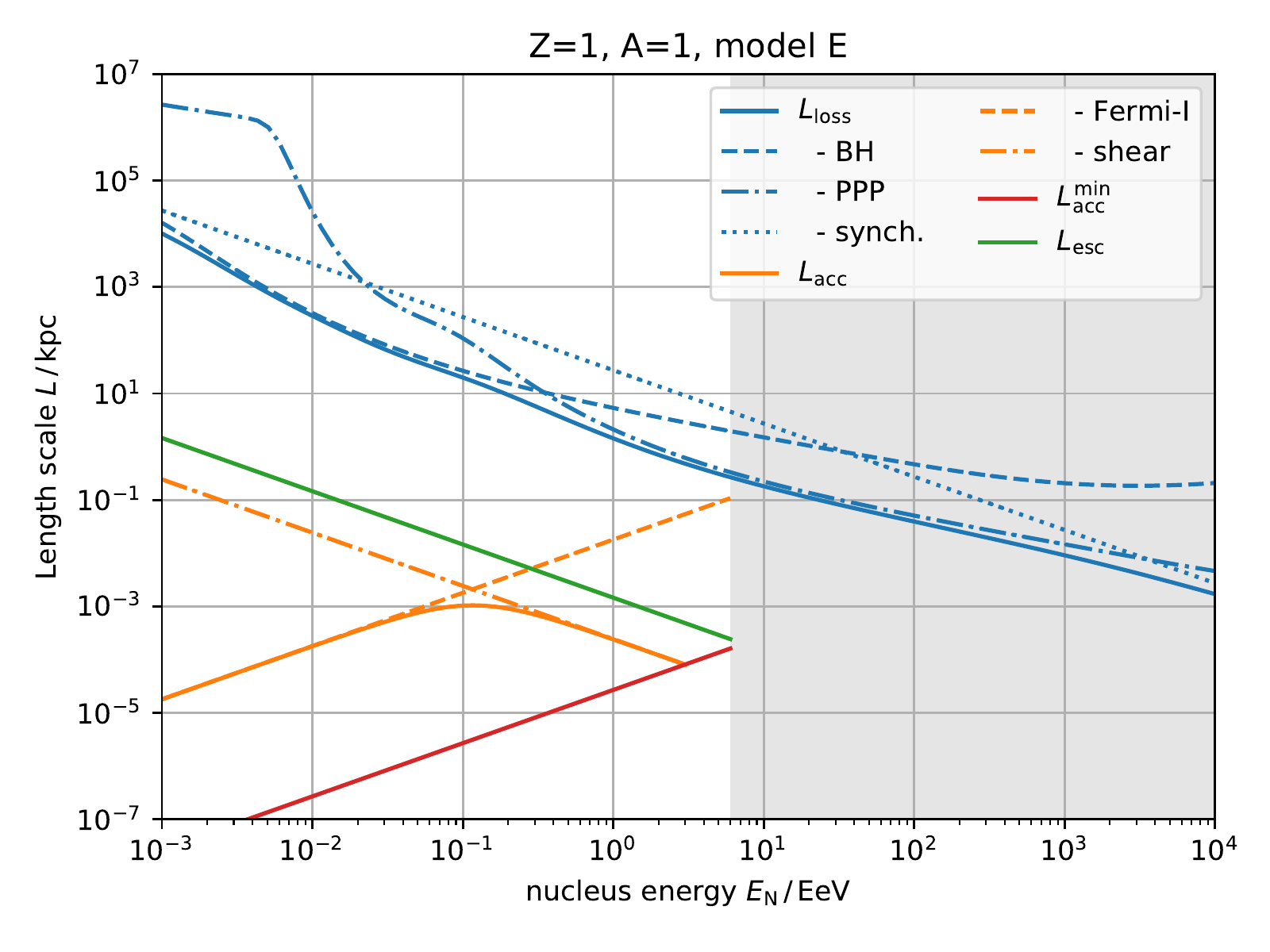}
        \includegraphics[width=\textwidth]{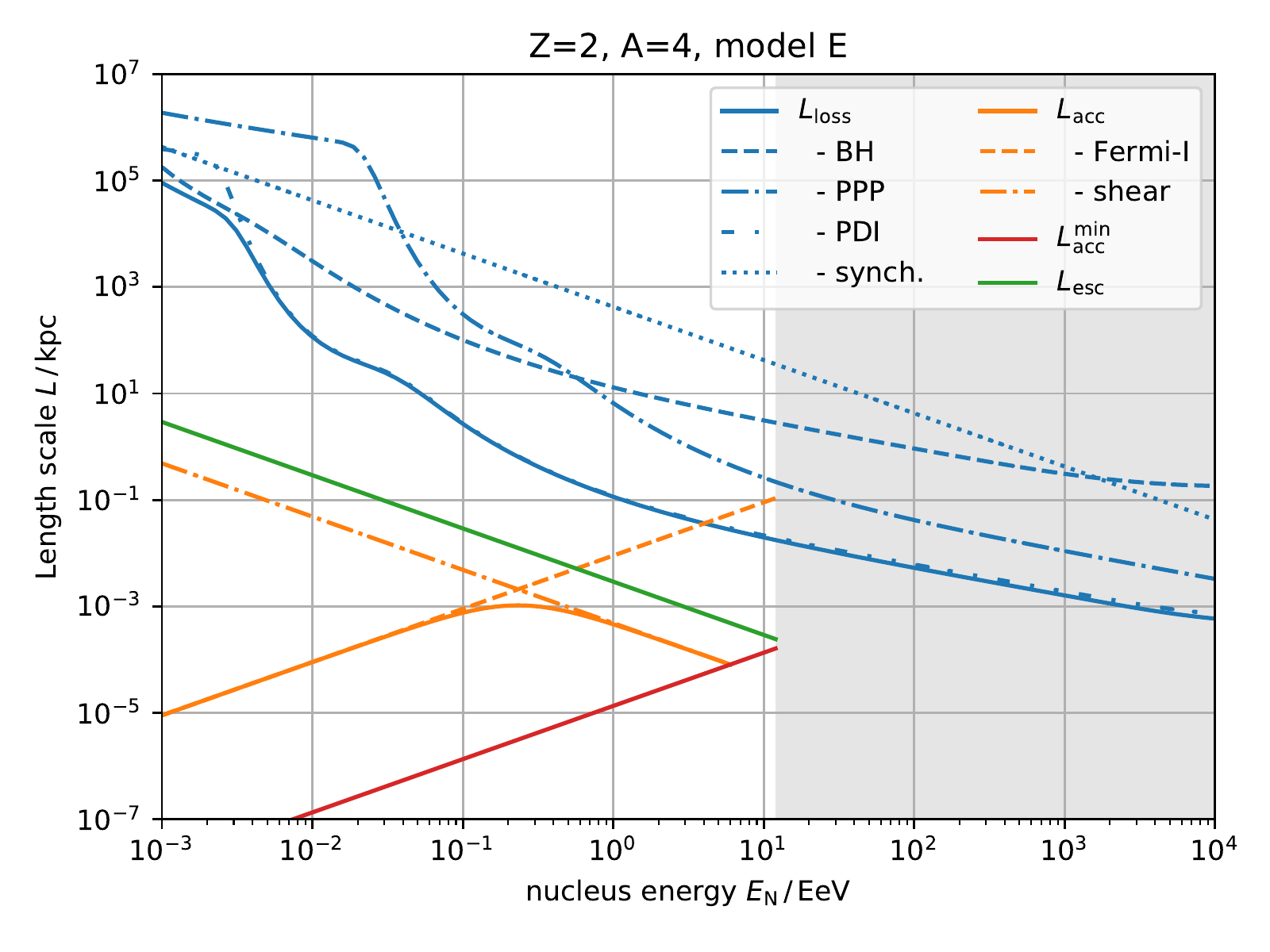}
        \includegraphics[width=\textwidth]{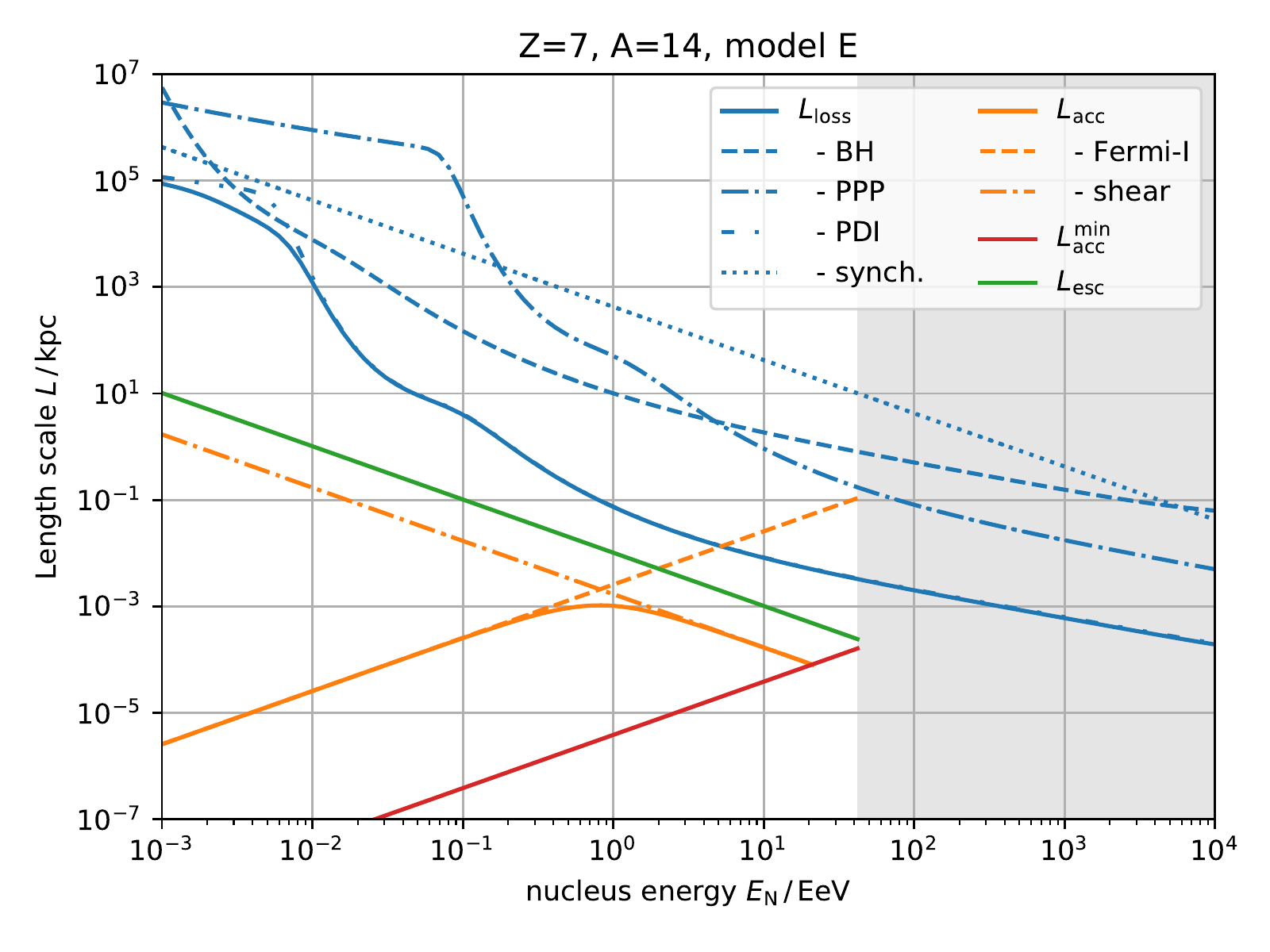}
        \includegraphics[width=\textwidth]{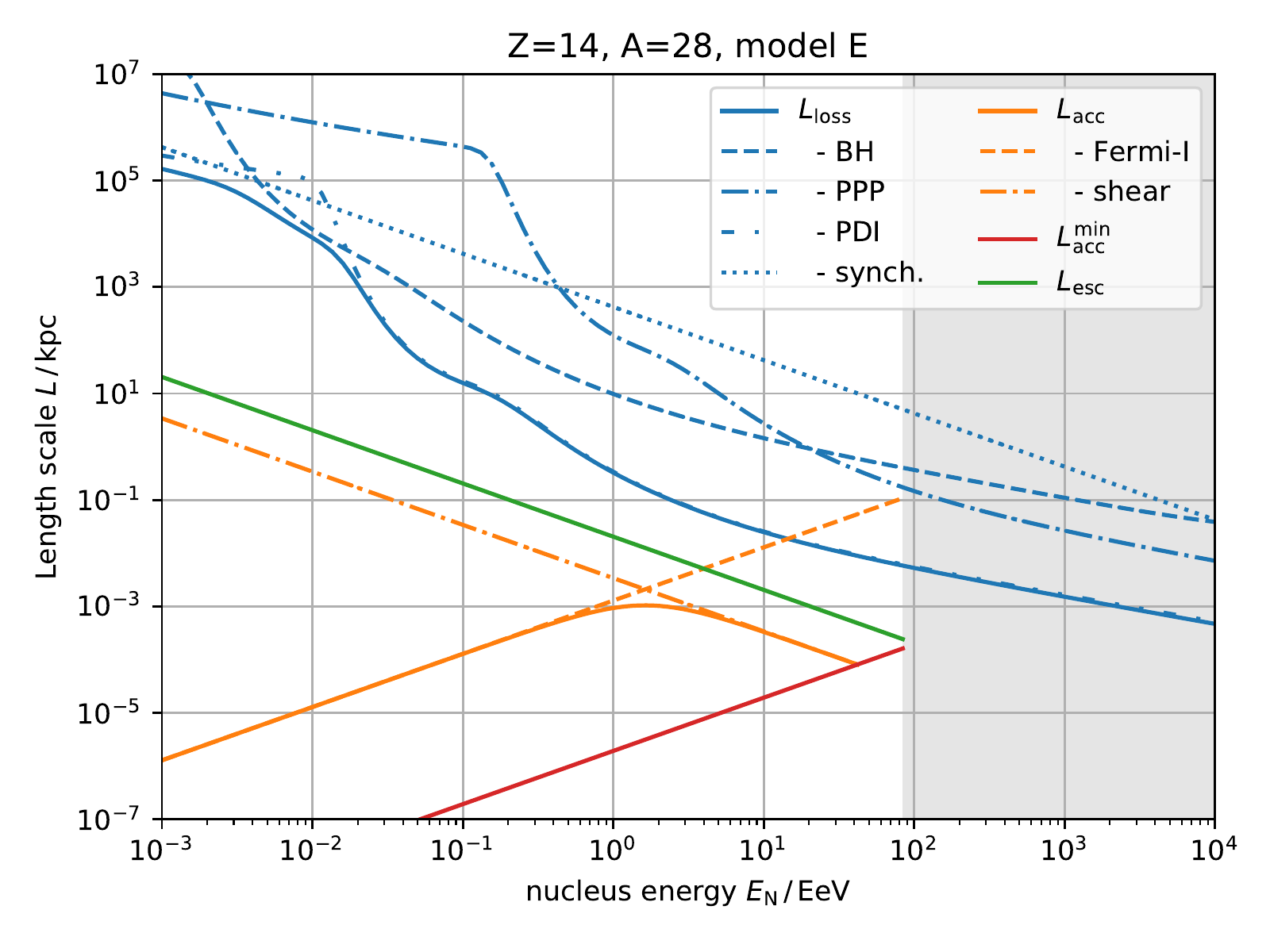}
        \includegraphics[width=\textwidth]{./figures/LengthScale_Model_E_1000260560_synch_Lmin}
    \end{minipage}
    \begin{minipage}{.33\textwidth}
        \includegraphics[width=\textwidth]{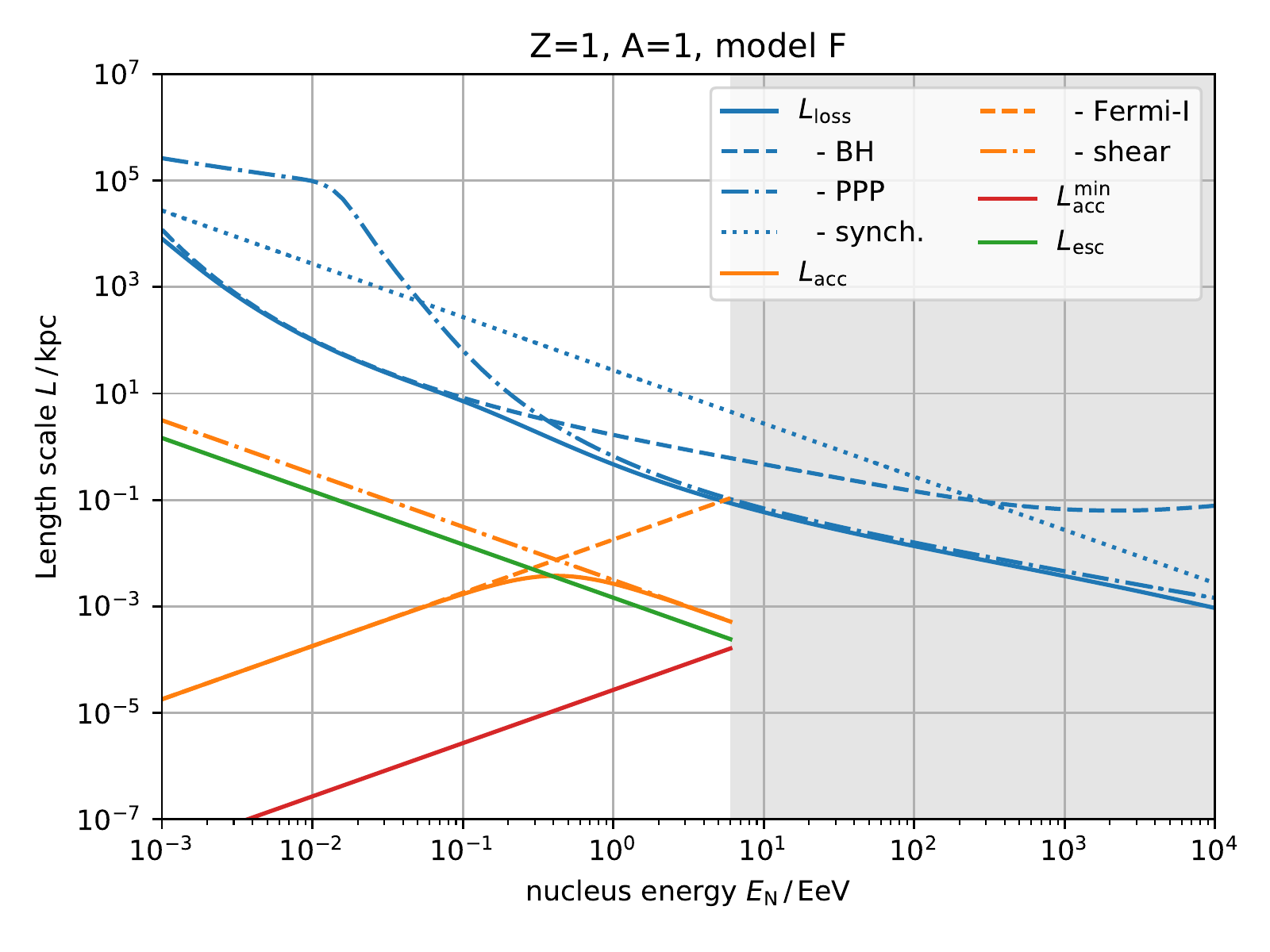}
        \includegraphics[width=\textwidth]{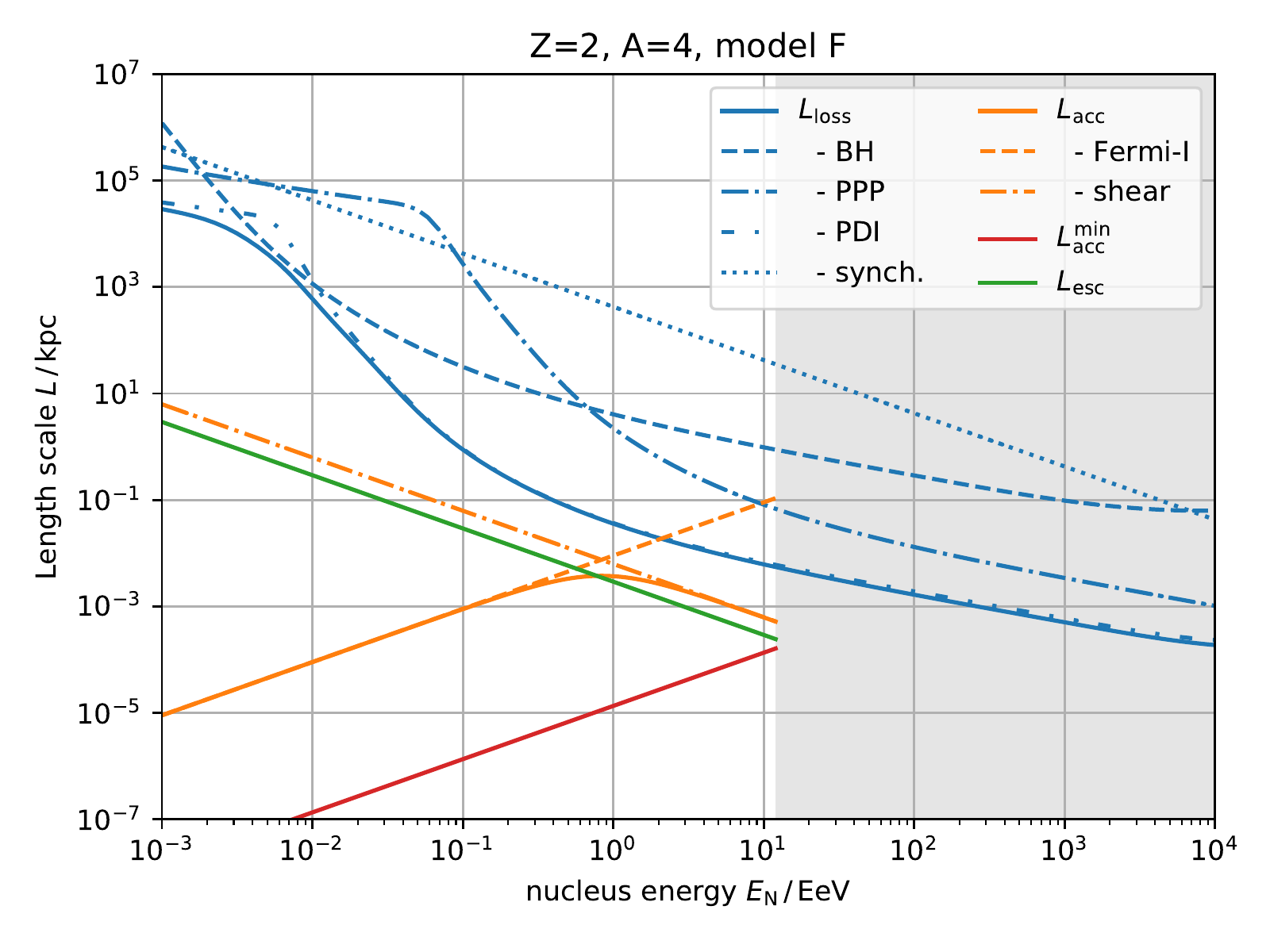}
        \includegraphics[width=\textwidth]{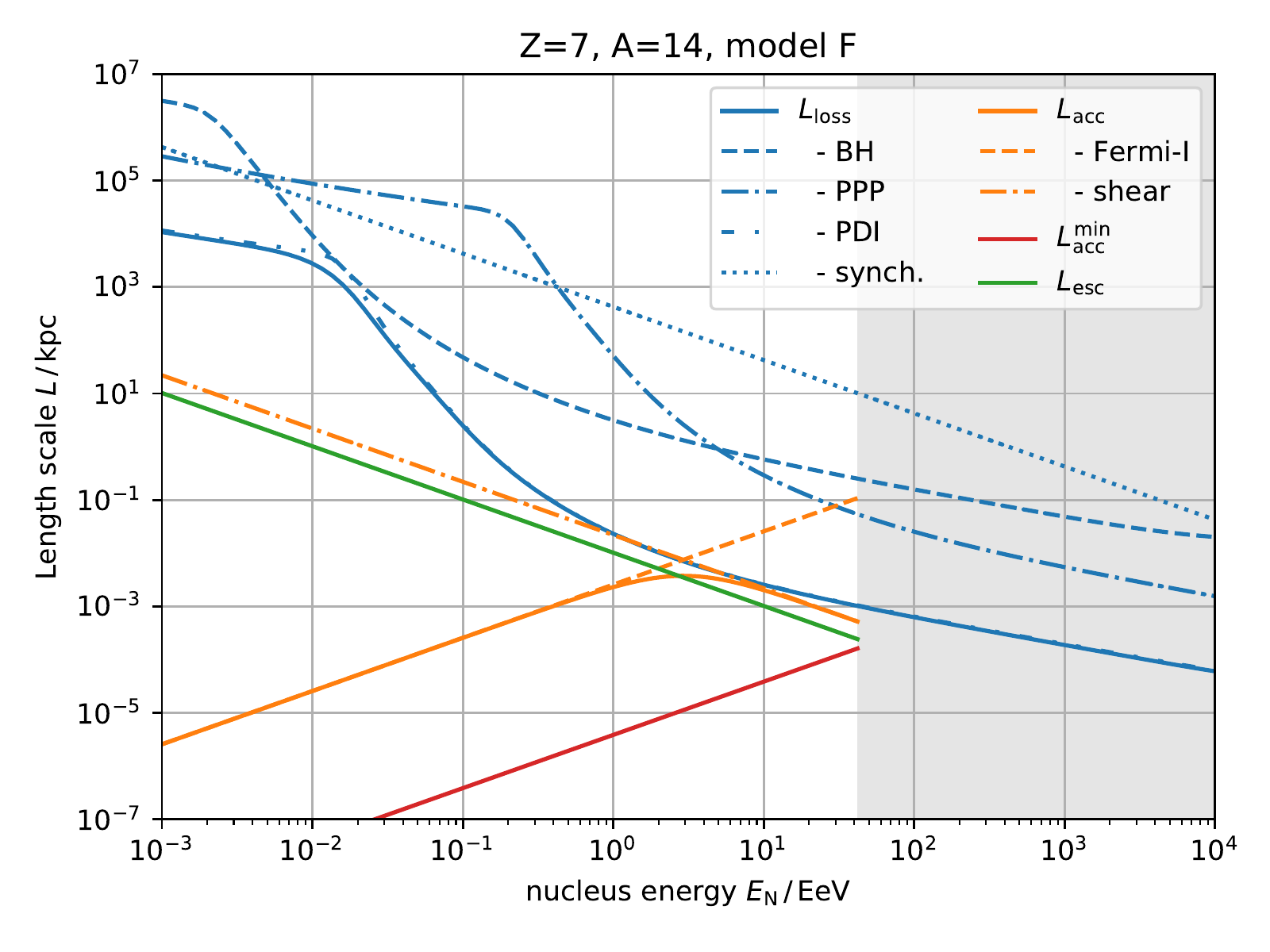}
        \includegraphics[width=\textwidth]{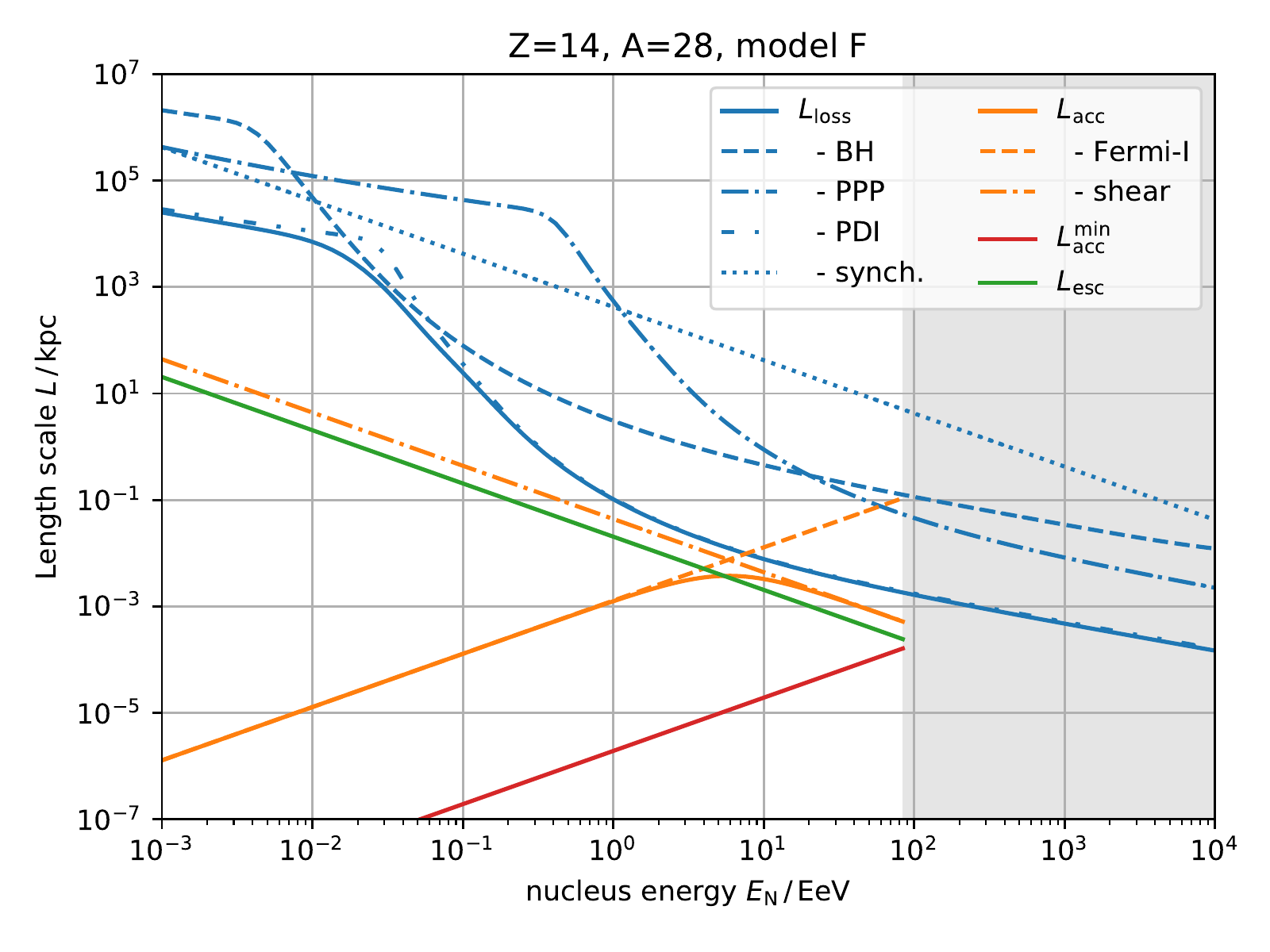}
        \includegraphics[width=\textwidth]{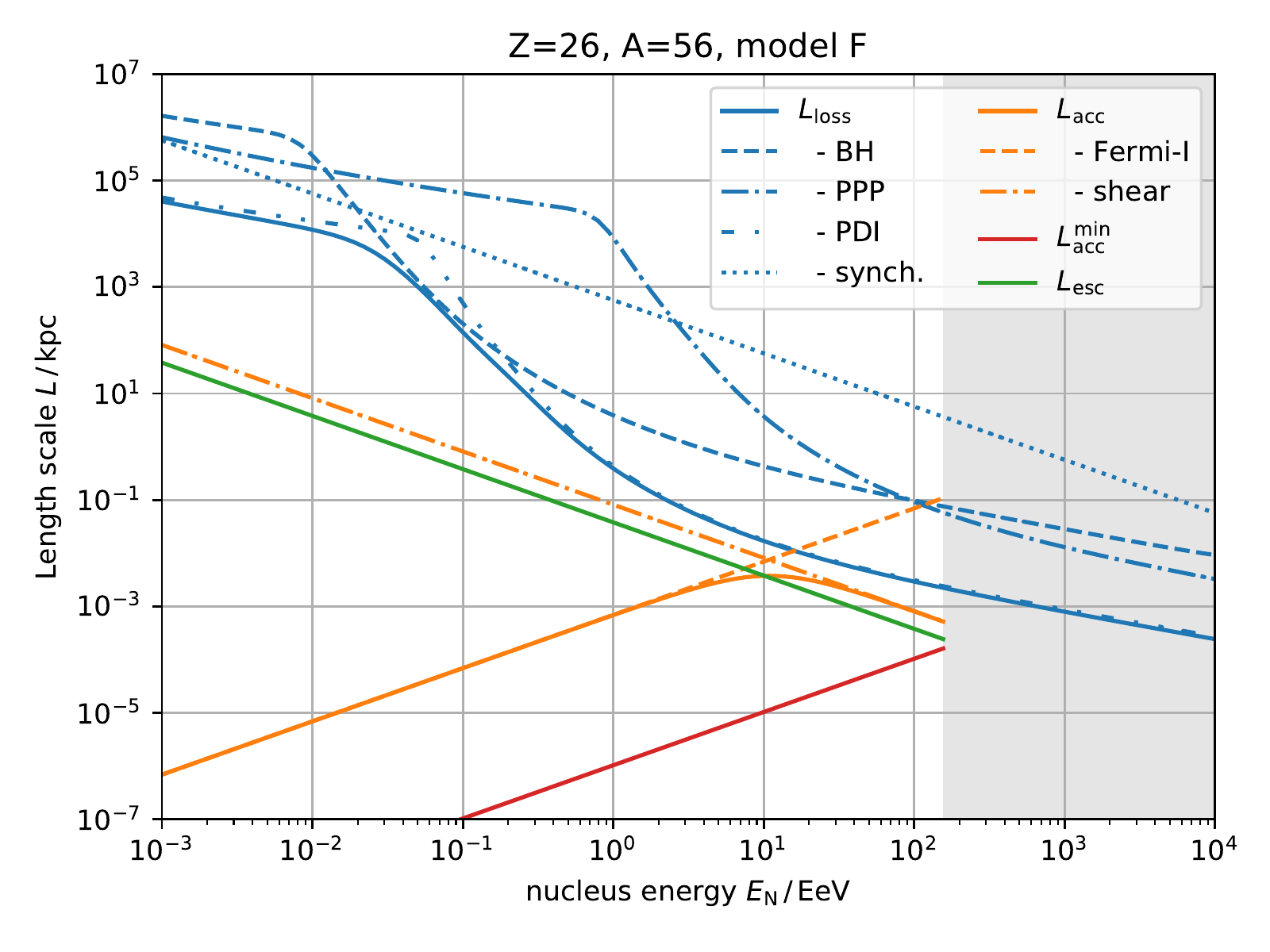}
    \end{minipage}
    \caption{Comparison of the lengths scales for all relevant processes based on Bohm diffusion is shown. From left to right the columns show models E and F, respectively. From top to bottom the tracer elements hydrogen, helium, nitrogen, silicon, and iron are shown.}
    \label{fig:AllLengths2}
\end{figure}

\begin{figure}[htbp]
\centering
    \begin{minipage}{.31\textwidth}
        \includegraphics[width=\textwidth]{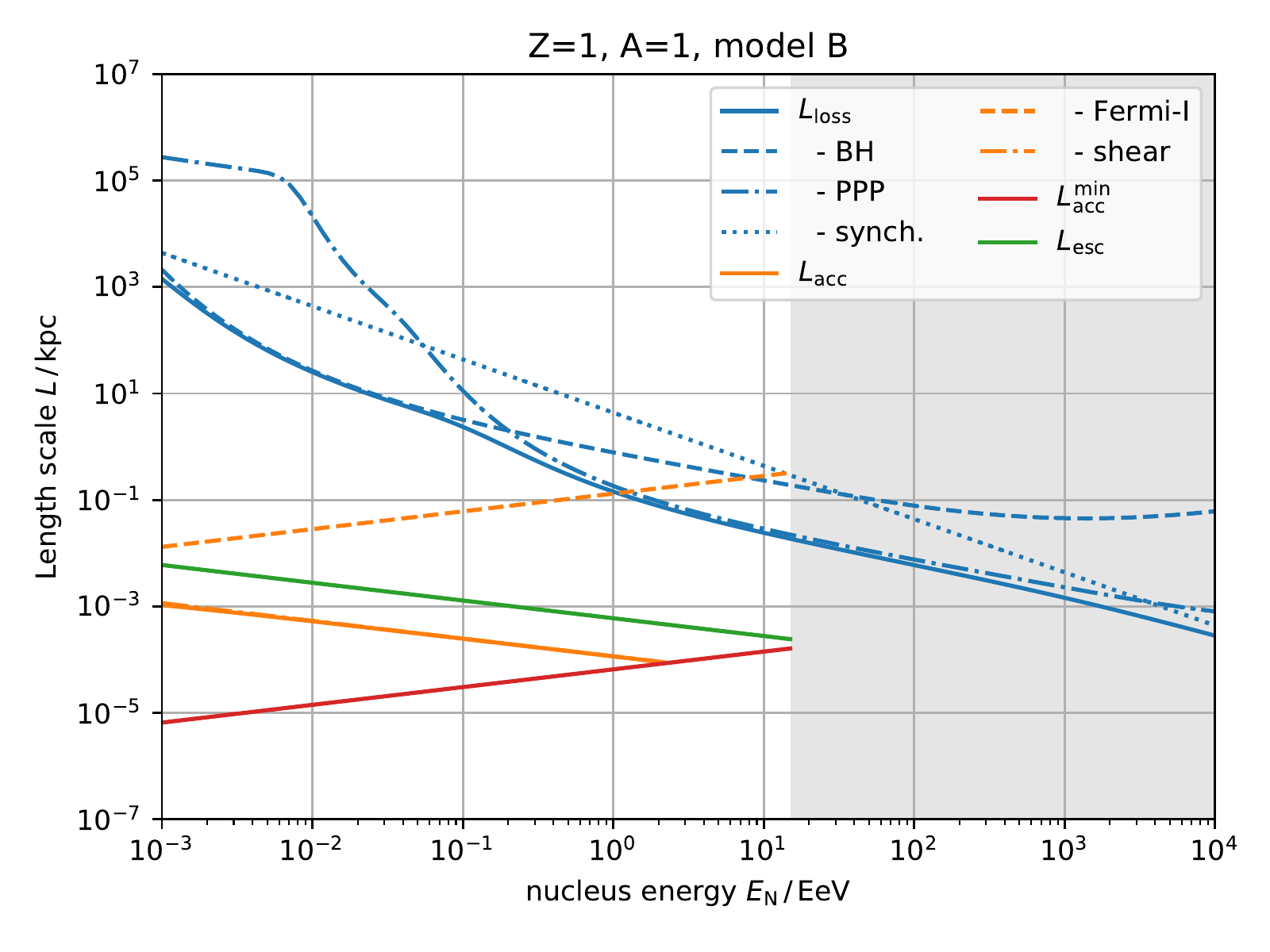}
        \includegraphics[width=\textwidth]{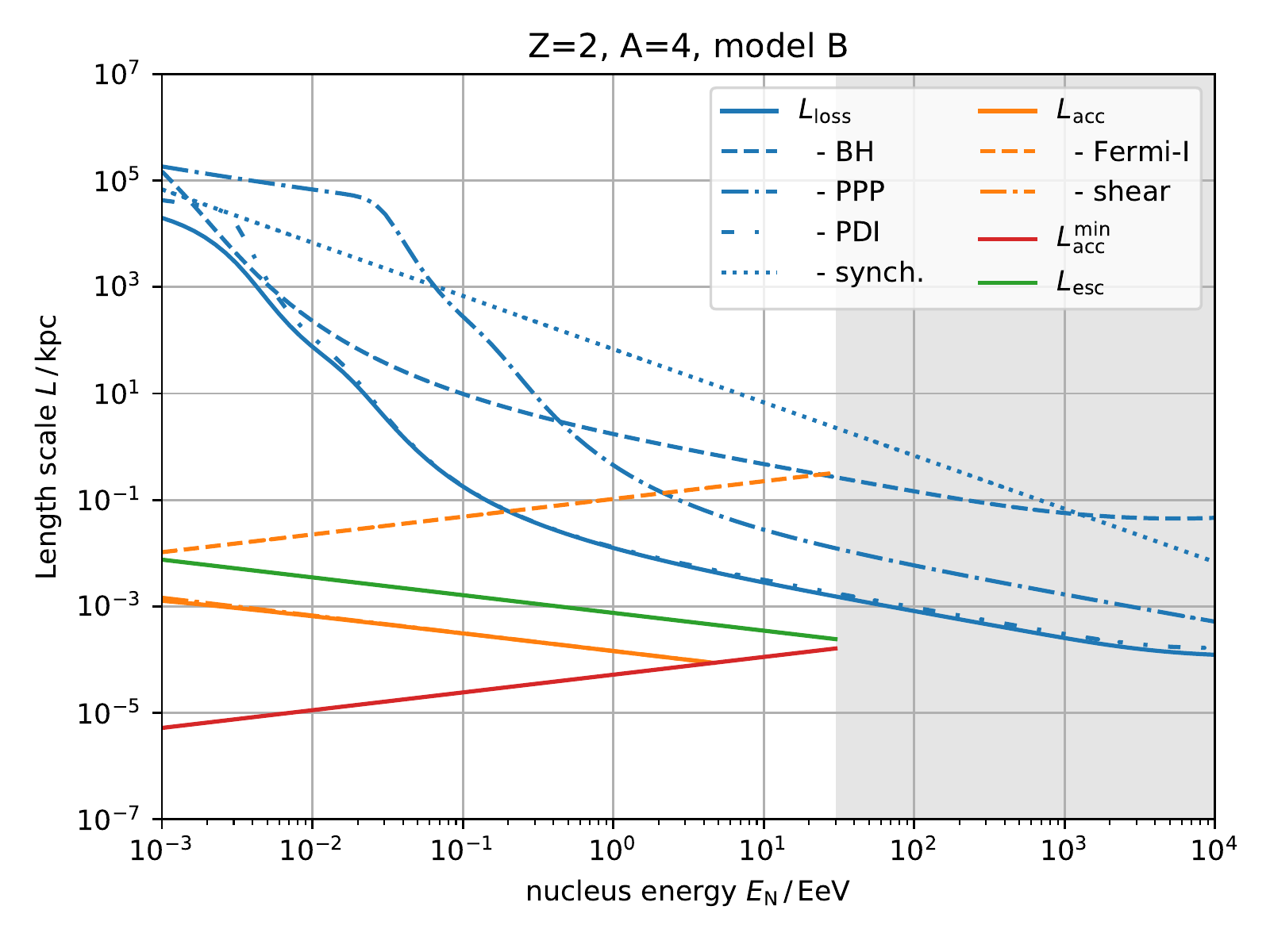}
        \includegraphics[width=\textwidth]{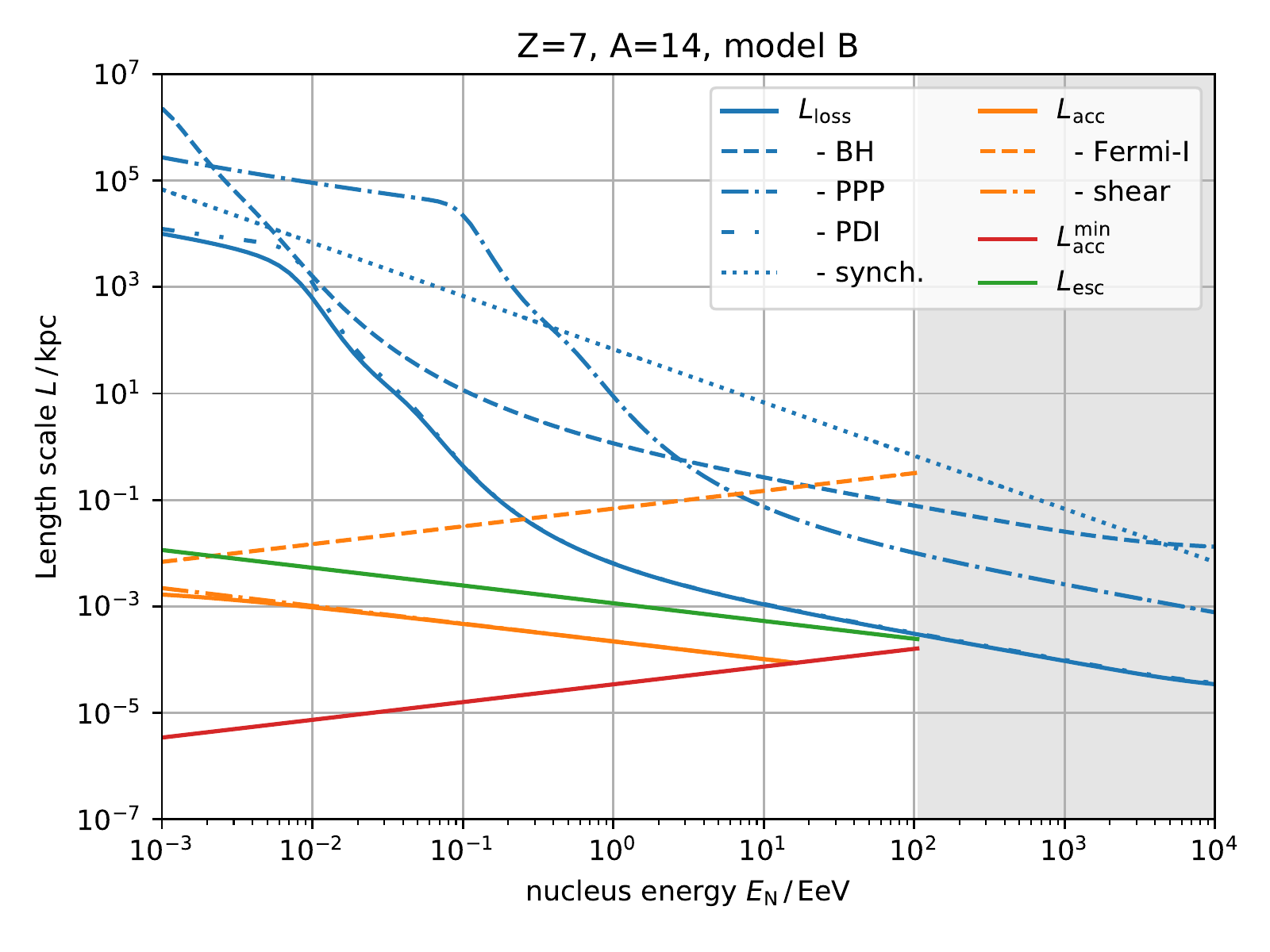}
        \includegraphics[width=\textwidth]{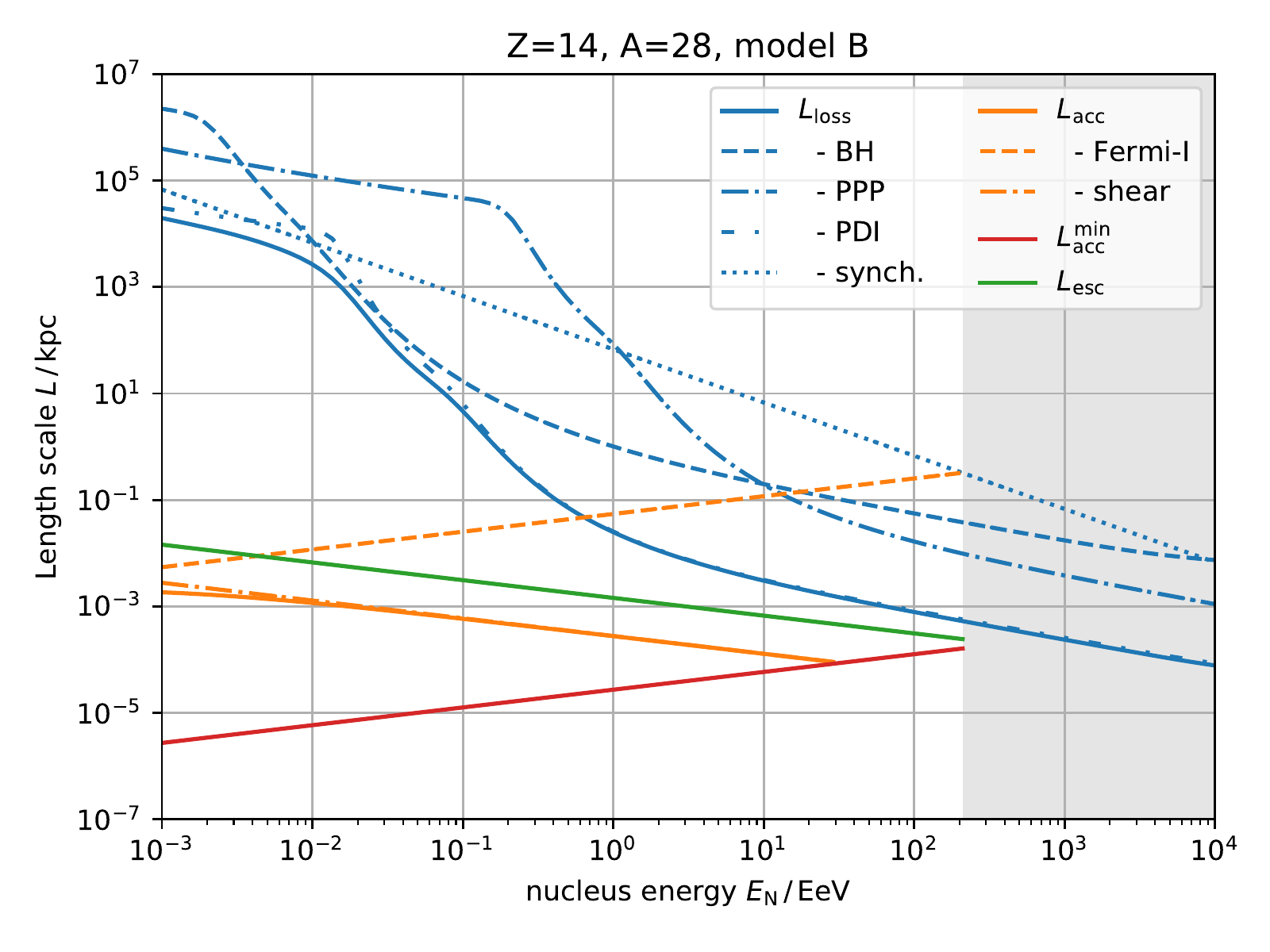}
        \includegraphics[width=\textwidth]{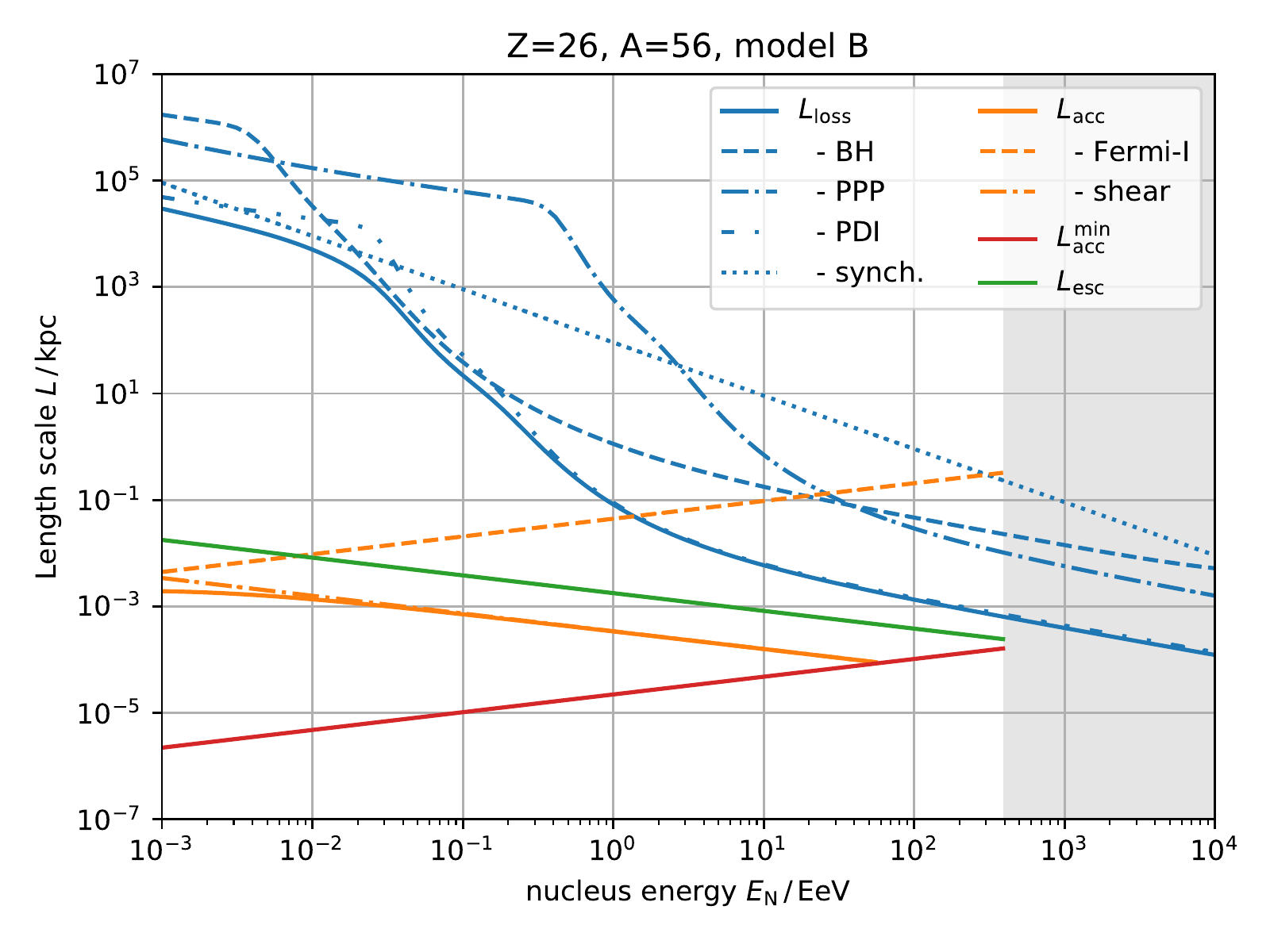}
    \end{minipage}
    \begin{minipage}{.31\textwidth}
        \includegraphics[width=\textwidth]{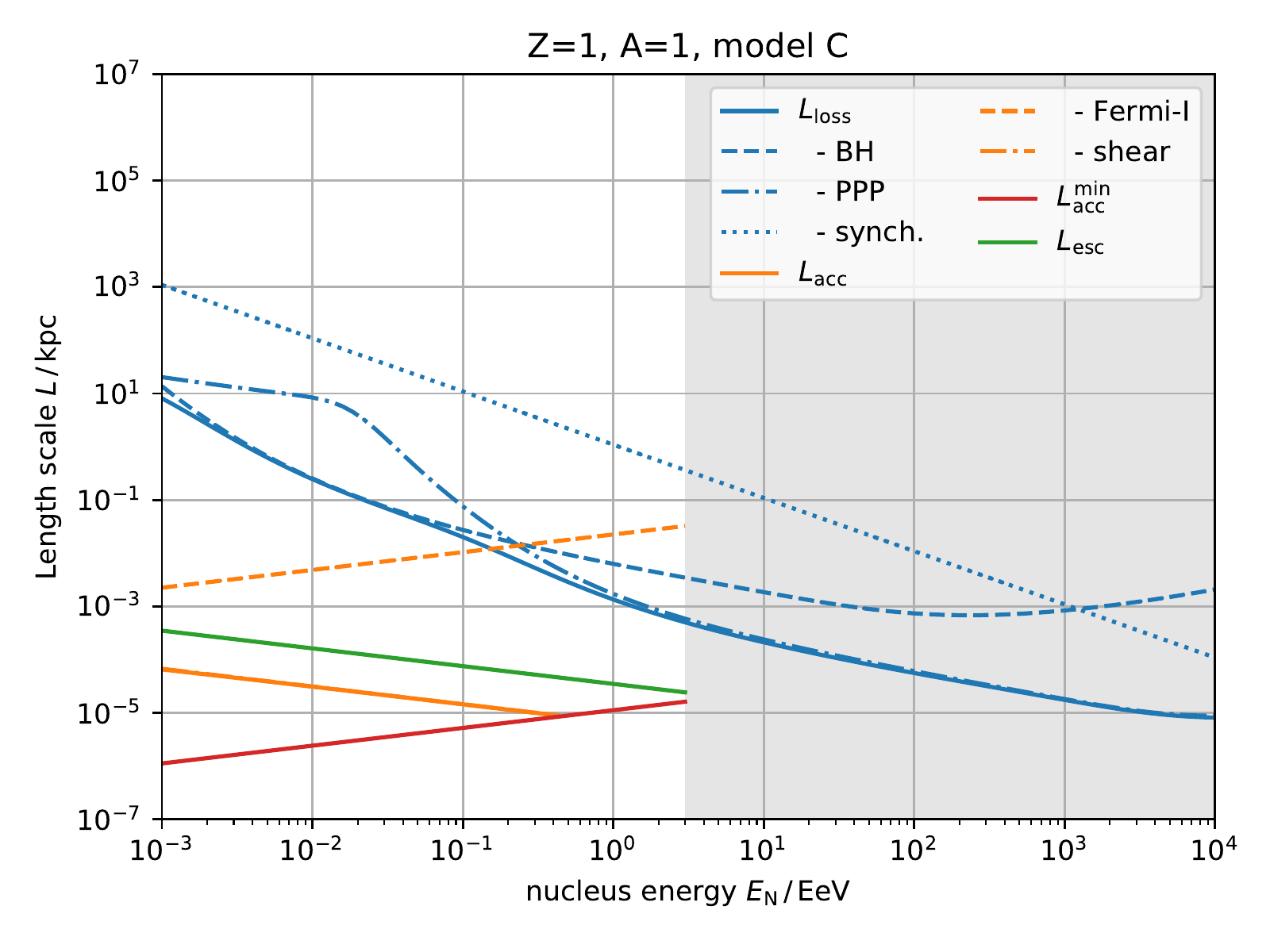}
        \includegraphics[width=\textwidth]{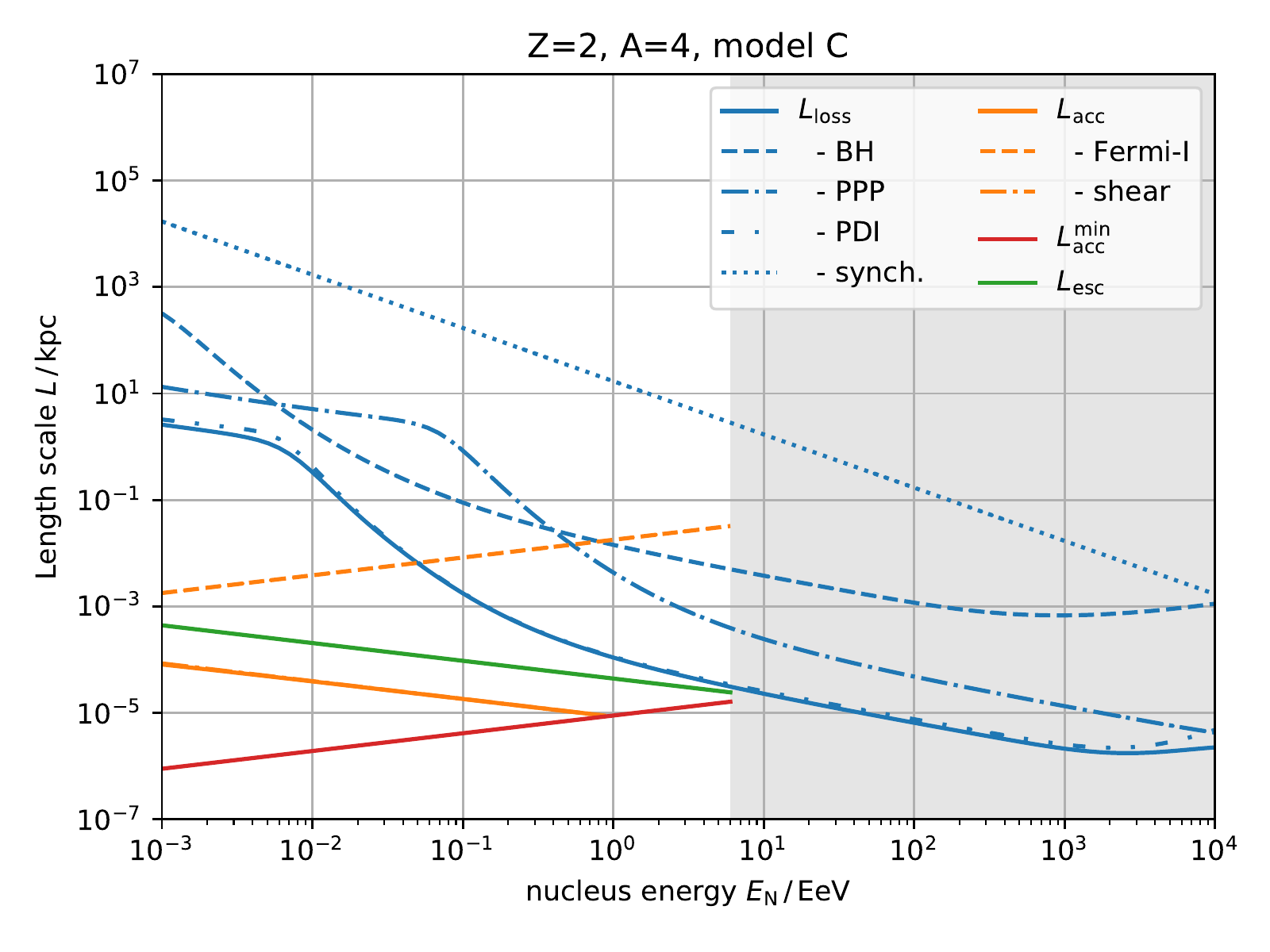}
        \includegraphics[width=\textwidth]{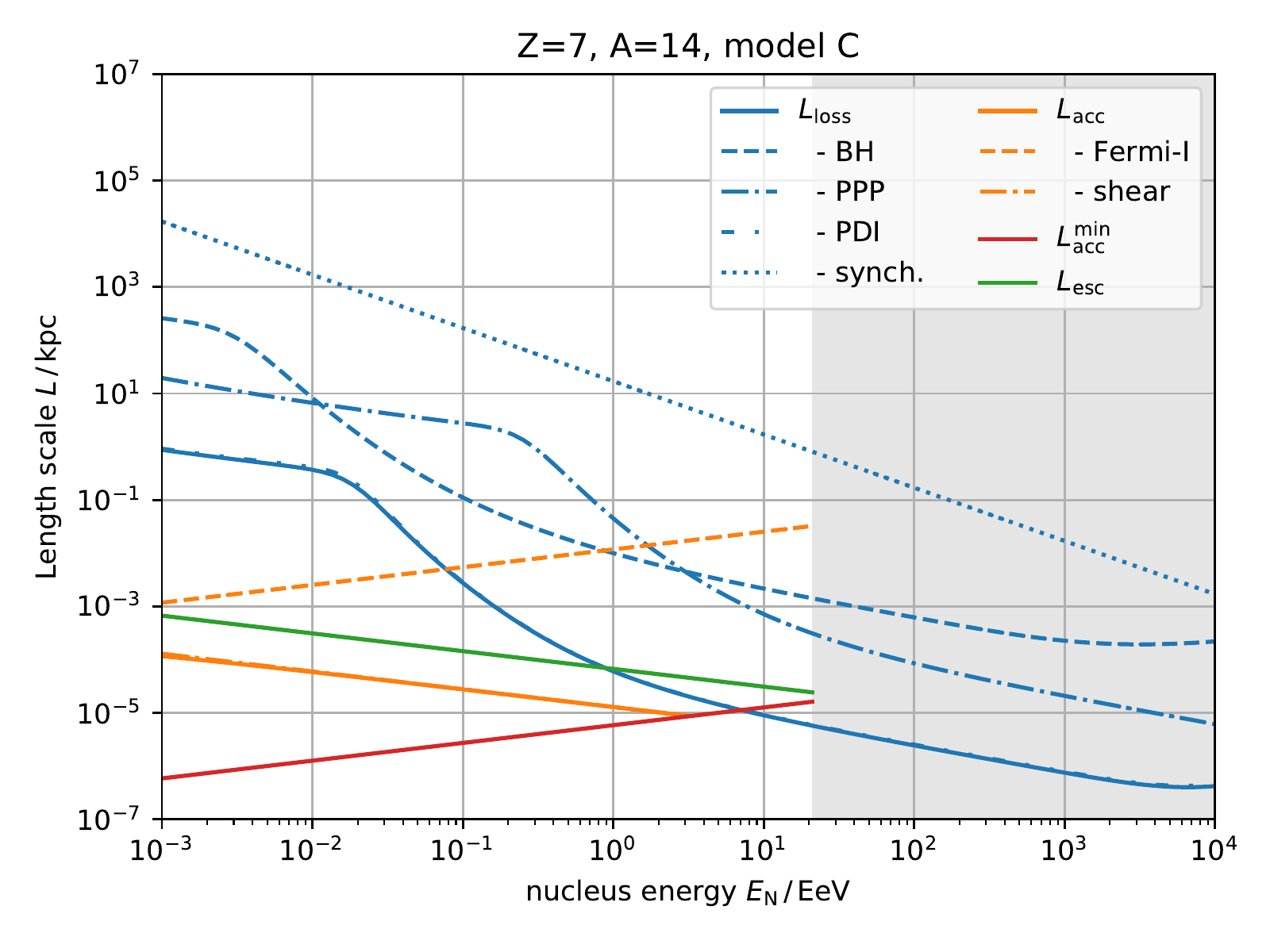}
        \includegraphics[width=\textwidth]{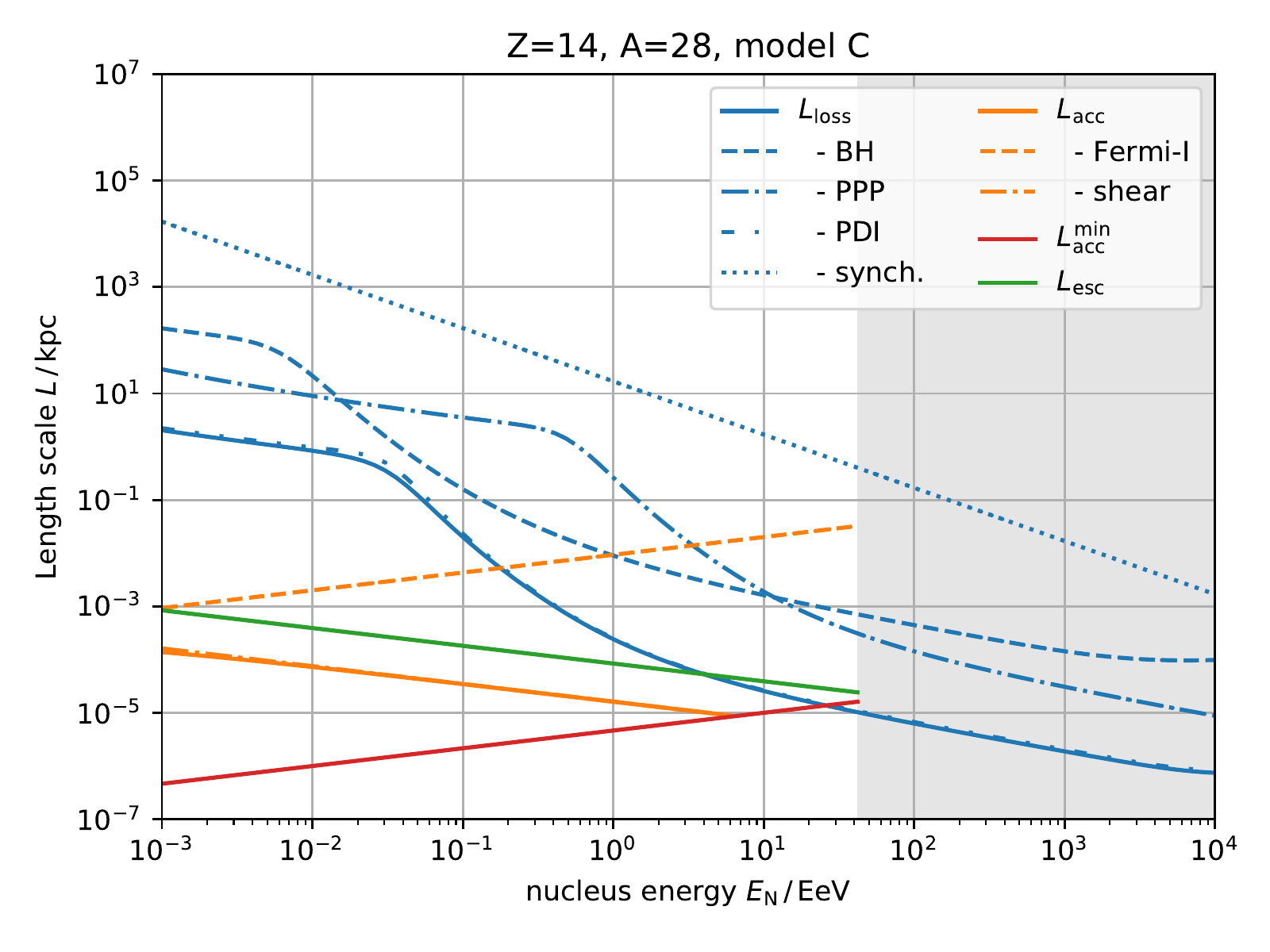}
        \includegraphics[width=\textwidth]{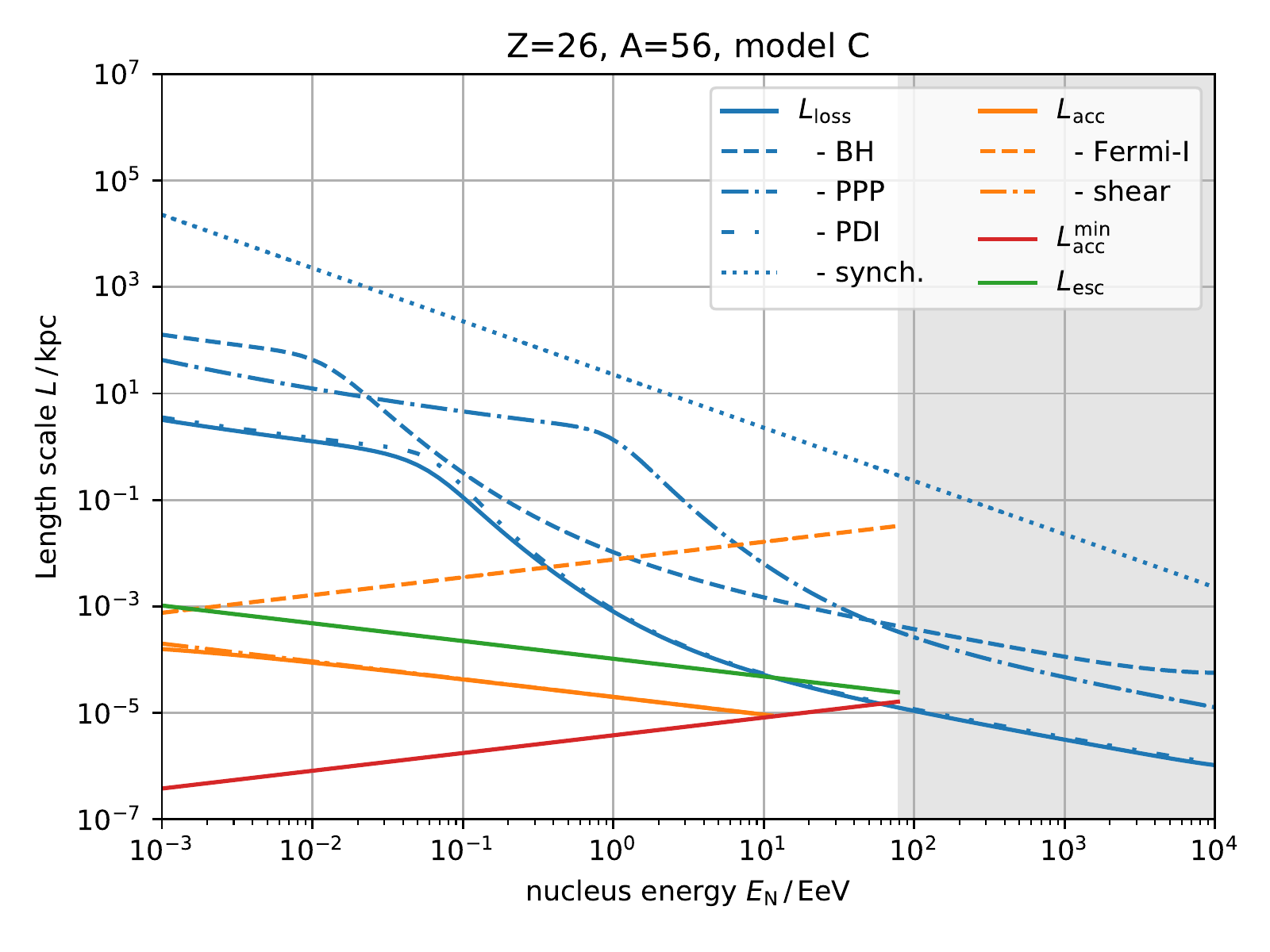}
    \end{minipage}
    \begin{minipage}{.31\textwidth}
        \includegraphics[width=\textwidth]{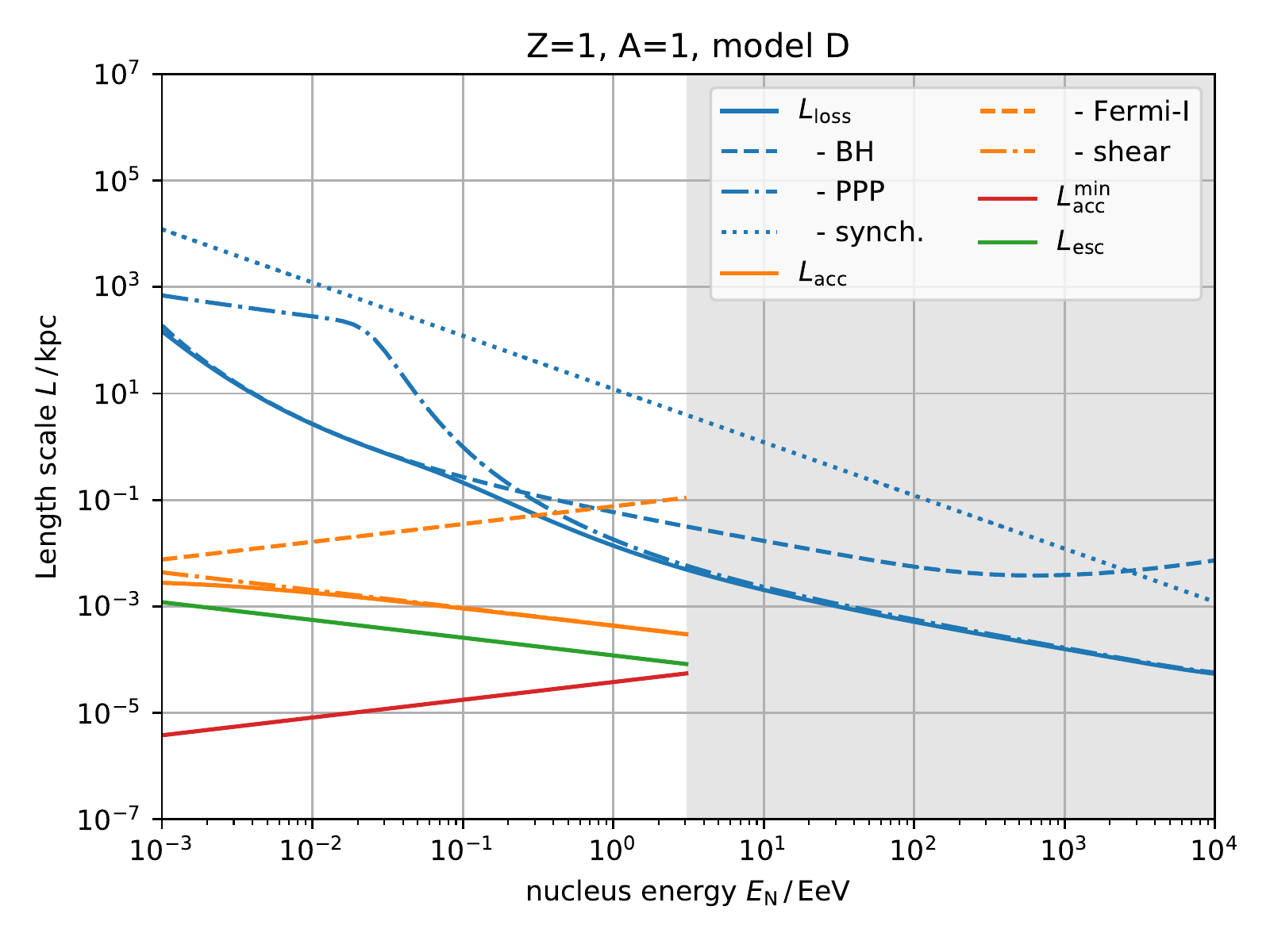}
        \includegraphics[width=\textwidth]{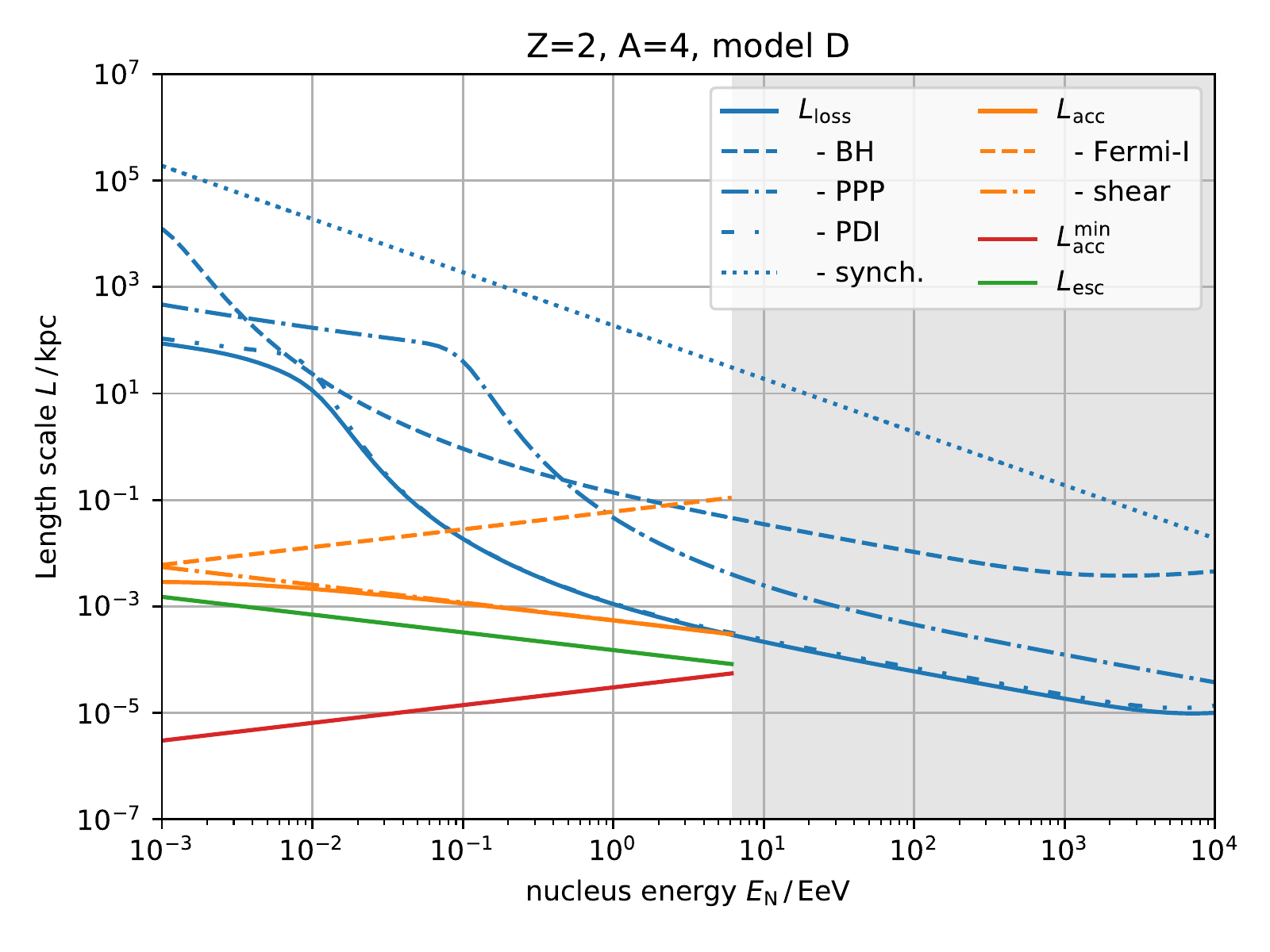}
        \includegraphics[width=\textwidth]{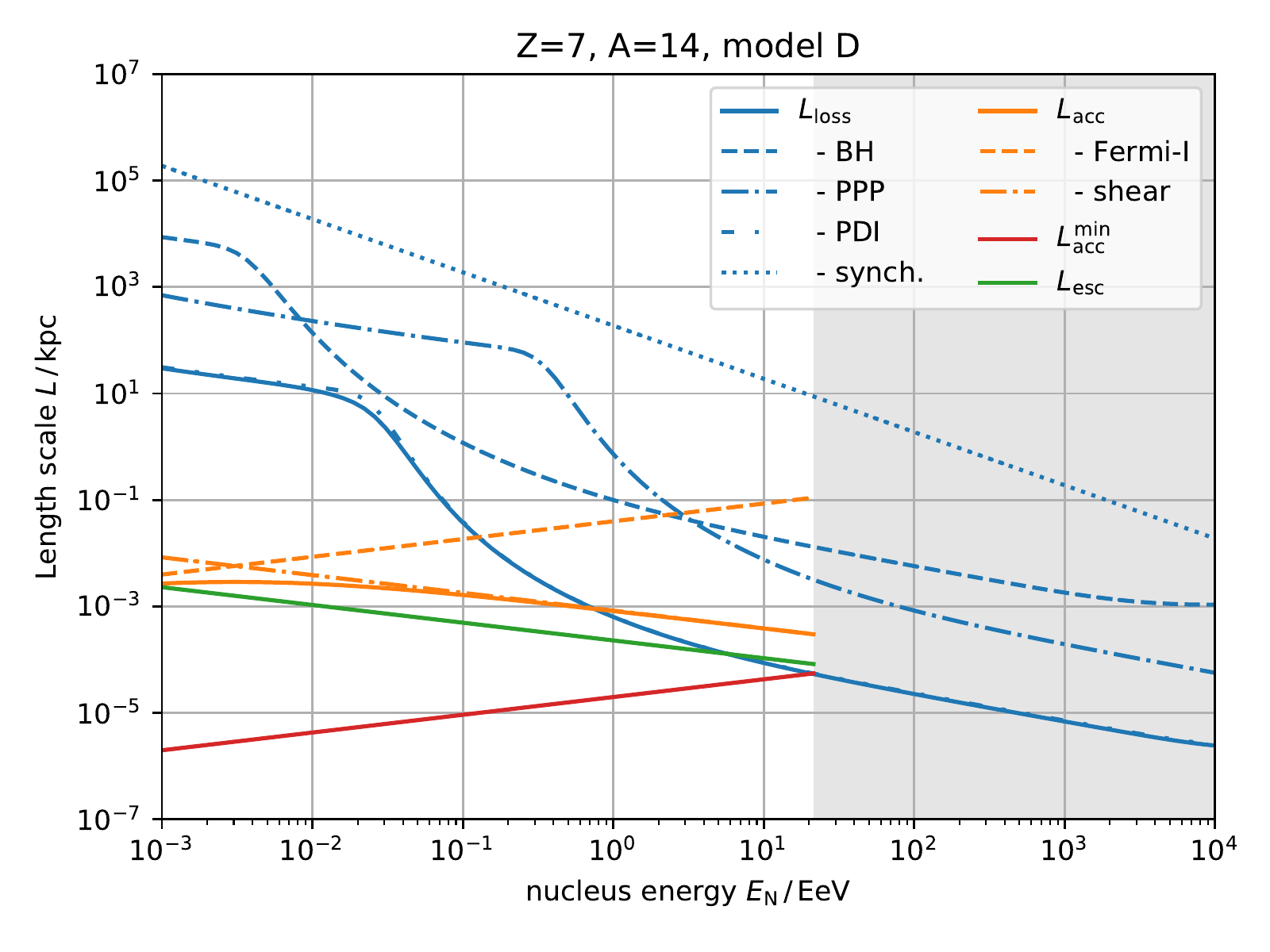}
        \includegraphics[width=\textwidth]{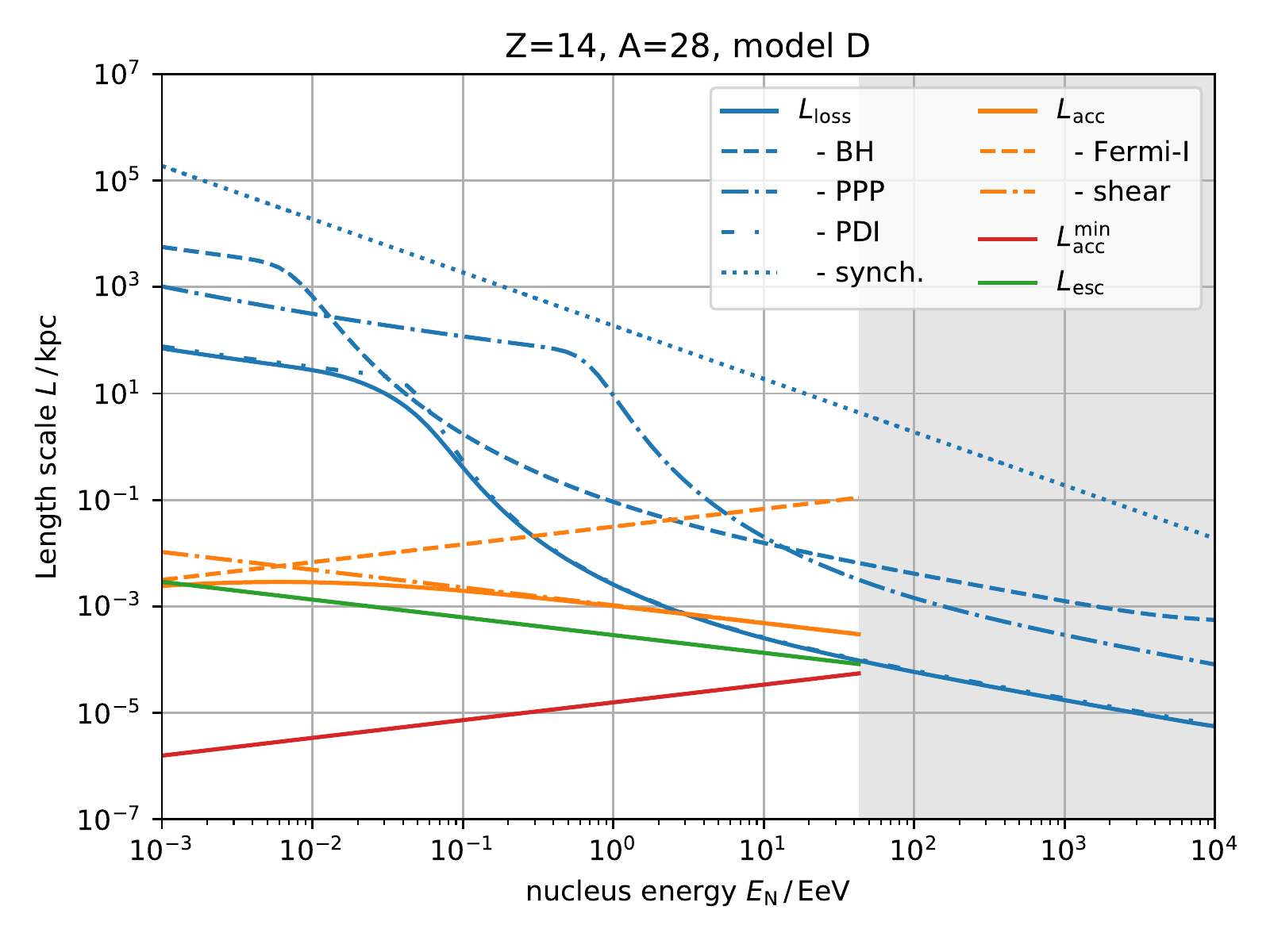}
        \includegraphics[width=\textwidth]{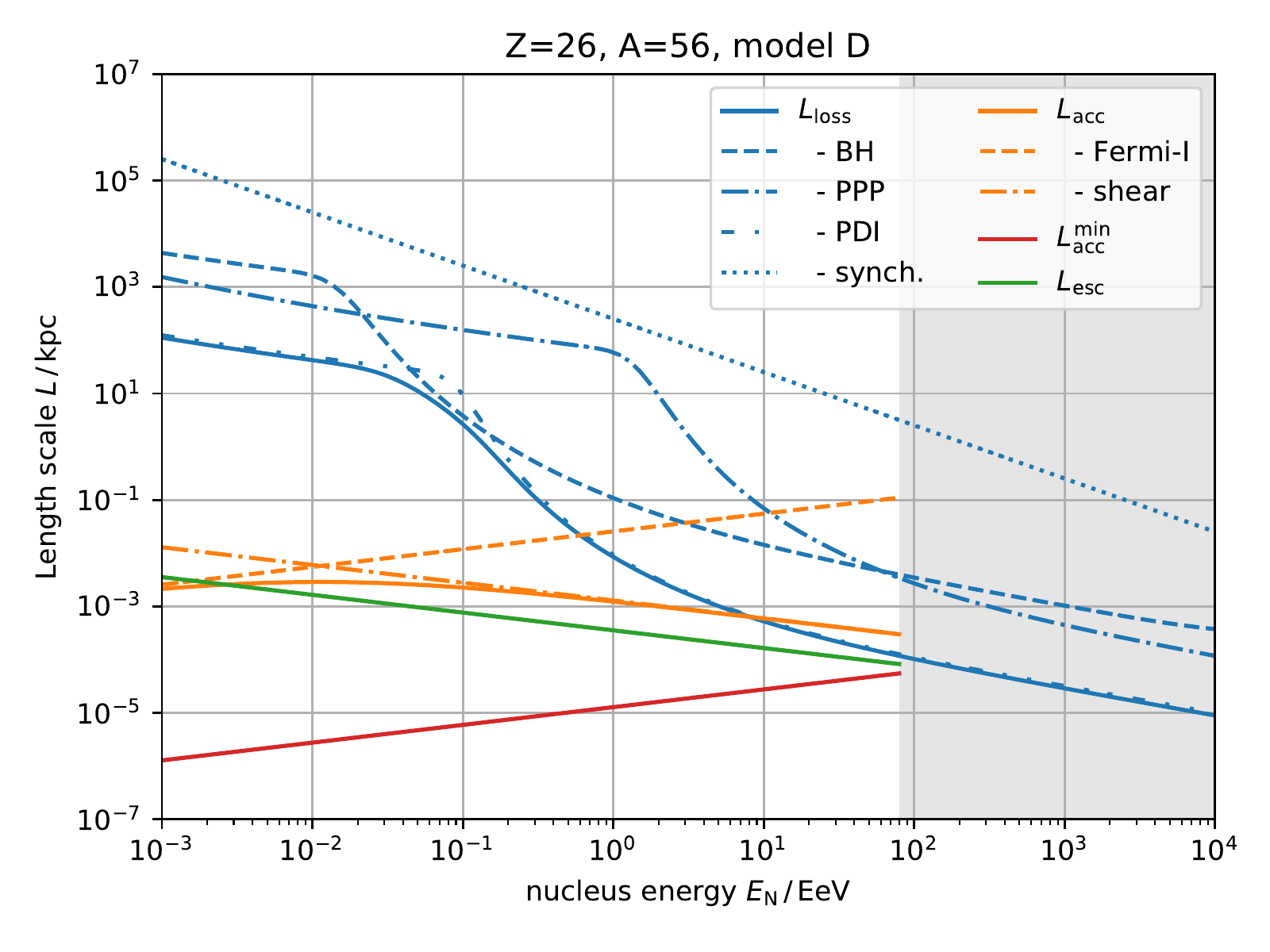}
    \end{minipage}
    \caption{Comparison of the lengths scales for all relevant processes based on Kolmogorov diffusion is shown. From left to right the columns show models B, C, and D respectively. From top to bottom the tracer elements hydrogen, helium, nitrogen, silicon, and iron are shown.}
    \label{fig:AllLengths1_kolm}
\end{figure}

\begin{figure}[htbp]
\centering
    \begin{minipage}{.33\textwidth}
        \includegraphics[width=\textwidth]{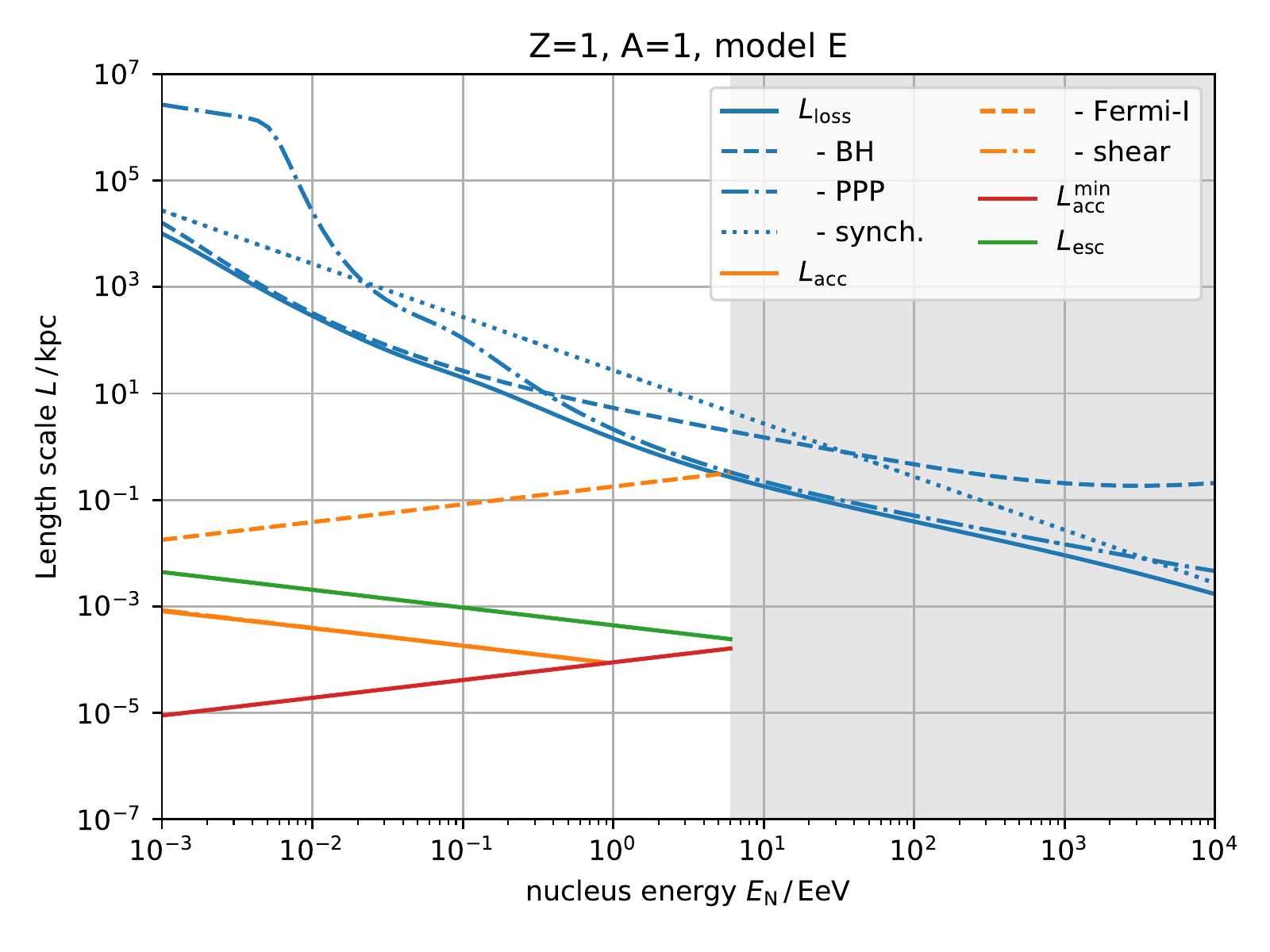}
        \includegraphics[width=\textwidth]{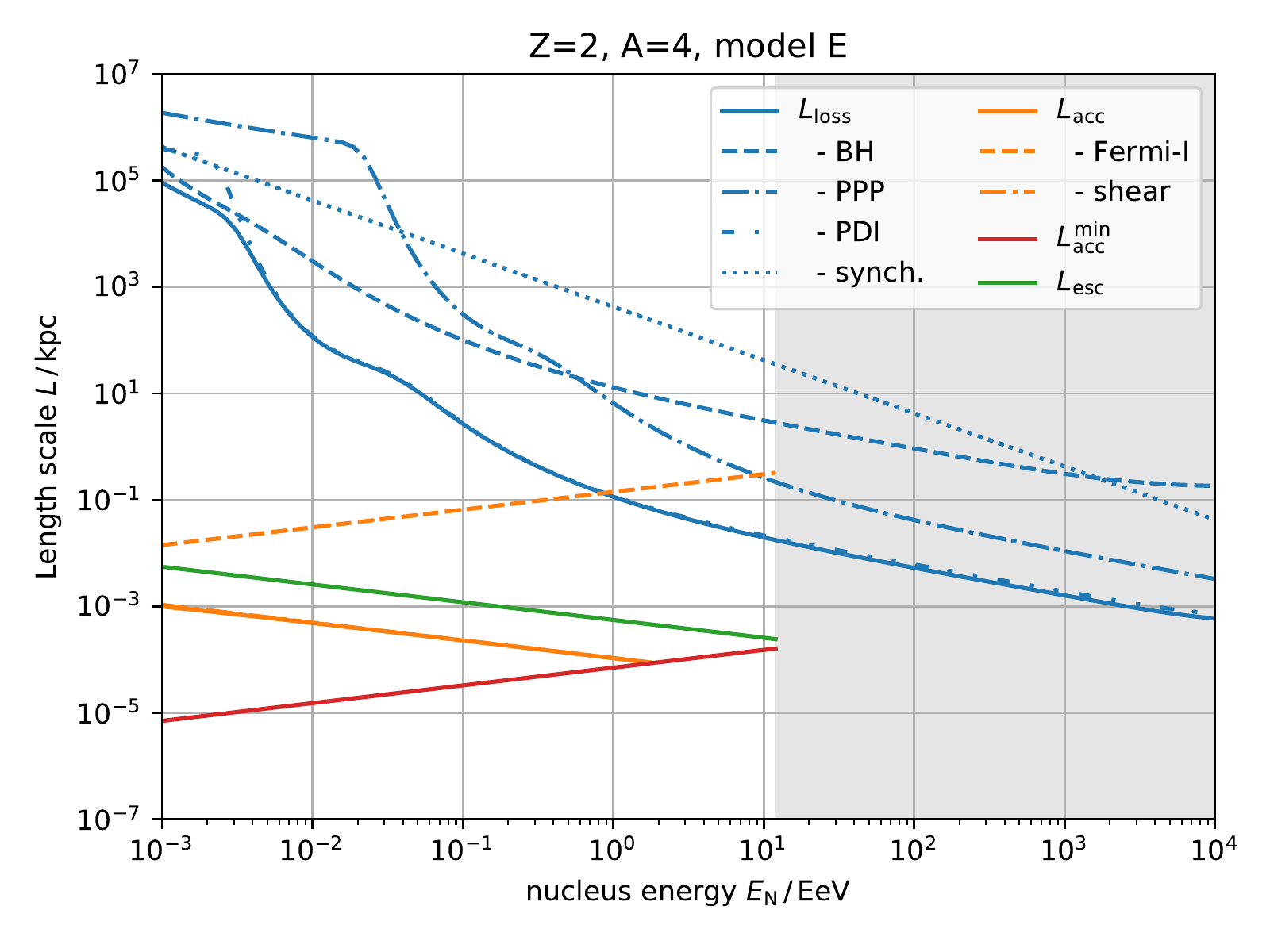}
        \includegraphics[width=\textwidth]{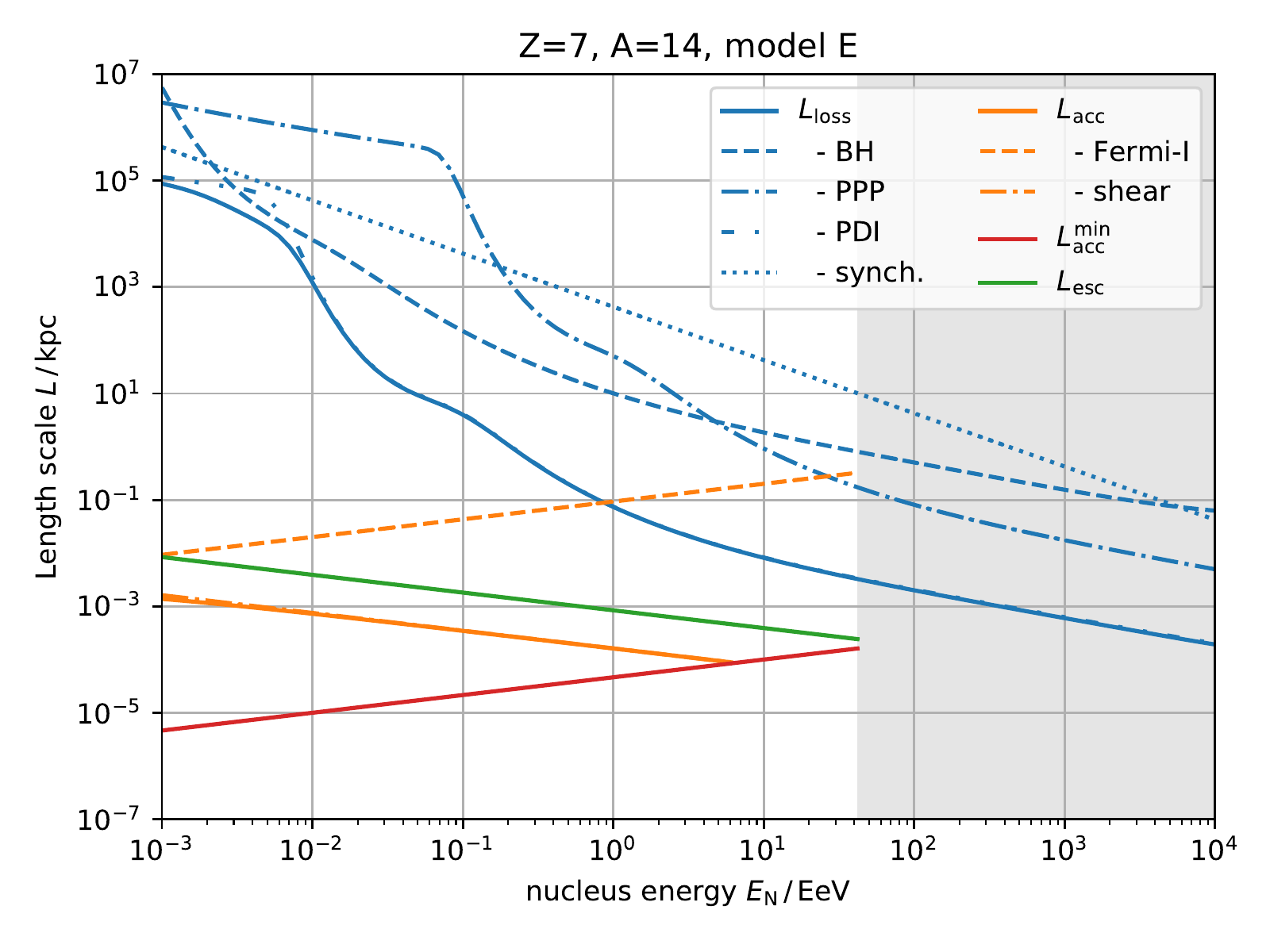}
        \includegraphics[width=\textwidth]{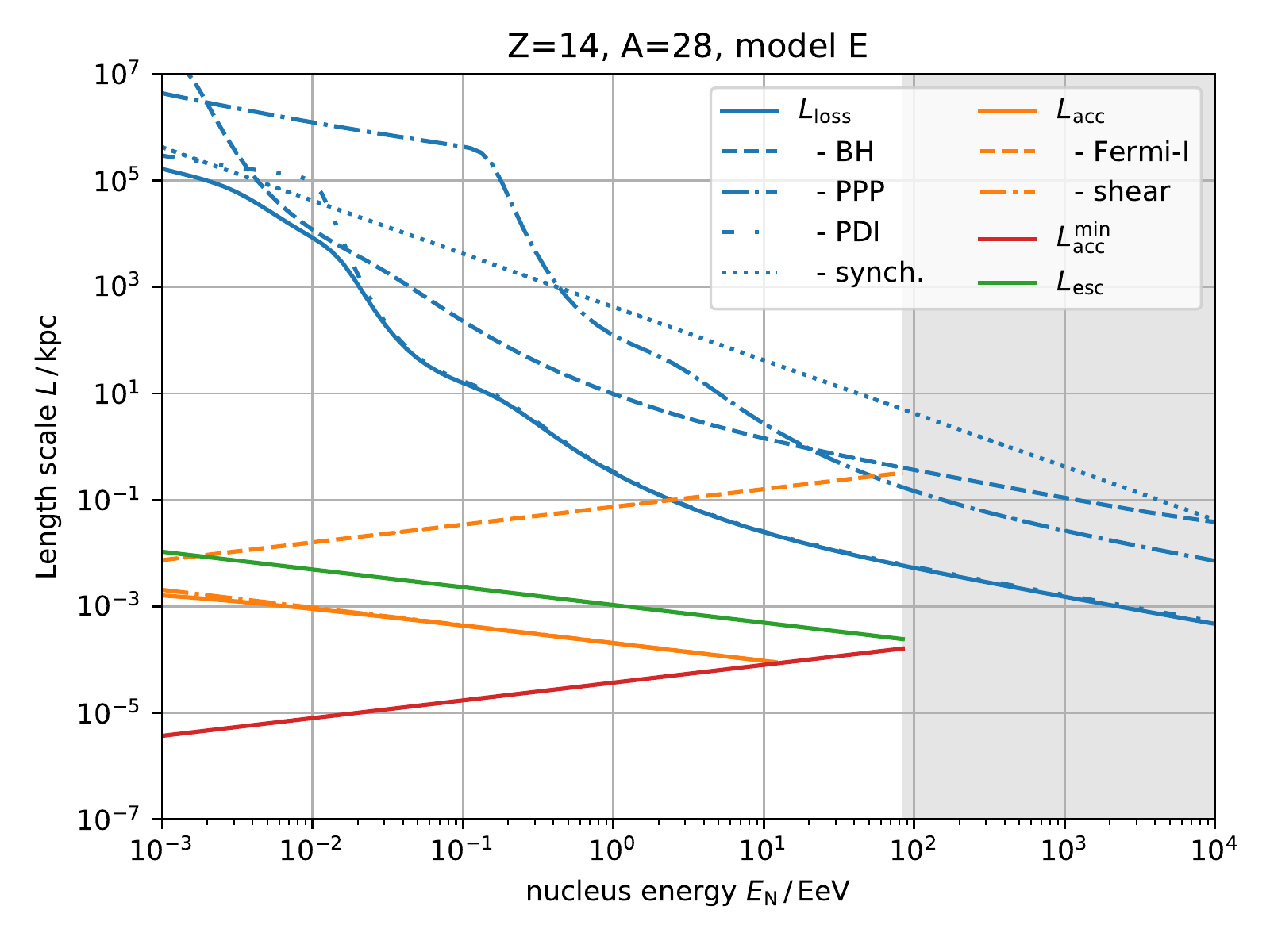}
        \includegraphics[width=\textwidth]{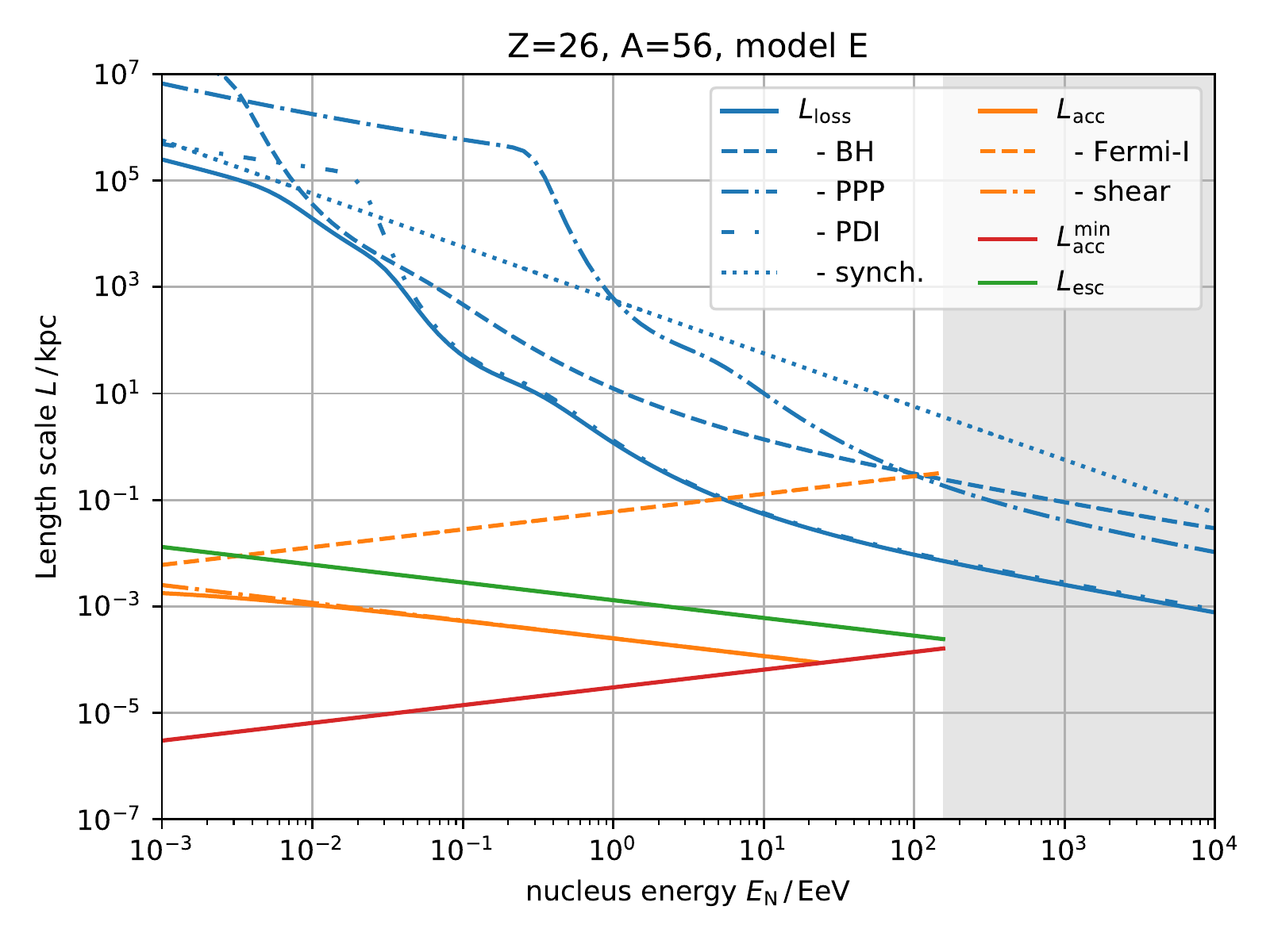}
    \end{minipage}
    \begin{minipage}{.33\textwidth}
        \includegraphics[width=\textwidth]{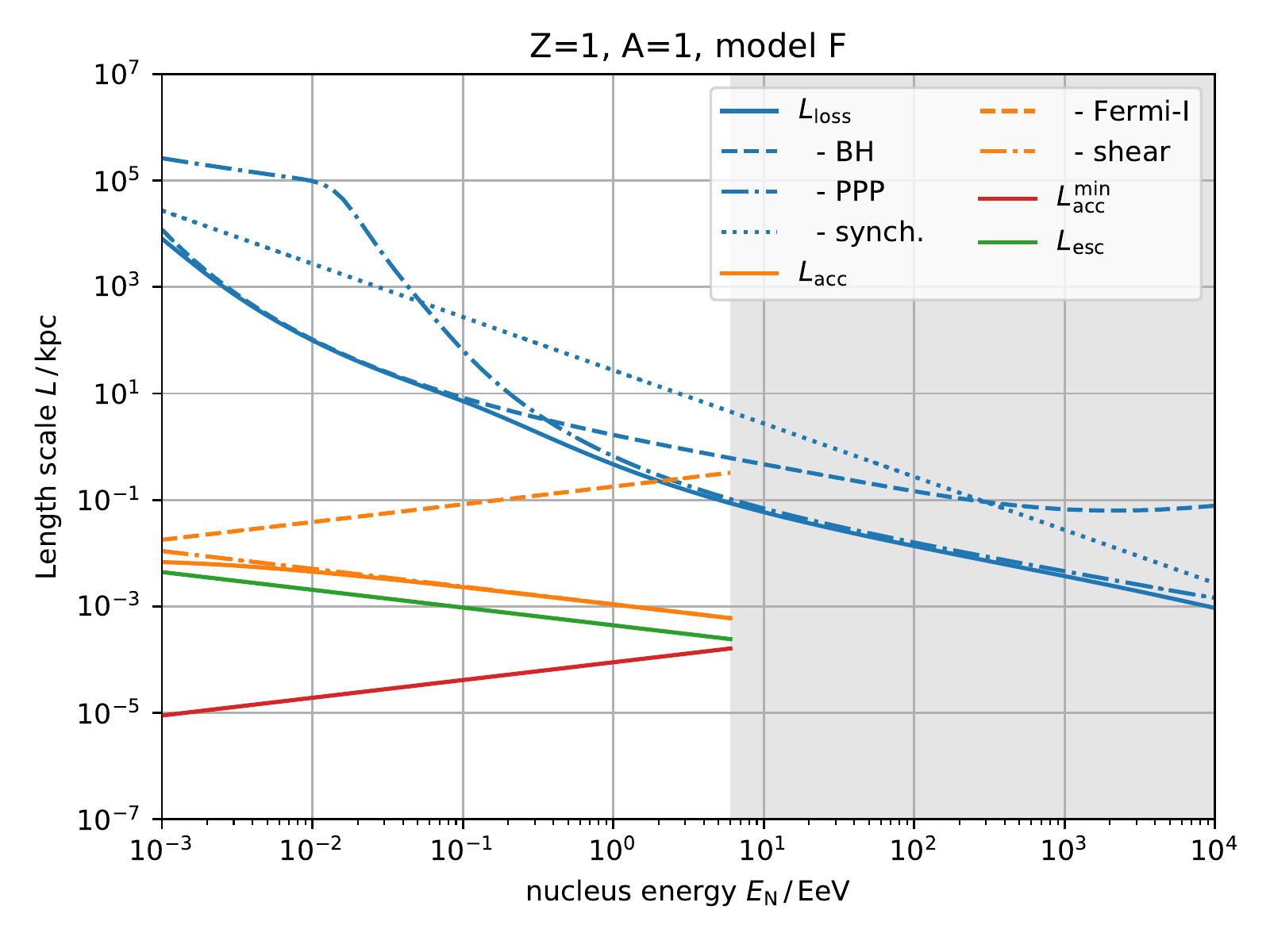}
        \includegraphics[width=\textwidth]{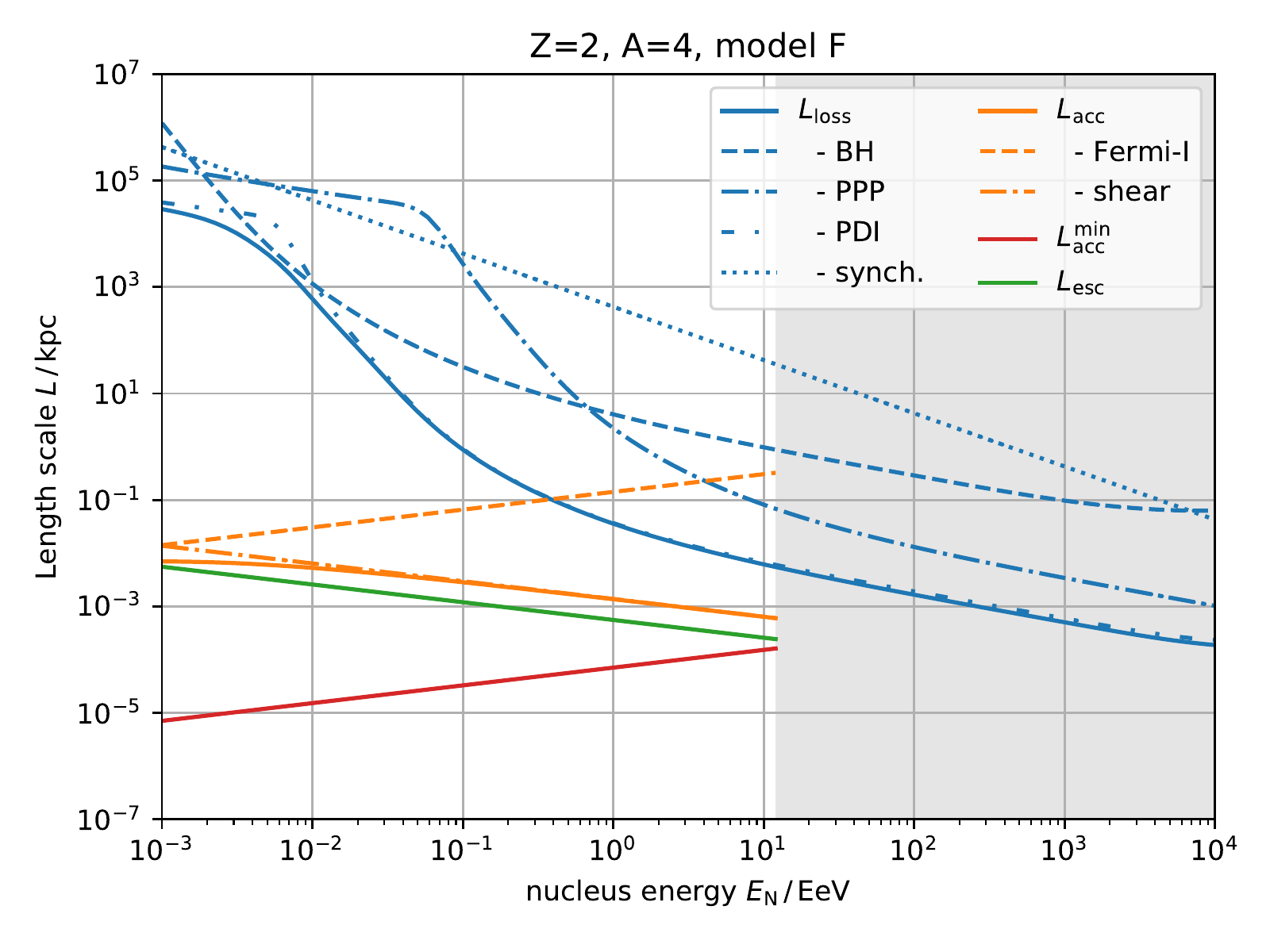}
        \includegraphics[width=\textwidth]{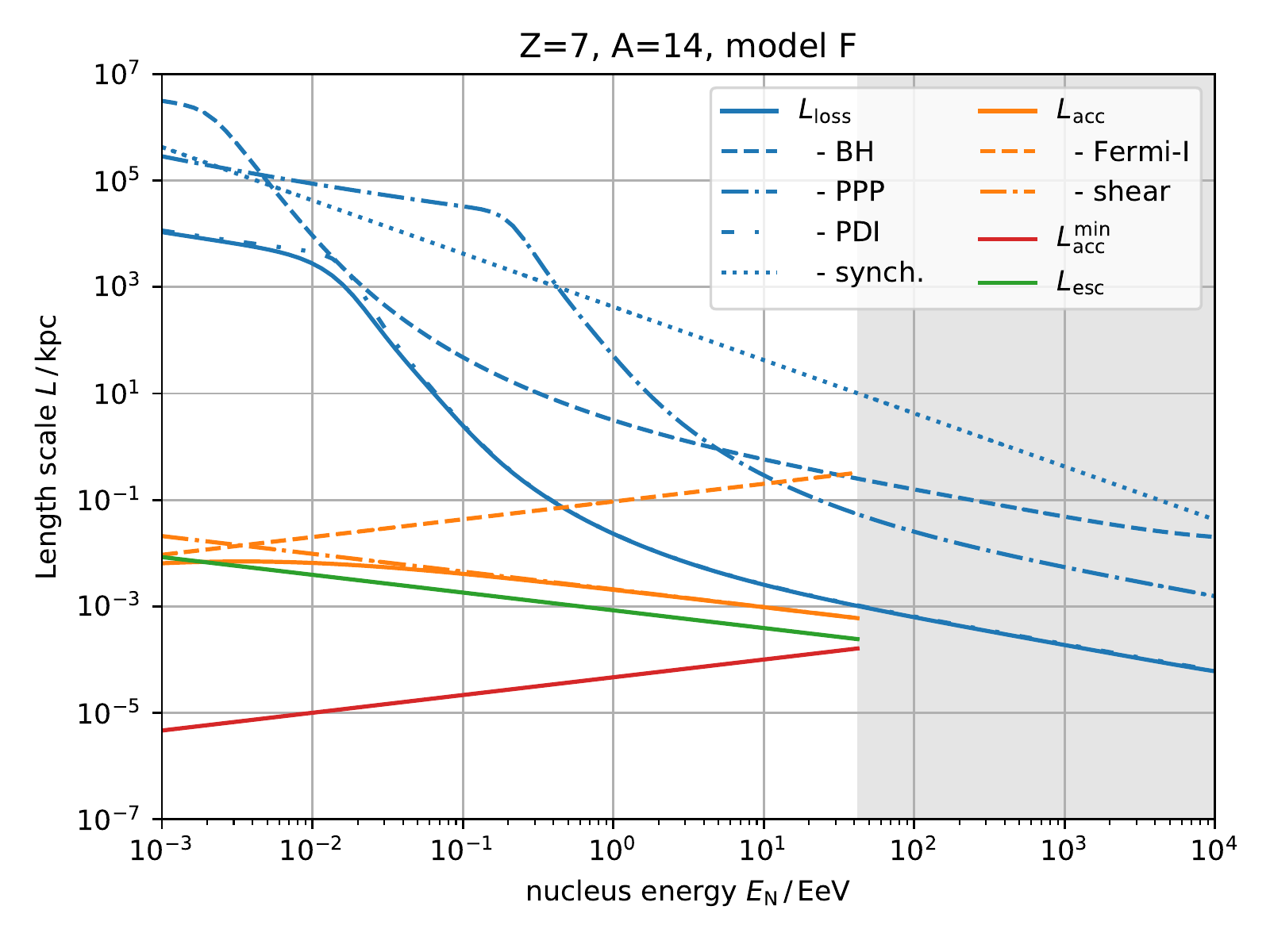}
        \includegraphics[width=\textwidth]{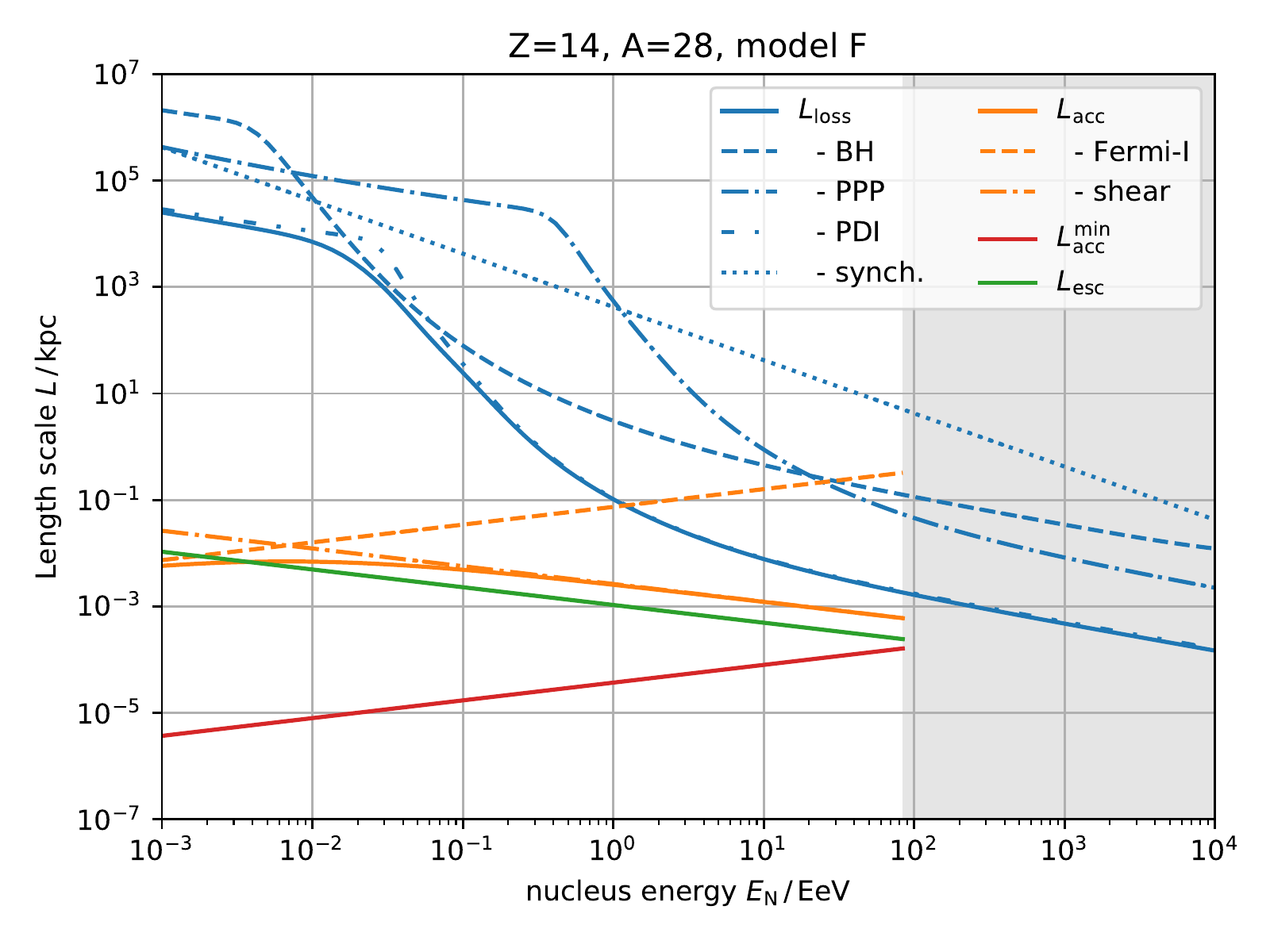}
        \includegraphics[width=\textwidth]{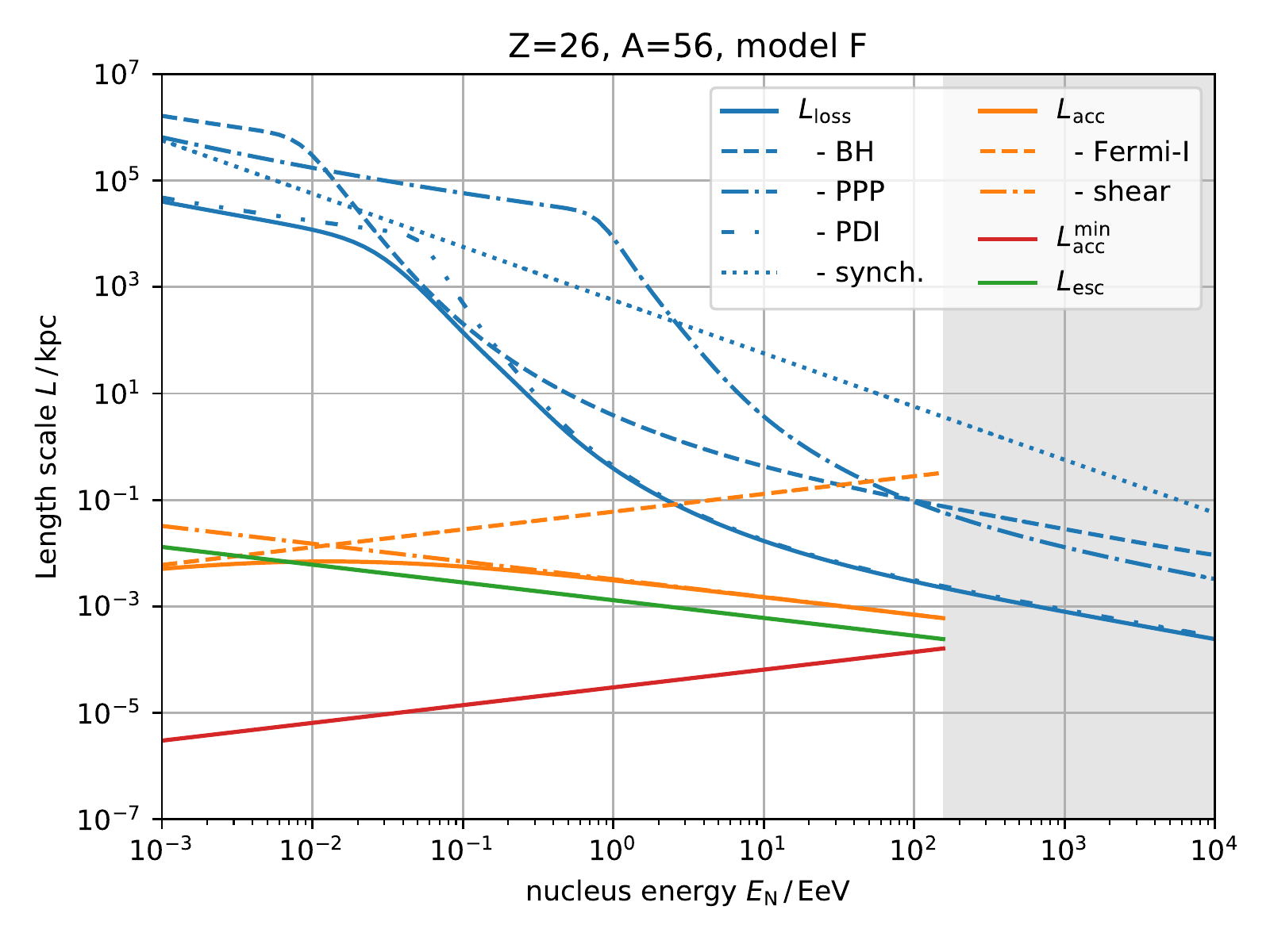}
    \end{minipage}
    \caption{Comparison of the lengths scales for all relevant processes based on Kolmogorov diffusion is shown. From left to right the columns show \emph{model E} and \emph{model F}, respectively. From top to bottom the tracer elements hydrogen, helium, nitrogen, silicon, and iron are shown.}
    \label{fig:AllLengths2_kolm}
\end{figure}

\section{Acceleration Probability}
\begin{figure}[h!]
    \begin{minipage}{.49\textwidth}
        \includegraphics[width=\textwidth]{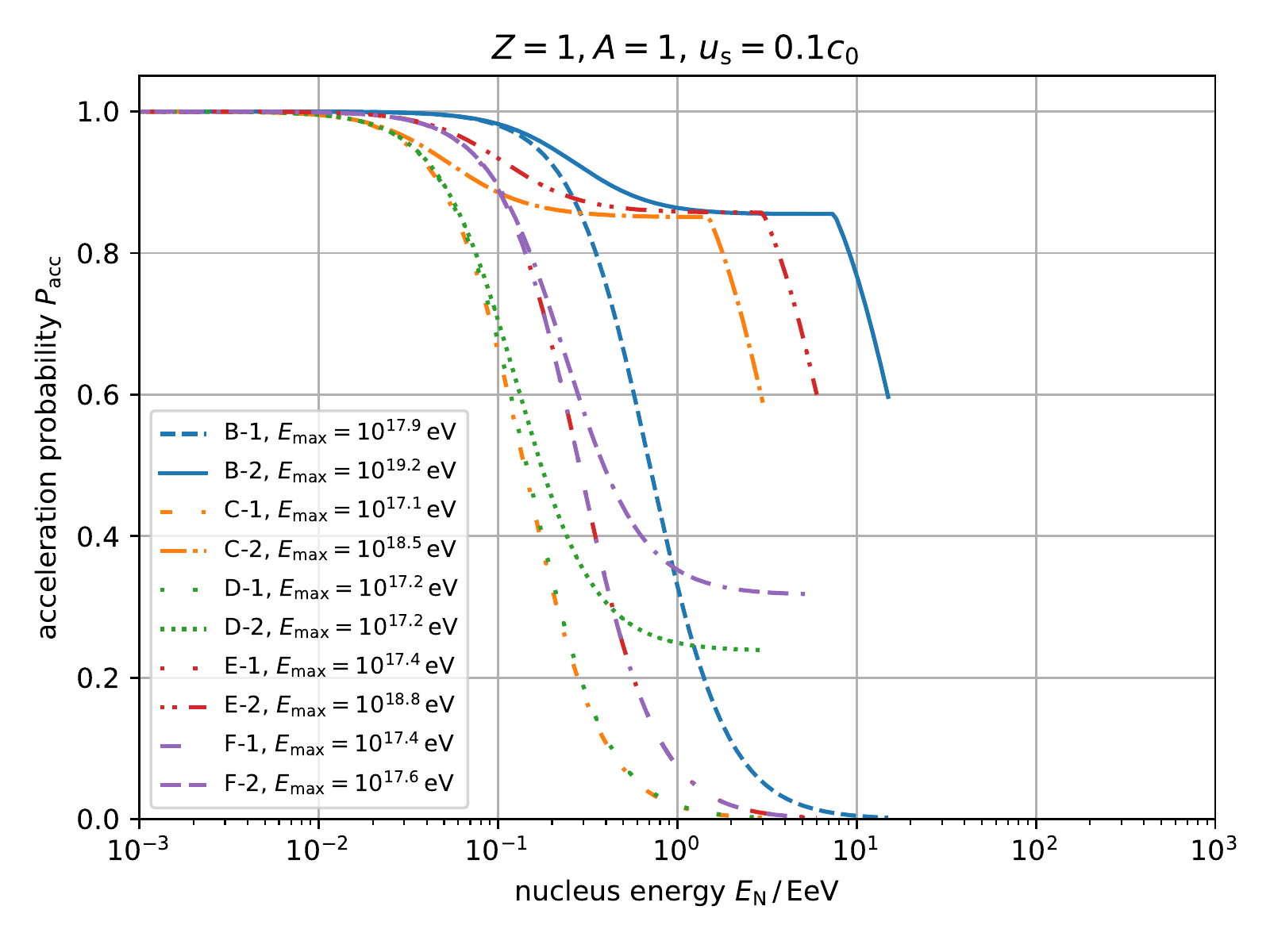}
        \includegraphics[width=\textwidth]{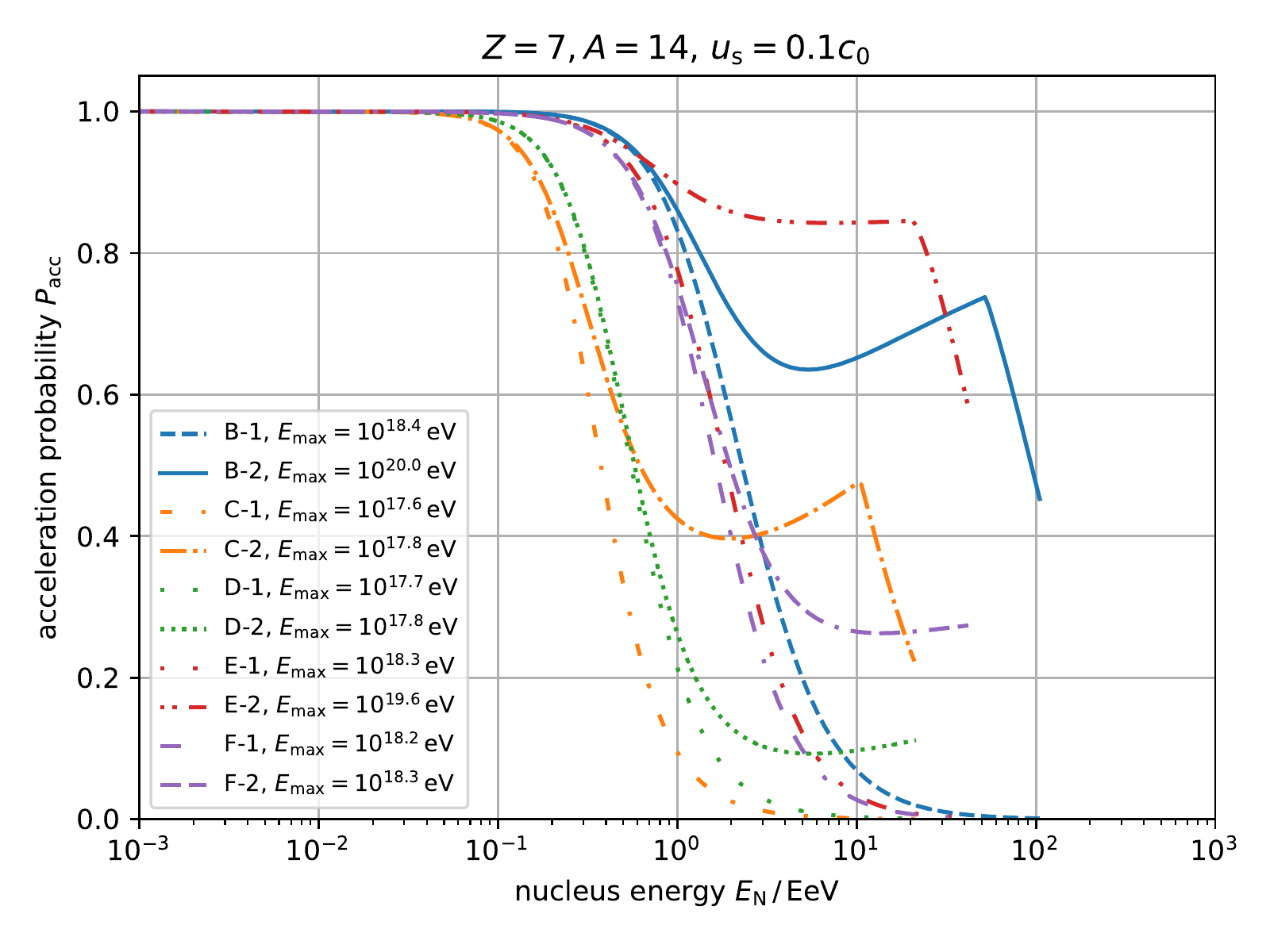}
    \end{minipage}
    \begin{minipage}{.49\textwidth}
        \includegraphics[width=\textwidth]{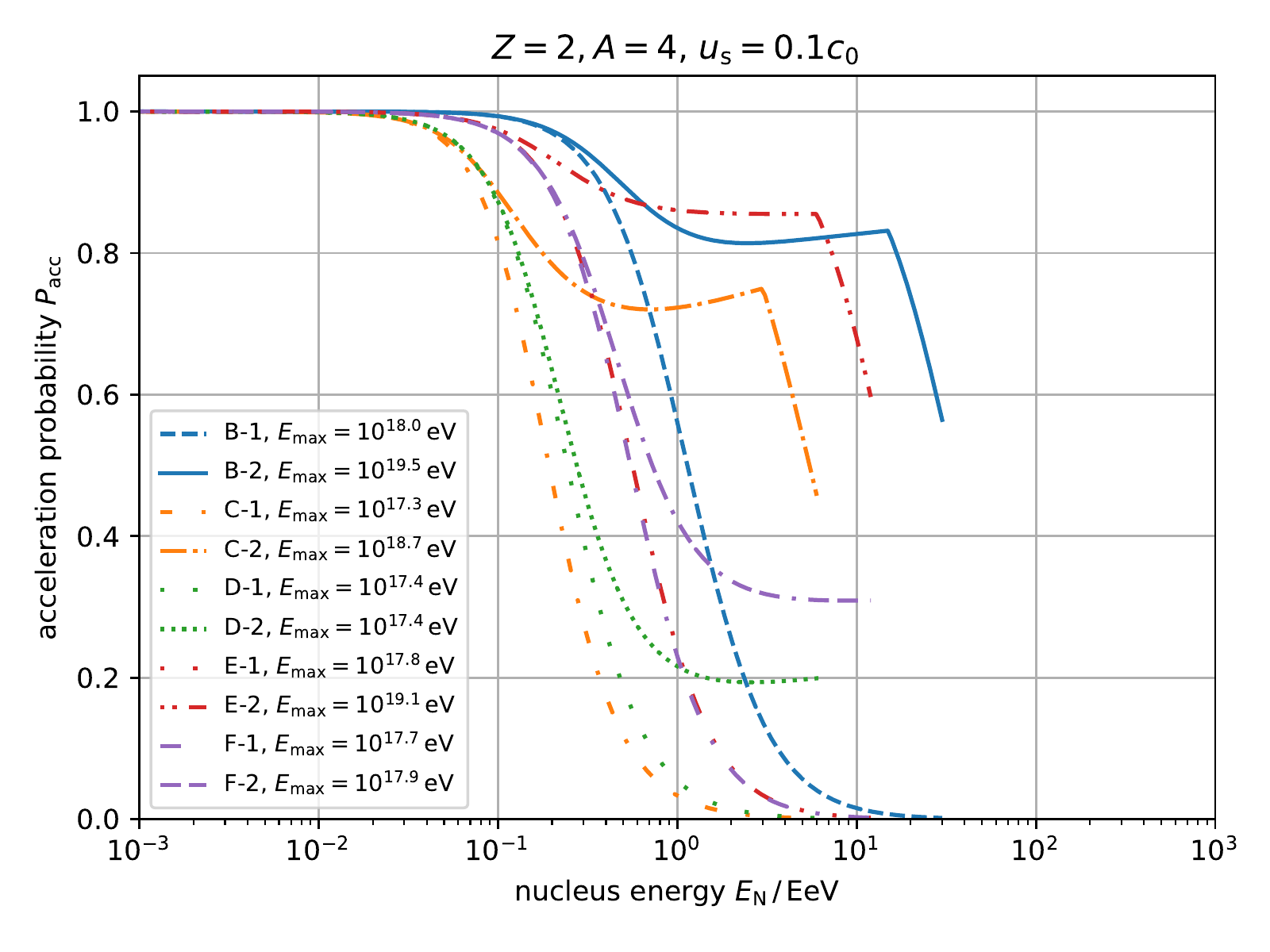}
        \includegraphics[width=\textwidth]{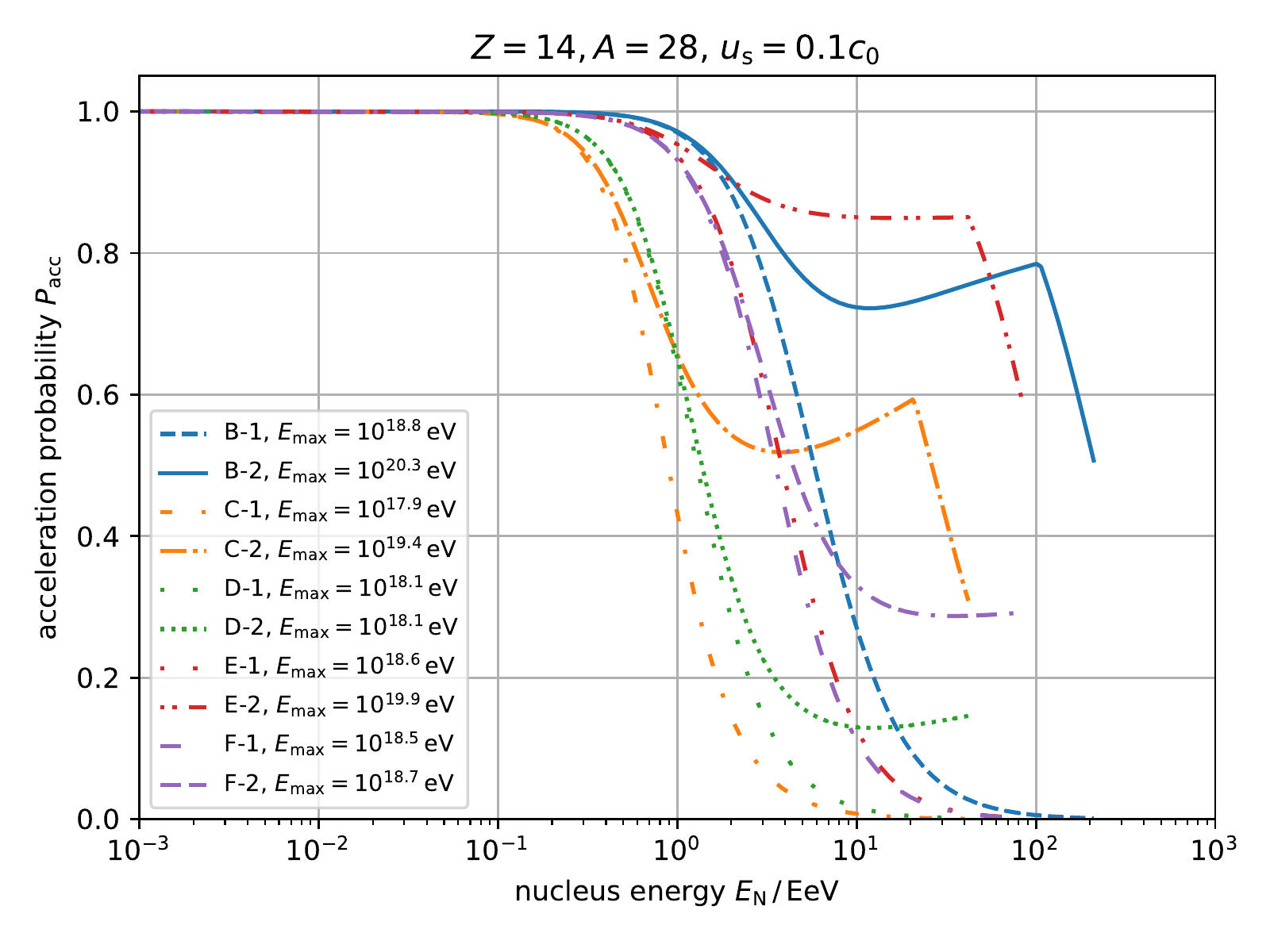}
    \end{minipage}
    \caption{Acceleration probability for hydrogen, helium, nitrogen, and silicon based on Bohm diffusion is shown. Apart from a shift approximately corresponding to the nuclei's charge, the curves are similar to each other. The higher the influence of the photo-disintegration resonances the more pronounced the local minimum is for the acceleration probability; see, e.g., the \emph{model~C-2} for nitrogen.}
    \label{fig:Pacc_all}
\end{figure}

\begin{figure}[h!]
\centering
    \begin{minipage}{.49\textwidth}
        \includegraphics[width=\textwidth]{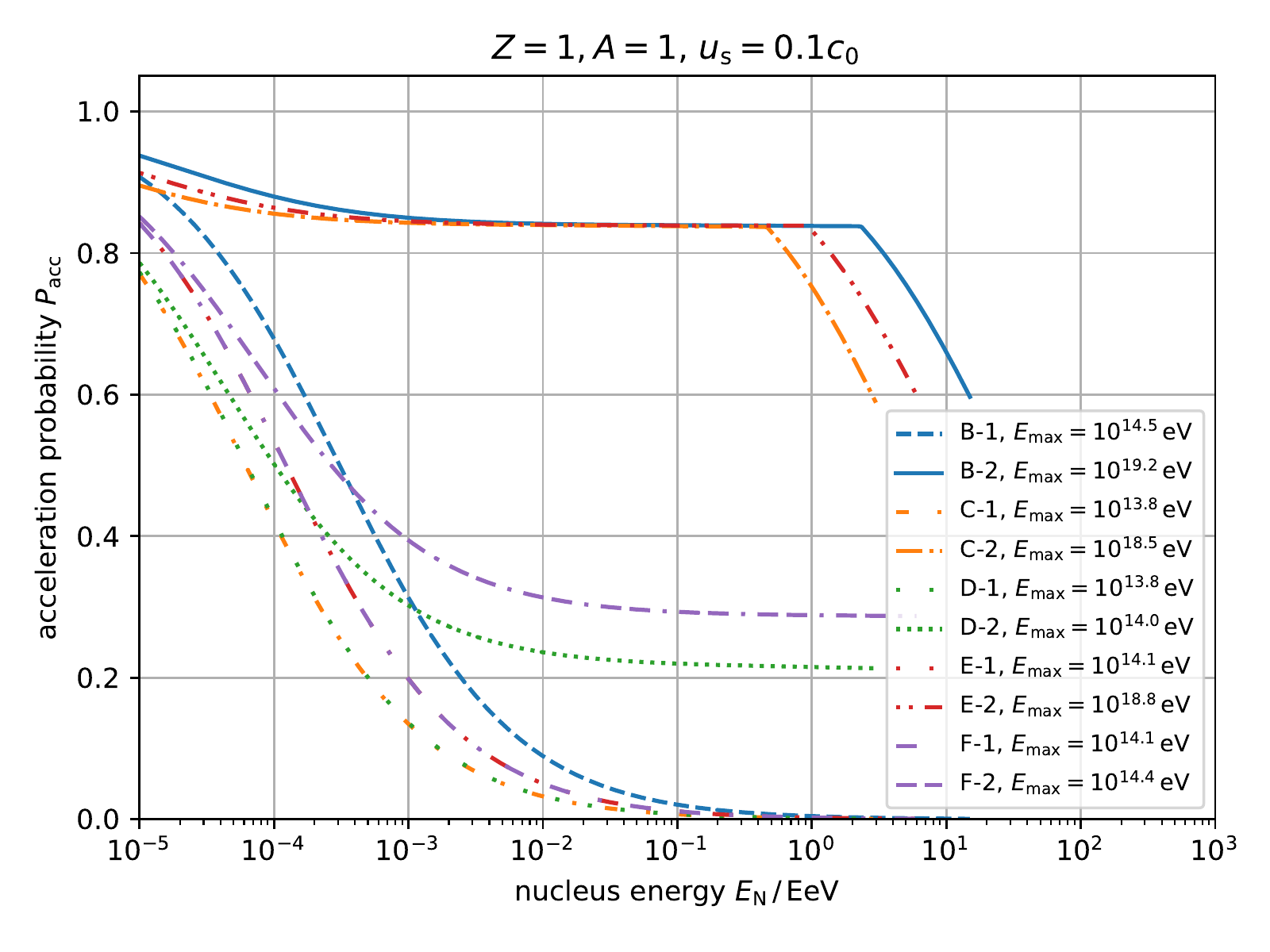}
        \includegraphics[width=\textwidth]{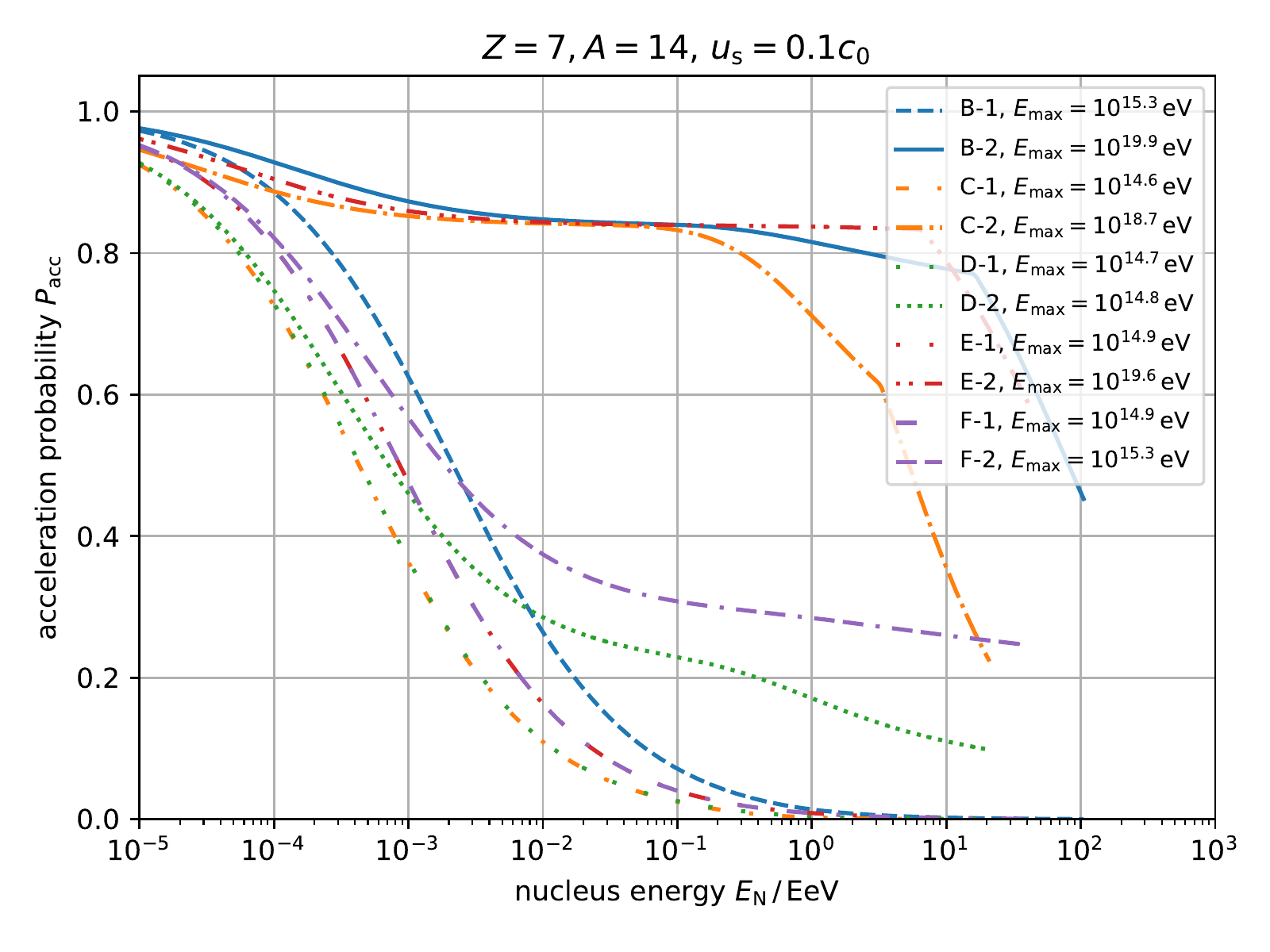}
    \end{minipage}
    \begin{minipage}{.49\textwidth}
        \includegraphics[width=\textwidth]{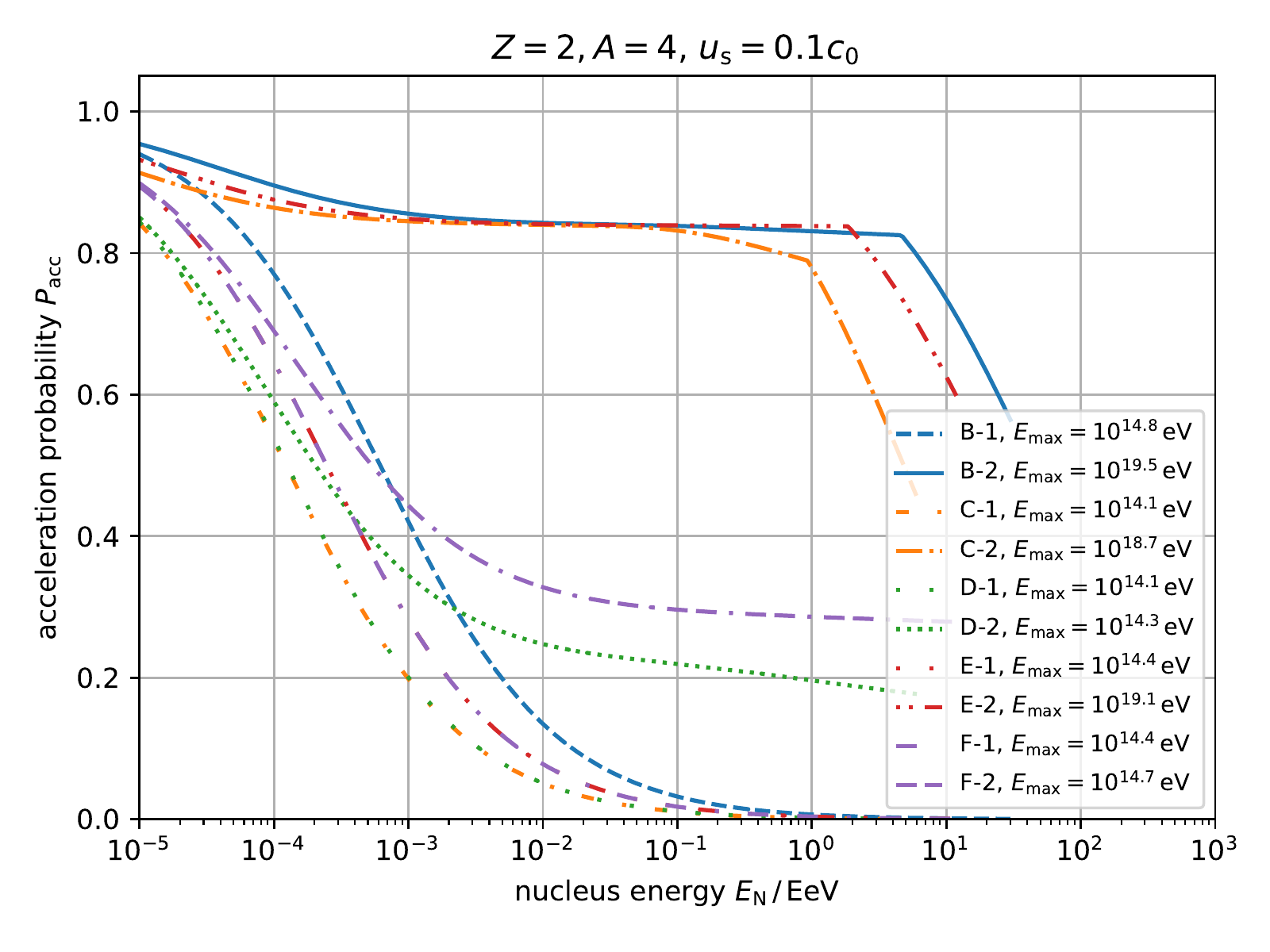}
        \includegraphics[width=\textwidth]{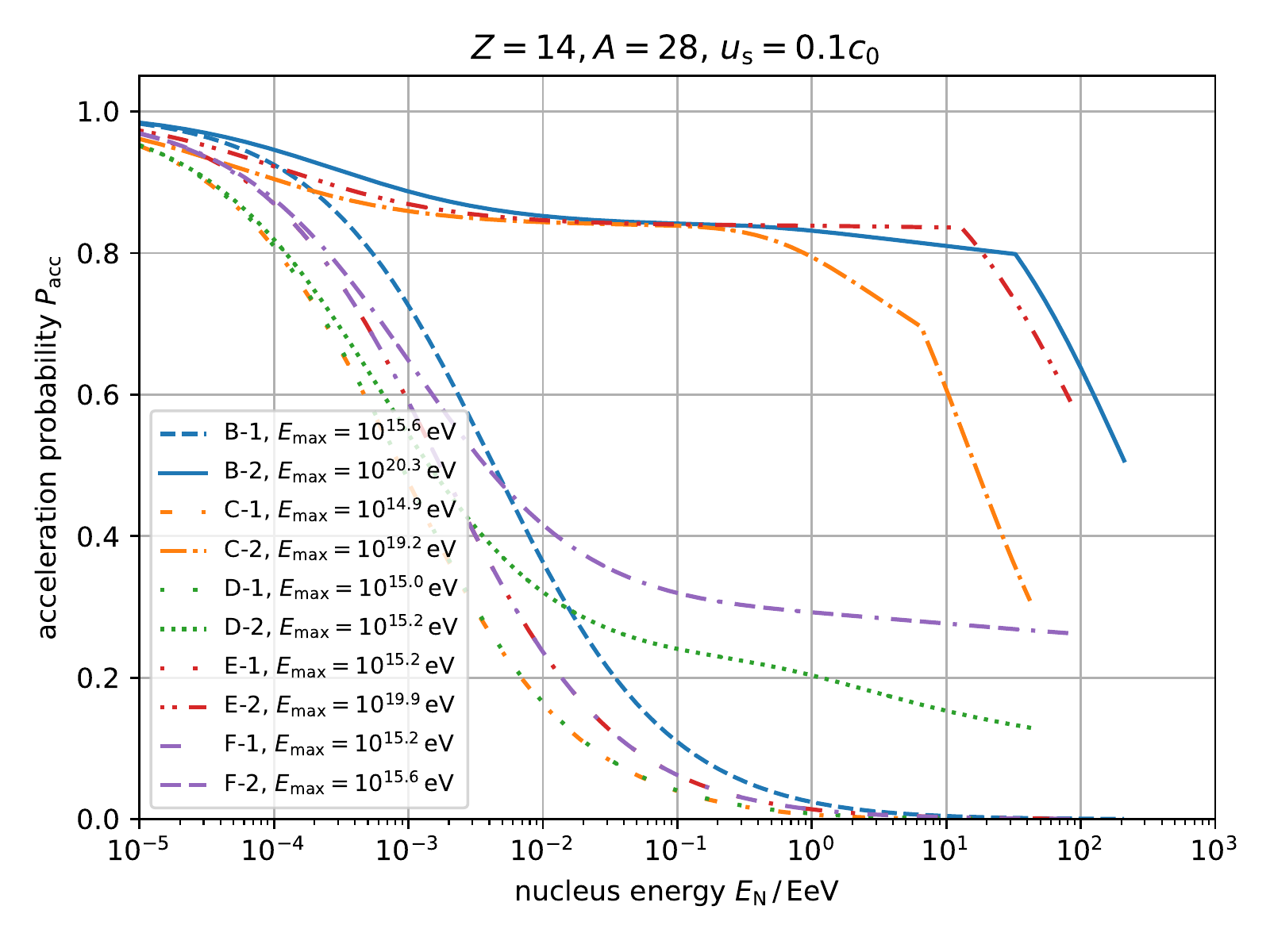}
    \end{minipage}
     \includegraphics[width=.5\textwidth]{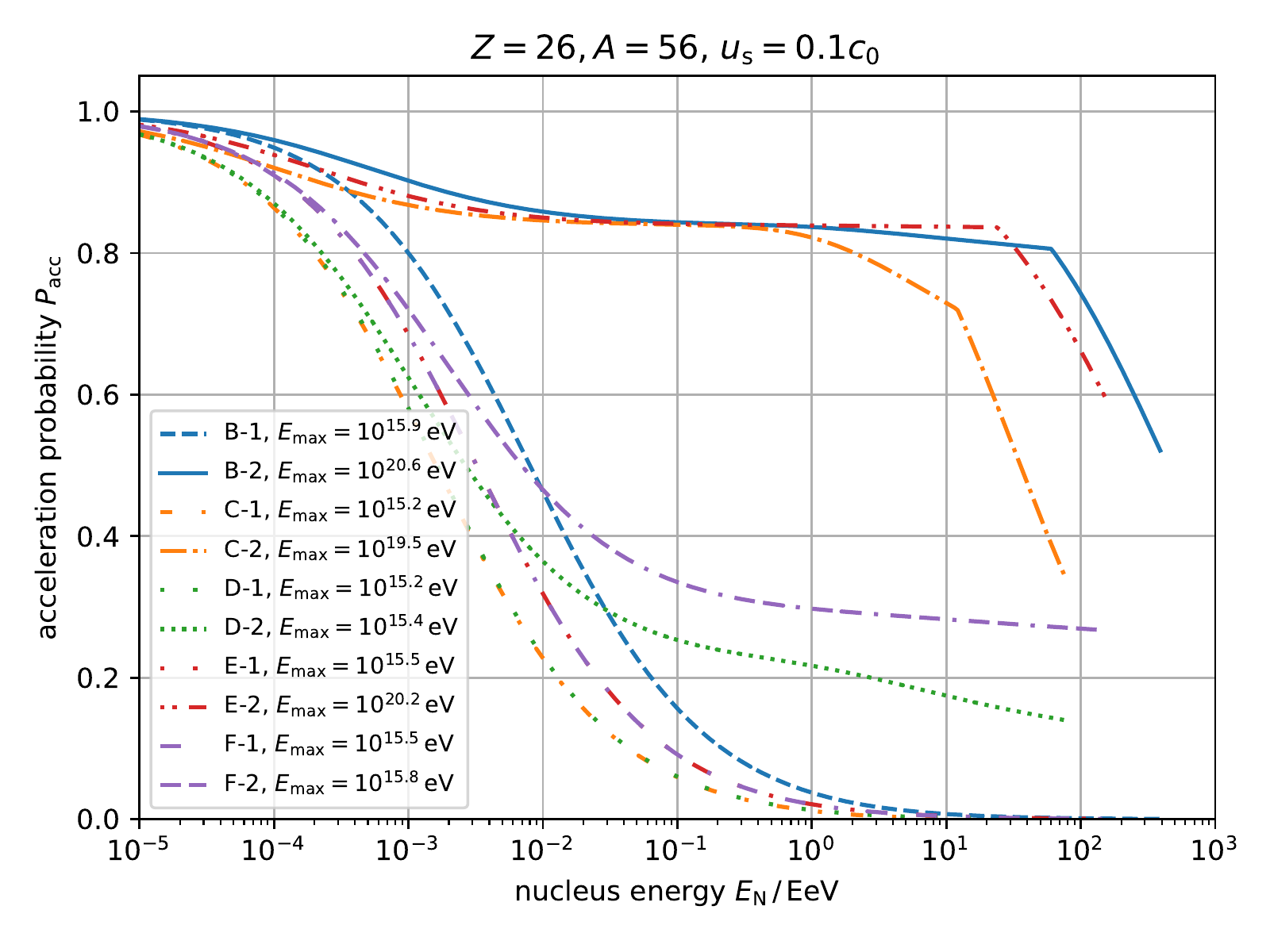}
    \caption{Acceleration probability for hydrogen, helium, nitrogen, silicon, and iron for Kolmogorov diffusion is shown. Compared with the Bohm diffusion approach the acceleration probability is much smoother.}
    \label{fig:Pacc_all_kolm}
\end{figure}

\clearpage
\section{Maximum energy for Kolmogorov diffusion}

\begin{table}[h!]
    \centering
    \caption{Maximum energy in the jet frame. Here the Kolmogorov diffusion scenario ($\alpha=1/3$) is considered where the X-1 models X-1 are based on pure Fermi-I acceleration, and the X-2 models use the hybrid approach. The shock speed was fixed at $u_s=0.1\,c$.}
        \begin{tabular}{lrrrrrr}
              \toprule
            model & \multicolumn{5}{c}{$\log_{10}(E_\mathrm{max}'/\mathrm{eV})$} & \multicolumn{1}{l}{$\langle \log_{10}(\zeta'/ \mathrm{V})\rangle$\tnote{a}} \\
                   & \multicolumn{1}{l}{p} & \multicolumn{1}{l}{He} & \multicolumn{1}{l}{N} & \multicolumn{1}{l}{Si} & \multicolumn{1}{l}{Fe} & \\ 
            \midrule
             model B-1 & $14.5$ & $14.8$ & $15.3$ & $15.6$ & $15.9$ & $14.48\pm0.02$\\
             model C-1 & $13.8$ & $14.1$ & $14.6$ & $14.9$ & $15.2$ & $13.78\pm0.02$\\
             model D-1 & $13.8$ & $14.1$ & $14.7$ & $15.0$ & $15.2$ & $13.82\pm0.03$\\
             model E-1 & $14.1$ & $14.4$ & $14.9$ & $15.2$ & $15.5$ & $14.08\pm0.02$\\
             model F-1 & $14.1$ & $14.4$ & $14.9$ & $15.6$ & $15.5$ & $14.16\pm0.15$\\ 
             \midrule
             model B-2 & $19.2$ & $19.5$ & $19.9$ & $20.3$ & $20.6$ & $19.16\pm0.05$\\
             model C-2 & $18.5$ & $18.7$ & $18.7$ & $19.2$ & $19.5$ & $18.18\pm0.24$\\
             model D-2 & $14.0$ & $14.3$ & $14.9$ & $15.2$ & $15.4$ & $14.01\pm0.03$\\
             model E-2 & $18.8$ & $19.1$ & $19.6$ & $19.9$ & $20.2$ & $18.78\pm0.02$\\
             model F-2 & $14.4$ & $14.7$ & $15.3$ & $15.6$ & $15.8$ & $14.42\pm0.03$\\
             \bottomrule
        \end{tabular}
    \label{tab:Emax_kolm}
\end{table}

\end{document}